\documentclass[12pt]{article}
\usepackage{amsfonts,amsmath,amssymb,epsf,framed}
\usepackage{amsmath,braket}
\usepackage{graphicx,hyperref,color,comment}
\usepackage{cite}

\usepackage{subfigure}

\usepackage{mathrsfs}
\usepackage{calligra}
\usepackage{bm,latexsym,amsthm}

\newcommand{\bcen}{\begin{center}}
\newcommand{\ecen}{\end{center}}
\newcommand{\bflr}{\begin{flushright}}
\newcommand{\eflr}{\end{flushright}}
\newcommand{\bfll}{\begin{flushleft}}
\newcommand{\efll}{\end{flushleft}}
\newcommand{\beq}{\begin{equation}}
\newcommand{\eeq}{\end{equation}}
\newcommand{\beqa}{\begin{eqnarray}}
\newcommand{\eeqa}{\end{eqnarray}}
\newcommand{\bite}{\begin{itemize}}
\newcommand{\eite}{\end{itemize}}
\newcommand{\benu}{\begin{enumerate}}
\newcommand{\eenu}{\end{enumerate}}

\usepackage[normalem]{ulem}

\newcommand{\A}{{\cal A}}

\newcommand{\h}{{\tilde h}}

\newcommand{\Tr}{{\mbox{Tr}}}
\newcommand{\cO}{{\cal O}}
\newcommand{\eq}[1]{(\ref{#1})}
\newcommand{\p}{\partial}

\newcommand{\e}{{e}}

\allowdisplaybreaks

\hypersetup{
    bookmarks=true,         
    unicode=false,          
    pdftoolbar=true,        
    pdfmenubar=true,        
    linktocpage=true,       
    pdffitwindow=false,     
    pdfstartview={FitH},    
    pdfnewwindow=true,      
    colorlinks=true,       
    linkcolor=blue,          
    citecolor=blue,        
    filecolor=blue,      
    urlcolor=blue           
}
\numberwithin{equation}{section}									

\begin{document}

\begin{titlepage}

\thispagestyle{empty}

\vspace*{-2cm}
\begin{flushright}
YITP-22-125
\\
RIKEN-iTHEMS-Report-22 
\\
\end{flushright}

\bigskip


\begin{center}
\noindent{{\Large \textbf{High-energy properties of the graviton scattering in quadratic gravity}}}\\
\vspace{1cm}

\quad 
Yugo Abe$^{a}$, Takeo Inami$^{b,c}$, 
Keisuke Izumi$^{d,e}$ 
\vspace{1cm}\\


{\it $^a$ National Institute of Technology, Miyakonojo College, Miyakonojo 885-8567, Japan}\\
\vspace{1mm}
{\it $^b$ iTHEMS Program, RIKEN, Wako 351-0198, Japan}\\
\vspace{1mm}
{\it $^c$Yukawa Institute for Theoretical Physics, Kyoto University,\\
Kitashirakawa Oiwakecho, Sakyo-ku, Kyoto 606-8502, Japan}\\
\vspace{1mm}
{\it $^d$Kobayashi-Maskawa Institute, Nagoya University, Nagoya 464-8602, Japan}\\
\vspace{1mm}
{\it $^{e}$Department of Mathematics, Nagoya University, Nagoya 464-8602, Japan}\\

\bigskip \bigskip
\vskip 2em
\end{center}

\begin{abstract}
We obtain the matter-graviton scattering amplitude  in the gravitational theory of quadratic curvature, 
which has $R_{\mu\nu}^2$ term in the action. 
Unitarity bound is not satisfied because of the existence of negative norm states, 
while an analog of unitarity bound for $S$-matrix unitarity holds 
due to the cancelation among the positive norm states and negative norm ones in the unitarity summation in the optical theorem. 
The violation of unitarity bound is a counter example of  Llewellyn Smith's conjecture on the relation between tree-level unitarity and
renormalizability. 
We have recently proposed a new conjecture that an analog of the unitarity bound for $S$-matrix unitarity gives the equivalent conditions to those for  renormalizability.
We show that the gravitational theory of quadratic curvature is a nontrivial example consistent with our conjecture. 
\end{abstract}

\end{titlepage}

\newpage

\setcounter{tocdepth}{2}
\tableofcontents


\section{Introduction}

There are a few basic properties which quantum field theories (QFT) are supposed to possess, causality, locality, analyticity, $S$-matrix unitarity. 
The last of these governs the high-energy behavior of scattering amplitudes in renormalizable QFT. 
In particular, more than half a century ago Froissart and Martin used unitarity together with dispersion relation to derive high-energy bound on elastic two-two amplitudes~\cite{Froissart:1961ux,Martin:1962rt}. 
Later the tree-level unitarity bound on two-two amplitudes at high energy called the {\it tree unitarity} was introduced~\cite{Bell:1973ex,Cornwall:1973tb,LlewellynSmith:1973yud,Cornwall:1974km,Fujimori:2015wda,Fujimori:2015mea}. 
Since the unitarity bound makes a relation among the quantities of different orders in the coupling constants, 
the tree unitarity inductively exposes full order properties of the perturbation with respect to the coupling constants~\cite{Cornwall:1974km},
and it is expected to be a very powerful tool in determining whether a given QFT is renormalizable or not.
The tree unitarity fails for QFT with higher derivative kinetic terms, including Lee-Wick type QFT~\cite{Lee:1969fy,Lee:1970iw} and $R_{\mu\nu}^2$ gravity~\cite{Stelle:1976gc,Ostrogradsky:1850fid}, because of negative norm field (ghost).
The tree unitarity can be extended to perturbative $S$-matrix unitarity ($SS^\dagger =1$) for higher-derivative QFT~ \cite{Abe:2018rwb,Abe:2020ikj}. 

The important question then arises whether the same holds true for the scattering amplitude in gravity theories. 
In this paper we address ourselves to this question, and to this end we compute the matter-graviton scattering amplitudes in two gravity theories, Einstein gravity%
\footnote{The amplitudes in Einstein gravity have been known since long time ago~\cite{DeWitt:1967uc,Berends:1974gk}.} 
and $R_{\mu\nu}^2$ gravity~\cite{Stelle:1976gc}, or quadratic gravity. 
The reason why we consider both is that the former is a non-renormalizable theory while the latter is a renormalizable one. 
A part of our results has been reported briefly in~\cite{Abe:2020ikj}. 
In the present paper we obtain the matter-graviton scattering amplitudes in full,
and study how the {\it perturbative $S$-matrix unitarity} is obeyed, even though the {\it tree unitarity} fails, in quadratic gravity. 
Some of the amplitudes have been computed in earlier papers by considering leading terms in the power of Mandelstam's $s$ (CMS energy squared) or for some helicity states~\cite{Holdom:2021hlo,Dona:2015tra}.

It was shown that in gauge theories the unitarity bound at tree-level called the tree unitarity implies similar conditions for renormalizability~\cite{Bell:1973ex,Cornwall:1973tb,LlewellynSmith:1973yud,Cornwall:1974km}. 
A similar situation should hold for the $S$-matrix unitarity.  
This expectation has been verified at tree level in scalar field theories with higher-order derivatives\cite{Abe:2018rwb}. 
Hence we focus on the $S$-matrix unitarity for matter-graviton scattering amplitudes at the tree-level (figure \ref{Fig:cha}) in $R_{\mu\nu}^2$ gravity. 

$S$-matrix unitarity is a key element in many processes involving gravitons; 
some studies of particle radiation from black holes have been made referring to unitarity~\cite{Saini:2015dea,tHooft:1996rdg,Susskind:1993if}. 
It was pointed out that the quadratic curvature changes the fate of black holes~\cite{Holdom:2002xy,Held:2022abx} 
and the spacetime causality~\cite{Izumi:2014loa,Reall:2014pwa,Camanho:2014apa,Reall:2014sla,Papallo:2015rna,Edelstein:2021jyu}.
Unitarity should come into play in graviton scattering at Planckian energies~\cite{Holdom:2021hlo,Dona:2015tra,tHooft:1987vrq,Giddings:2009gj,Pottel:2020iuz}. 
The explicit evaluation of the matter-graviton amplitudes (of all combinations of spin/helicity states including ghost field) in quadratic gravity should show us the basic properties of QFT at Planckian energies,  which are so far unknown to us. 

One intriguing property of the quadratic gravity is that the gravity field $h_{\mu\nu}$ contains massive negative norm field (ghost) in addition to the usual massless graviton field and a scalar component. 
The appearance of ghost field is common to higher-derivative field theories~\cite{Ostrogradsky:1850fid} and it implies negative probability.
This apparent difficulty of the higher-derivative field theory has been investigated repeatedly in relation to unitarity 
(see for instance a review article~\cite{Salvio:2018crh}) since the early works of Lee and Wick~\cite{Lee:1969fy,Lee:1970iw}.
In this paper we are concerned with ghost fields only in connection with the question of how it comes to rescue the perturbative $S$-matrix unitarity.

A recent work~\cite{Donoghue:2019fcb} studies how the ghost field may be treated as an unstable
resonance maintaining the unitarity.
More recently there has been a new development~\cite{Tokuda:2020mlf,Alberte:2020jsk,Noumi:2021uuv}, called gravitational
positivity bound, which aims at relating, by use of unitarity, dispersion relation and
graviton exchange, particle physics at low energies to Planckian energies. The
basic notion and tools are developed earlier in~\cite{Adams:2006sv}. We wish to learn from our
evaluation of the scattering amplitudes in the quadratic gravity some hint on the
question whether this theory may provide the UV completion of quantum gravity.

The matter-matter scattering amplitude in $R_{\mu\nu}^2$ gravity has  earlier been investigated~\cite{Abe:2017abx}; the authors of this paper have shown that
the unitarity bound is satisfied. 
The matter considered there have no negative norm states, and thus, 
the discussion of $S$-matrix unitarity in matter-matter scattering cannot  uncover the issue of  negative norm fields.
In this paper, we analyse the matter-graviton scattering, which is the simplest situation that involves the negative norm gravitons in the initial and/or final states. 

We introduce a scalar field as matter. 
The renormalized action of $R_{\mu\nu}^2$ gravity with a scalar field that we consider is~\cite{Elizalde:1994gv}
\begin{eqnarray}
&&S= S_{g} + S_{m}, \label{action} \\
&&S_{g}=\int d^4x \sqrt{-g}\left( \Lambda + \frac{1}{\kappa^2} R + \alpha R^2 + \beta R_{\mu \nu} ^2 \right), \label{GA}\\
&&S_{m}=\int d^4x \sqrt{-g}\left(-\frac12 g^{\mu \nu} \partial_{\mu}\phi\partial_{\nu}\phi - \frac12 m^2 \phi^2
-\frac{1}{4!} \lambda \phi^4 + \xi \phi^2 R 
\right).
\label{MA}
\end{eqnarray}
We call $S_{g}$ the gravitational action and $S_{m}$  the matter action. 
We analyse scatterings on the  Minkowski background and thus $\Lambda$ is set to vanish. 
Moreover, in tree-level amplitude of matter-graviton scattering, 
the quartic-order term of $\phi$ ({\it i.e.} $\lambda \phi^4$) does not give any contribution.

Our (classical) background is the vacuum in Minkowski spacetime, and graviton $h_{\mu\nu}$ is defined by the deviation of the physical metric $g_{\mu\nu}$ from  Minkowski metric $\eta_{\mu\nu}$,
\begin{eqnarray}
h_{\mu\nu}:=g_{\mu\nu}-\eta_{\mu\nu}.
\label{definitionh}
\end{eqnarray}
We expand the action \eq{action} with respect to $h_{\mu\nu}$ 
and study the quantum field theory on the background metric $\eta_{\mu\nu}$.

 It is appropriate to see at this early stage what kinds of vertices of $h_{\mu\nu}$'s  and $\phi$'s we need to compute the tree amplitudes of matter-graviton scattering. 
From all tree graphs contributing to the scattering in figure \ref{Fig:cha}, we find that we need vertices of only three kinds, $h_{\mu\nu}\phi^2$, $h_{\mu\nu}^3$, $h_{\mu\nu}^2\phi^2$. 

The computation of the gravitation scattering is bound to be quite involved and lengthy, particularly so for the case of quadratic gravity. To avoid reading the lengthy calculations on the readers’ side, much of the lengthy calculations is relegated to a few appendices, Appendices~\ref{PEV}, \ref{SUPPLE}, \ref{AppSA} and \ref{AppUV}. Appendix~\ref{mq} is a minimal necessary account of the canonical quantization of the massive gravity. In the text only the results of the computation are reported. 

This paper is organized as follows.
In section \ref{2nd}, the canonical quantization is done. 
We see that the asymptotic degrees of freedom of graviton are decomposed into three parts, 
massless gravitons, a massive scalar graviton and massive gravitons. 
The first two have positive norm, while the last is negative norm excitation, that is ghost modes. 
In section \ref{int}, we show the propagators and vertex functions, which are required in the calculation of scattering amplitude. 
In section \ref{SA}, we give the derivation of scattering amplitude. 
In section \ref{SAUV}, we show the UV limit of scattering amplitude. 
We confirm that perturbative $S$-matrix unitarity is satisfied in the quadratic gravity, in section \ref{SMU}.
Section~\ref{Summary} is devoted to a summary and discussion.  
Appendix~\ref{mq} is given for the detailed derivation of canonical quantization of massive graviton.
In Appendix~\ref{PEV}, we give a calculation technic in gravitational perturbation.
Appendices~\ref{SUPPLE}, \ref{AppSA} and \ref{AppUV} are given to show the detailed calculations.

We use the following notation. 
$R_{\mu\nu}^2$ represents $R_{\mu\nu}R^{\mu\nu}$ and tensor squared appearing in this paper means similar.
The symmetrization of indices is expressed as 
\begin{eqnarray}
A^{\mu (\alpha_1 \alpha_2| \nu| \alpha_3| \lambda| \alpha_4\alpha_5\dots \alpha_n)}
:=\frac1{n!} \sum_{\sigma} A^{\mu \alpha_{\sigma_1} \alpha_{\sigma_2} \nu \alpha_{\sigma_3} \lambda \alpha_{\sigma_4}\alpha_{\sigma_5}\dots \alpha_{\sigma_n}},
\end{eqnarray}
where the sum is computed over all permutations $\sigma=\{\sigma_1,\cdots,\sigma_n\}$ of the set $\{1,\cdots, n\}$.


\section{Canonical quantization}\label{2nd}

Two alternative ways of quantizing higher derivative gravity have been given in the past, the path integral method~\cite{Stelle:1976gc} and the canonical quantization~\cite{Kawasaki:1981zw}. The latter method seems more suitable for our purpose of computing scattering amplitudes and of dealing with negative norm states. We recapitulate the minimal account of canonical quantization for this purpose, particularly introducing the creation and annihilation operators. 

For the scalar field, it is done by analysing the quadratic parts of the scalar field action \eq{action},
\begin{eqnarray}
S_{m}^{(2)}=  \frac12 \int d^4 x    \left[ -  \eta^{\mu\nu} \partial_\mu \phi \partial_\nu \phi - m^2  \phi^2  \right]. 
\label{MA2}
\end{eqnarray}
The canonical  quantization is done in the usual way,
\begin{eqnarray}
\phi = \int \frac{d^3p}{\sqrt{(2\pi)^3 2 p_0}} \left\{ a_\phi({\bm p}) e^{-ipx}+a_\phi^\dagger({\bm p}) e^{ipx} \right\},  \hspace{5mm} \left( p_0:=\sqrt{{\bm p}^2 +m^2}\right), 
\end{eqnarray} 
where $a_\phi({\bm p})$ and $a_\phi^\dagger({\bm p})$ are the creation and anihilation operators, respectively. 

The second-order action for graviton is obtained by the expansion of the gravitational action \eq{GA} with respect to  $h_{\mu\nu}$, 
\begin{eqnarray}
S_{g2}=
\frac14 \int d^4 x \, \left[
\frac1{\kappa^2} h_{\mu\nu} {\cal L}^{\mu\nu,\alpha\beta} h_{\alpha\beta}
+\alpha \left( \eta_{\mu\nu} {\cal L}^{\mu\nu,\alpha\beta} h_{\alpha\beta}\right)^2
+\beta \left(  {\cal L}^{\mu\nu,\alpha\beta} h_{\alpha\beta}\right)^2
\right],
\label{ga1}
\end{eqnarray}
where ${\cal L}^{\mu\nu,\alpha\beta}$ is a differential operator defined as
\begin{eqnarray}
{\cal L}^{\mu\nu,\alpha\beta}:= 
\Box \eta^{\mu(\alpha}\eta^{\beta)\nu} 
- \p^\mu \p^{(\alpha}\eta^{\beta)\nu} 
- \p^\nu \p^{(\alpha}\eta^{\beta)\mu} 
-\Box \eta^{\mu\nu}\eta^{\alpha\beta}
+ \p^\mu \p^\nu\eta^{\alpha\beta} 
+ \eta^{\mu\nu}\p^\alpha \p^\beta.
\end{eqnarray}
The last two terms of Eq.\eq{ga1} have the higher-order derivatives in time, which give negative norm states~\cite{Ostrogradsky:1850fid}. 
This type of theory can be quantized directly, 
but it involves many constraints, which makes the analysis complicated. 
An easy way of quantizing such a theory is to transform
action \eq{ga1} into that with  second-order derivatives by introducing Lagrange multiplier $\lambda_{\mu\nu}$,
\begin{eqnarray}
S_{g2}=
\frac14 \int d^4 x \, \left[
\frac1{\kappa^2} h^{\mu\nu}U_{\mu\nu}
+\alpha U^2 
+\beta U_{\mu\nu}^2
+ \lambda_{\mu\nu} \left( U^{\mu\nu} - {\cal L}^{\mu\nu,\alpha\beta}h_{\alpha\beta} \right)
\right], \label{ga2}
\end{eqnarray}
where 
$U=U^\mu{}_\mu$.
The variation of this action with respect to $\lambda_{\mu\nu}$ gives 
\begin{eqnarray}
U^{\mu\nu} - {\cal L}^{\mu\nu,\alpha\beta}h_{\alpha\beta}=0,
\end{eqnarray}
and substituting back this into action \eq{ga2}, 
we go back to the original action \eq{ga1} . 

Any tensor $A_{\mu\nu}$ can be decomposed into the trace part and traceless part by
%
\begin{eqnarray}
A_{\mu\nu}= \frac14 A \eta_{\mu\nu} + \bar A_{\mu\nu}.
\end{eqnarray}
The traceless part $\bar A_{\mu\nu}$, satisfying $\bar A_{\mu}{}^\mu=0$,
 is obtained by operating the projection tensor 
\begin{eqnarray}
\bar G_{\mu\nu,\alpha\beta} = \eta_{\mu(\alpha}\eta_{\beta)\nu}- \frac14 \eta_{\mu\nu}\eta_{\alpha\beta},
\end{eqnarray}
that is 
\begin{eqnarray}
\bar A_{\mu\nu}=\bar G_{\mu\nu,\alpha\beta}A^{\alpha\beta}.
\end{eqnarray}
Then, the action \eq{ga2} can be written as
\begin{eqnarray}
&&S_{g2}=
\frac14 \int d^4 x \, \left[
\beta \bar U_{\mu\nu}{}^2
+\frac1{\kappa^2} \bar h_{\mu\nu} \bar U^{\mu\nu}
+\bar\lambda_{\mu\nu} \bar U^{\mu\nu}  \right.
\nonumber\\
&& \hspace{40mm}\left.
+\left(\alpha+ \frac\beta4\right) U^2 
+\frac1{4\kappa^2}hU
+ \frac14 \lambda U 
- \lambda_{\mu\nu}  {\cal L}^{\mu\nu,\alpha\beta}h_{\alpha\beta}  \right]
\label{ga3}
\end{eqnarray}
The constraint equations on $U$ and $\bar U_{\mu\nu}$ follow from the variation of action \eq{ga3} 
with respect to $U$ and $\bar U_{\mu\nu}$.
Substituting back these constraint equations into action \eq{ga3}, we have 
\begin{eqnarray}
S_{g2}=
\frac14 \int d^4 x \, \left[
- \frac1{4\beta} \left(\frac1{\kappa^2} \bar h_{\mu\nu}+\bar \lambda_{\mu\nu} \right)^2 \!\!
-\frac1{64\alpha+16\beta}\left(\frac1{\kappa^2}h+\lambda\right)^2 \!\!
- \lambda_{\mu\nu}  {\cal L}^{\mu\nu,\alpha\beta}h_{\alpha\beta}  \right] \! .\label{ga4}
\end{eqnarray}
The field redefinition 
\begin{eqnarray}
&&h_{\mu\nu} = H_{\mu\nu} + I_{\mu\nu} \label{h=H+I}\\
&&\lambda_{\mu\nu}= -\frac1{\kappa^2}H_{\mu\nu}+\frac1{\kappa^2}I_{\mu\nu}
\end{eqnarray}
makes the action \eq{ga4} diagonal in $H_{\mu\nu}$ and $I_{\mu\nu}$,
\begin{eqnarray}
S_{g2}=
\frac1{4\kappa^2} \int d^4 x \, \left[
H_{\mu\nu} {\cal L}^{\mu\nu,\alpha\beta}H_{\alpha\beta} 
-\left( I_{\mu\nu} {\cal L}^{\mu\nu,\alpha\beta}I_{\alpha\beta}
+ \frac1{\beta \kappa^2} \left( I_{\mu\nu}^2- \frac{\alpha}{4\alpha+\beta}I^2\right) 
\right) 
\right],
\label{ga5}
\end{eqnarray}
where $I:= I^\mu{}_\mu$.
Since the quadratic action \eq{ga5} is composed of diagonalized gravitons with the second-order derivatives,
we quantize these fields in the usual manner.

The form of the second-order action \eq{ga5} for $H_{\mu\nu}$ is the  same as that for the linearized graviton in Einstein gravity, 
that is a masslass spin-2 mode,
and thus the same asymptotic states as those in Einstein gravity are obtained, 
\begin{eqnarray}
H_{\mu\nu} = \sum_{\sigma} \int {d^3p}\frac{\kappa}{\sqrt{(2\pi)^3 p_0}} \left\{ a_H^{(\sigma)}({\bm p}) e^{-ipx}+a_H^{(\sigma)}{}^\dagger({\bm p}) e^{ipx} \right\} e_{\mu\nu}^{(\sigma)}({\bm p}) , \label{2acH} 
\end{eqnarray} 
where $p_0:=\sqrt{{\bm p}^2}$ and  the sum is computed over all elements of a basis of asymptotic states. 
The same as in Einstein gravity, the number of independent components for massless graviton $H_{\mu\nu}$ is two.
$\sigma$ in Eq.\eqref{2acH} represents two helicity states.

There are a few different conventions of defining graviton polarization. 
Berends and Gastmans  use $\sigma=+2,  -2$ as basis elements~\cite{Berends:1974gk}. 
We use those constructed from the vector elements~\cite{Abe:2020ikj,Aubert:2003je}, given in Eqs.\eq{tildee} of Appendix~\ref{mq}.

The second-order action \eq{ga5}  for $I_{\mu\nu}$ is that for massive graviton~\cite{Fierz:1939ix} but with the negative sign. 
The mass term does not have Fierz-Pauli form unless $3\alpha+\beta\neq 0$. 
Therefore,  this massive graviton generically includes not only spin-$2$ but also spin-$0$ (scalar) degrees of freedom. 
Because of the over all negative sign of the action, the massive spin-$2$ graviton gives negative norm states, 
while the positive-norm spin-$0$ graviton appears. 
The way of quantization is shown in Appendix \ref{mq}, and the result is
\begin{eqnarray}
&&I_{\mu\nu} = \sum_{\sigma} \int d^3p \frac{\kappa}{\sqrt{(2\pi)^3p_0}} \left\{ a_T^{(\sigma)} ({\bm p}) e^{-ipx}+a_T^{(\sigma)}{}^\dagger ({\bm p}) e^{ipx} \right\} e^{(\sigma)}_{\mu\nu} \nonumber \\
&& \hspace{25mm} +
 \int d^3p \frac{\kappa}{\sqrt{(2\pi)^3p_0}} \left\{ a_S ({\bm p}) e^{-ipx}+a_S^\dagger ({\bm p}) e^{ipx} \right\} \frac1{\sqrt{3}} \theta_{\mu\nu},
 \label{2acI}
\end{eqnarray} 
where
\begin{eqnarray}
\theta_{\mu\nu} := \eta_{\mu\nu}+  \hat p_\mu  \hat p_\nu, \qquad  \hat p_\mu := \frac{p_\mu}{\sqrt {\left| p^2 \right|}}.
\label{deftheta}
\end{eqnarray} 
%
The polarization basis $e^{(\sigma)}_{\mu\nu}$ appearing in Eq.~\eqref{2acH} and that in Eq.~\eqref{2acI} are the same except that the mass in the former is set to zero while that in the latter nonzero. This observation makes our computation of a few different amplitudes substantially easier. 

The commutation relations for annihilation and creation operators are
\begin{eqnarray}
&&[a_\phi ({\bm k}) , a_\phi^\dagger ({\bm p})] = \delta^3 ({\bm k}-{\bm p}), \qquad
[a_H^{(\sigma)} ({\bm k}) , a_H^{(\tau)}{}^\dagger ({\bm p})] = \delta_{\sigma\tau} \delta^3 \bigl({\bm k}-{\bm p}\bigr) ,
\nonumber \\ 
&&[a_T^{(\sigma)} ({\bm k}) , a_T^{(\tau)}{}^\dagger ({\bm p})] = - \delta_{\sigma\tau} \delta^3 \bigl({\bm k}-{\bm p}\bigr) , \qquad
[a_S ({\bm k}) , a_S^\dagger ({\bm p})] = \delta^3 ({\bm k}-{\bm p}).
\end{eqnarray} 
The negative sign in the third commutation relation means that the particle states of $a_{T}^{(\sigma)} ({\bm p})$ lead to negative norms.


\section{Propagators and Vertex functions}\label{int}

\subsection{Propagators}

\begin{figure}[tb]
  \begin{center}
    \includegraphics[clip,width=3.0cm]{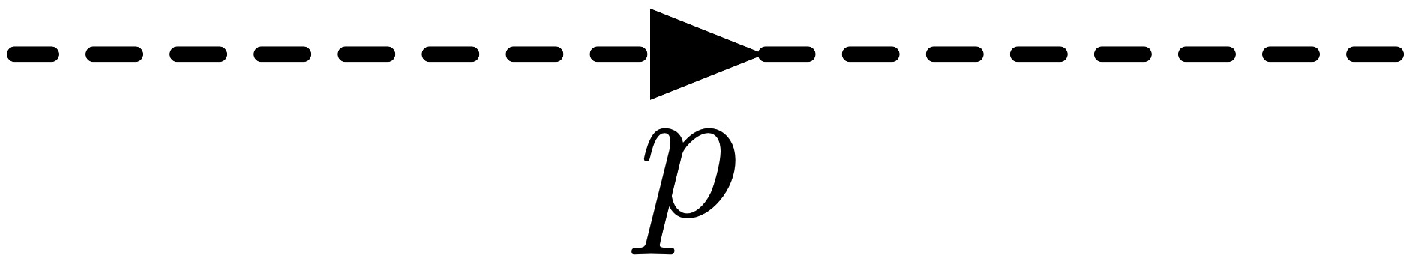}
    \hspace{10mm}
    \includegraphics[clip,width=3.0cm]{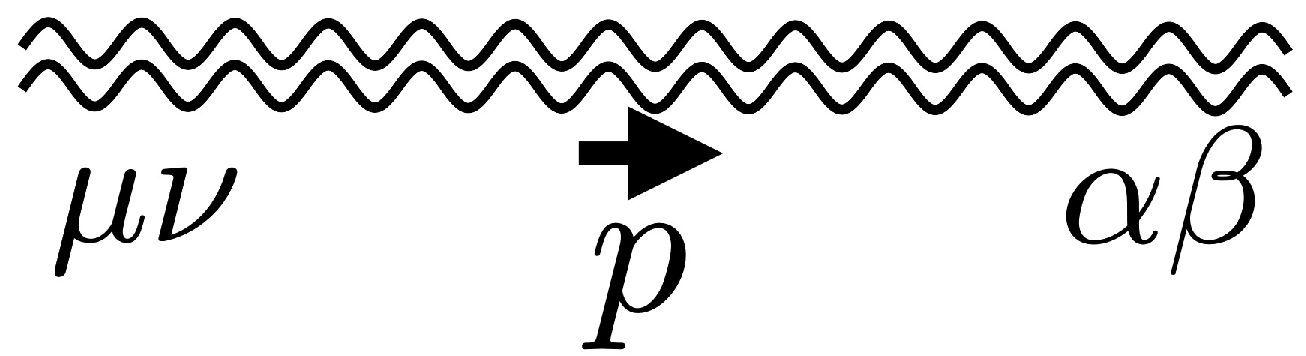}
    \caption{Scalar (left) and graviton (right) propagators.}
   \label{Fig:pro}
  \end{center}
\end{figure}

The propagators for the scalar field $\phi$ and graviton $h_{\mu\nu}$ are obtained  from the quadratic actions \eq{MA2} and \eq{ga1}. 
For scalar field, it becomes 
\begin{eqnarray}
G_\phi =  \frac{-i}{p^2+m^2}.
\end{eqnarray}
The quadratic action for graviton \eq{ga1} is written as 
\begin{eqnarray}
S_{g2} = \int d^4x \,
h_{\alpha\beta} \left[
\frac14 \left( \frac1{\kappa^2} \Box + \beta \Box^2 \right) P^{(2)\, \alpha \beta, \mu \nu}
+  \left( -\frac1{2\kappa^2} \Box +  (3\alpha+\beta) \Box^2 \right) P^{(0)\, \alpha \beta, \mu \nu}
\right]
h_{\mu\nu}, \nonumber \\
\label{G2}
\end{eqnarray}
where $P^{(2)}_{\alpha \beta, \mu \nu}$ is the projection to the transverse-traceless component,
and $P^{(0)}_{\alpha \beta, \mu \nu}$ to the transverse-trace part, defined respectively by%
\footnote{
To be precise, the projection tensors are defined in Fourier space.
}
\begin{eqnarray}
&&P^{(2)}_{\alpha \beta, \mu \nu}:= \frac12\left( \theta_{\alpha \mu}\theta_{\beta\nu}+\theta_{\alpha \nu}\theta_{\beta\mu}\right)-\frac13 \theta_{\alpha \beta}\theta_{\mu\nu} , 
\label{propa2}
\\
&&P^{(0)}_{\alpha \beta, \mu \nu}:= \frac13 \theta_{\alpha \beta}\theta_{\mu\nu} .
\label{propa0}
\end{eqnarray}
Thus, the quadratic action for graviton is decomposed into the spin-2 (transverse-traceless)
and the spin-0 (transverse-trace) parts. 
The degrees of freedom other than the transverse ones are gauge modes. 
Theories with gauge degrees of freedom may be dealt with by 
Becchi-Rouet-Stora-Tyutin (BRST) method, where gauge and Faddeev-Popov (FP) ghost modes should be taken into account.
However, in the tree-level approximation that we take in this paper, gauge modes do not appear even in the internal lines. 
Hence, it is enough to derive the propagators and vertex functions only for transverse modes. 
Therefore, we take the harmonic gauge~\cite{Stelle:1976gc}
\begin{eqnarray}
\partial^\mu h_{\mu\nu}=0. \label{TV}
\end{eqnarray}
The graviton propagator is obtained by taking the inverse of each part in  the quadratic action for graviton \eq{G2}, 
\begin{eqnarray}
G_{\alpha\beta,\mu\nu}=\frac{2i}{\beta p^2 \left(p^2 +m_I^2\right)}P^{(2)}_{\alpha \beta, \mu \nu} 
+ \frac{i}{2(3\alpha+\beta)p^2 \left( p^2+ m_S^2\right) } P^{(0)}_{\alpha \beta, \mu \nu}. 
\label{gpro}
\end{eqnarray}
where $m_I$ and $m_S$ are masses of  the massive spin-2 and spin-0 modes, respectively, satisfying%
\footnote{
To avoid tachyonic instability for massive graviton $I_{\mu\nu}$, 
we might be required to impose $\beta\le 0$  and $3\alpha + \beta \ge0$.
}
\begin{eqnarray}
m_I^2= - (\beta \kappa^2)^{-1}, \hspace{20mm} m_S^2= \left( 2\kappa^2 (3\alpha +\beta)\right) ^{-1}.
\end{eqnarray}
The Feynman diagrams for propagators are shown in figure \ref{Fig:pro}.

\subsection{Vertex functions}

Vertex functions are obtained by further expanding the total action \eq{action} with respect to $h_{\mu\nu}$. 
As we have commented, imposing the transversality condition for graviton $h_{\mu\nu}$, Eq.\eq{TV} does not 
affect  the result in the tree-level approximation. 
Therefore, hereinafter the transversality condition~\eq{TV} is used without notice. 
The matter-graviton scattering is expessed by the Feynman diagrams with two $\phi$- and two $h_{\mu\nu}$-external lines, 
and then, only linear and quadratic orders for $h_{\mu\nu}$ of the matter action, and  cubic order of gravitational action are required. 
(See the Feynman diagrams appearing later in figure \ref{Fig:cha}.)

\begin{figure}[tb]
  \begin{center}
    \includegraphics[clip,width=3.6cm]{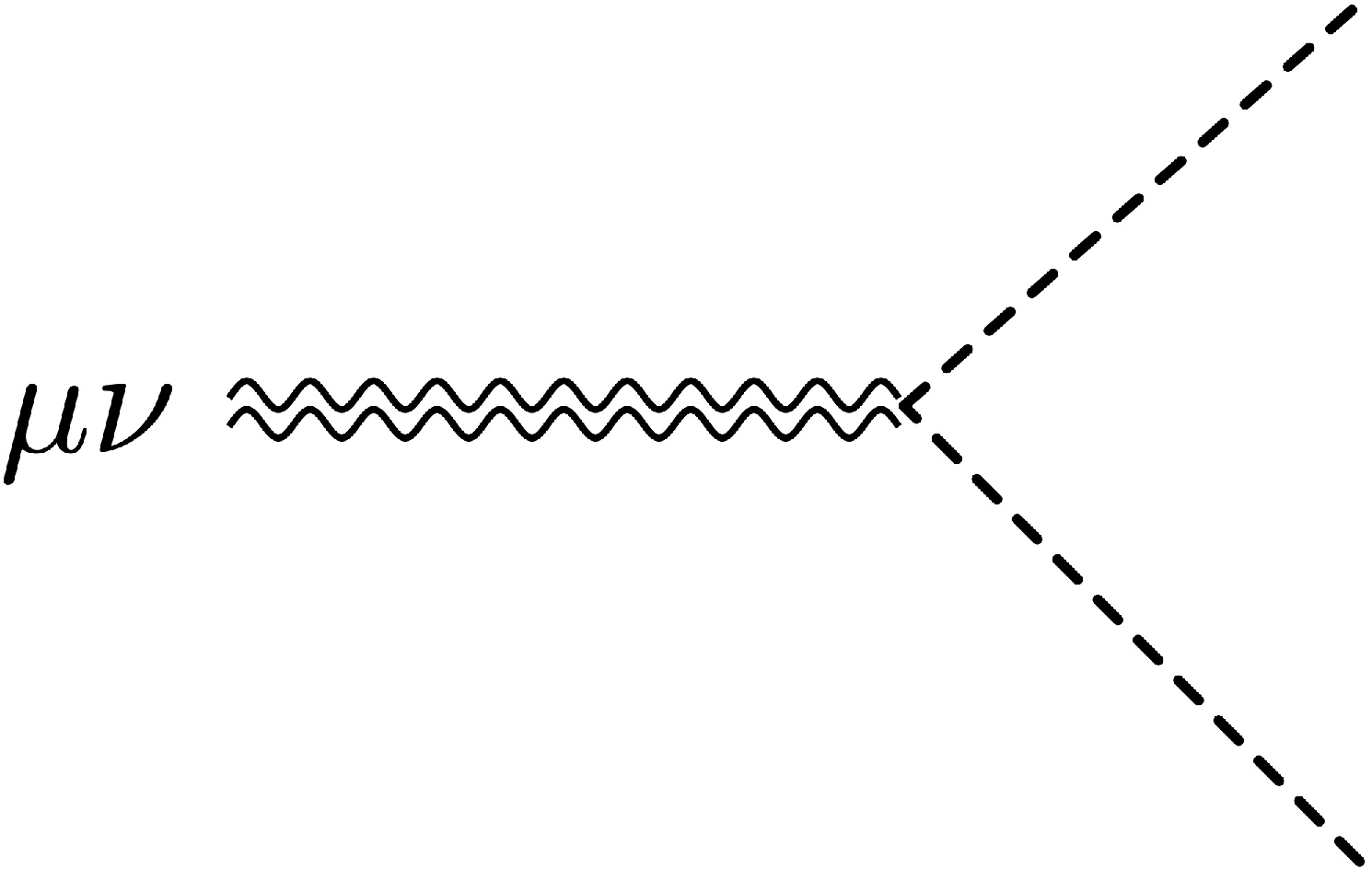}
    \hspace{5mm}
    \includegraphics[clip,width=3.0cm]{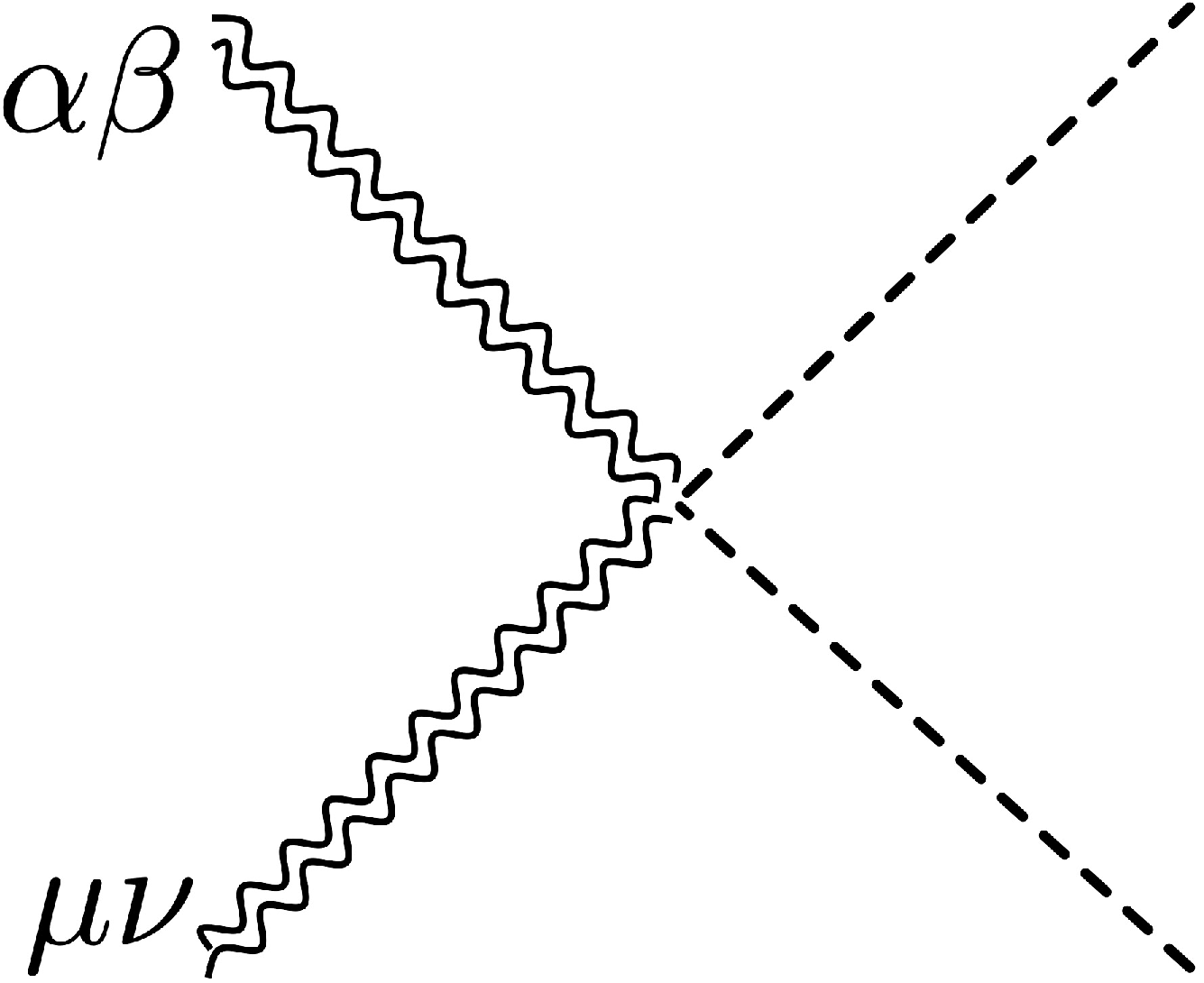}
    \hspace{5mm}
    \includegraphics[clip,width=3.6cm]{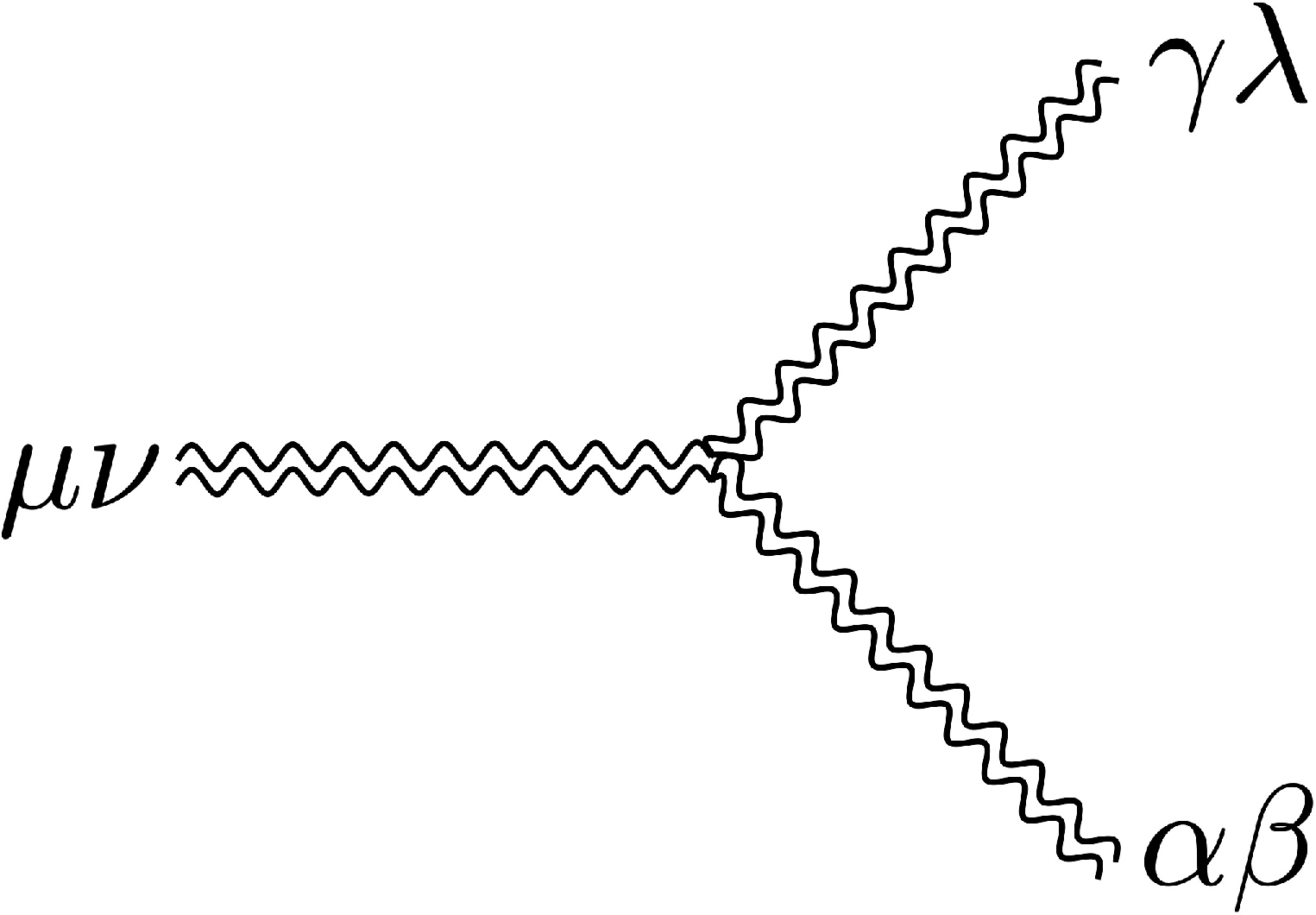}
  \caption{Graviton-matter three-point (left) and the graviton-matter four-point (middle), graviton three-point (right) vertex functions: 
  The double wiggly lines corresponds to graviton $h_{\mu\nu}$, while the scalar field $\phi$ is described by dotted lines.
   }
    \label{Fig:ver}
  \end{center}
\end{figure}

Let us further expand the matter action \eq{MA} in powers of $h_{\mu\nu}$,
\begin{eqnarray}
&&S_{m}
= \int d^4 x    \left\{ \frac12 \left[ -  \eta^{\mu\nu} \partial_\mu \phi \partial_\nu \phi - m^2  \phi^2  \right]
\right. 
\nonumber \\ &&  \hspace{10mm}
 + \frac14 h   \left[ -  \eta^{\mu\nu} \partial_\mu \phi \partial_\nu \phi - m^2  \phi^2  \right] 
+ \frac12 h^{\mu\nu} \partial_\mu \phi \partial_\nu \phi
- \xi \phi^2  \Box  h
\nonumber \\ &&  \hspace{10mm}
+ \left( \frac1{16} h^2 - \frac18 h^{\alpha\beta}h_{\alpha\beta} \right) \left[ -  \eta^{\mu\nu} \partial_\mu \phi \partial_\nu \phi - m^2  \phi^2  \right]
+\frac14 h h^{\mu\nu} \partial_\mu \phi \partial_\nu \phi 
- \frac12 h^{\mu\alpha} h_\alpha{}^\nu \partial_\mu \phi \partial_\nu \phi
\nonumber \\
&& \hspace{10mm}
+\xi \phi^2 \left( -\frac12  h \Box  h - \frac14 \left( \p_\mu  h \right) \left( \p^\mu  h \right) 
+  h^{\mu\nu} \p_\mu\p_\nu  h +  h^{\mu\nu} \Box  h_{\mu\nu} 
\right.
\nonumber \\
&&\left. \left.  \hspace{40mm}
+ \frac34 \left(\p_\alpha  h_{\mu\nu}\right) \left(\p^\alpha  h^{\mu\nu}\right) 
-\frac12 \left(\p_\alpha  h_{\mu\beta}\right) \left(\p^\beta  h^{\mu\alpha}\right) 
\right)\right\} 
+ \cO\bigl(\phi^2 h^3 \bigr).
\end{eqnarray}
The second line in the above equation gives the 3-point vertex functions for $\phi(p_1)\phi(p_2) h_{\mu\nu}(p_3)$.
We decompose it into $\xi$-independent and -dependent parts, 
\begin{eqnarray}
&&\lambda_{3}^{\mu\nu} = \tilde \lambda_{3}^{\mu\nu} + \xi \tilde \lambda_{3,\xi}^{\mu\nu},\label{sgvf3} \\
&&\tilde \lambda_{3}^{\mu\nu} =\frac 12 \left[ \left( p_{1\alpha}p_2^\alpha -m^2 \right) \eta^{\mu\nu} - p_1^\mu p_2^\nu -p_1^\nu p_2^\mu \right],
\\
&&\tilde \lambda_{3,\xi}^{\mu\nu} = 2 p_3^2 \eta^{\mu\nu}.
\end{eqnarray}
Similarly, the 4-point vertex functions for $\phi(p_1)\phi(p_2) h_{\mu\nu}(p_3)h_{\alpha\beta}(p_4)$ become
\begin{eqnarray}
&&\lambda_{4}^{\mu\nu,\alpha\beta} =  \tilde \lambda_{4}^{\mu\nu,\alpha\beta} + \xi \tilde \lambda_{4,\xi}^{\mu\nu,\alpha\beta}, \label{sgvf4} \\
&&\tilde \lambda_{4}^{\mu\nu,\alpha\beta} =\frac 14 \left[ \left( p_{1\gamma}p_2^\gamma -m^2 \right)\left( \eta^{\mu\nu}\eta^{\alpha\beta}-\eta^{\mu\alpha}\eta^{\nu\beta}-\eta^{\mu\beta}\eta^{\nu\alpha}\right)\right] 
\nonumber \\ &&  \hspace{15mm}
+ \frac14  \left[-\left(p_1^\mu p_2^\nu +p_1^\nu p_2^\mu \right) \eta^{\alpha\beta}
-\left(p_1^\alpha p_2^\beta +p_1^\beta p_2^\alpha \right) \eta^{\mu\nu} 
+\left(p_1^\mu p_2^\alpha +p_1^\alpha p_2^\mu \right) \eta^{\nu\beta}
\right. \nonumber \\ && \hspace{20mm} \left.
+\left(p_1^\mu p_2^\beta +p_1^\beta p_2^\mu \right) \eta^{\nu\alpha}
+\left(p_1^\nu p_2^\alpha +p_1^\alpha p_2^\nu \right) \eta^{\mu\beta}
+\left(p_1^\nu p_2^\beta +p_1^\beta p_2^\nu \right) \eta^{\mu\alpha} \right], 
\\
&&\tilde \lambda_{4,\xi}^{\mu\nu,\alpha\beta} =  \left(p_3^2+p_4^2 + p_3\cdot p_4\right) \eta^{\mu\nu}\eta^{\alpha\beta}
-\left(2p_3^2+2p_4^2 +3 p_3\cdot p_4\right) \eta^{\mu\alpha}\eta^{\nu\beta} 
\nonumber \\
&&\hspace{20mm}
-2 p_4^\mu p_4^\nu \eta^{\alpha\beta} -2 p_3^\alpha p_3^\beta \eta^{\mu\nu} 
+ \eta^{\mu(\alpha} p_3^{\beta)} p_4^{\nu}  +  \eta^{\nu(\alpha} p_3^{\beta)} p_4^{\mu} .
\end{eqnarray}
The corresponding Feynman diagrams are shown in figure \ref{Fig:ver}.

Let us move to the graviton three-point vertex function corresponding to the rightmost graph in figure \ref{Fig:ver}. 
It is obtained from the gravitational action \eq{GA} but it involves a lengthy calculation. 
The derivation is given in Appendix~\ref{PEV} and we show here only the result%
\footnote{
The result here is simplified by partial integration.
},
\begin{eqnarray}
&&S_{g3} =\int d^4 x \left\{-\frac1{\kappa^2} \left[\frac1{16}  h^2 \Box  h 
-\frac18   h  h^{\alpha\beta} \Box  h_{\alpha\beta} 
-\frac3{16}  h^{\alpha\beta} h_{\alpha\beta} \Box  h
+\frac14  h^{\alpha\beta}  h_{\beta}{}^{\gamma} \Box  h_{\alpha\gamma}
\right.\right.
\nonumber \\
&&\hspace{20mm}
-\frac14  h  h^{\alpha\beta} \partial_\alpha\partial_\beta  h 
+\frac14  h^{\alpha\beta}  h_{\alpha}{}^{\gamma} \partial_\beta\partial_ \gamma  h 
\left. 
- \frac12  h^{\alpha\gamma}  h^{\beta\lambda}  \partial_\alpha \partial_\beta  h_{\gamma \lambda} 
+\frac14  h^{\alpha\beta}  h^{\gamma\lambda} \partial_\alpha \partial_\beta  h_{\gamma\lambda} \right]
\nonumber \\ 
&&\hspace{10mm}
+\alpha\left[\frac14  h^2 \Box^2  h 
-\frac34  h^{\alpha\beta} h_{\alpha\beta} \Box^2  h 
- \frac12 h^{\alpha\beta} \Box  h_{\alpha \beta} \Box  h
-2  h   h^{\alpha\beta} \partial_\alpha\partial_\beta \Box  h
+ h^{\mu\alpha}  h_\mu{}^\beta \partial_\alpha \partial_\beta \Box  h
 \right]
\nonumber \\ 
&&\hspace{10mm}
+\beta\left[
-\frac1{16}  h  \Box  h \Box  h
+\frac3{32}  h^2 \Box^2  h
-\frac3{16}  h_{\alpha\beta} h^{\alpha\beta}\Box^2  h
+\frac18  h  \Box  h_{\alpha\beta}\Box  h^{\alpha\beta} \right.
\nonumber\\ && \hspace{20mm}
-\frac18 h^{\alpha\beta} \Box  h_{\alpha\beta} \Box  h   
-\frac14  h^{\alpha\beta}    h_{\beta}{}^{\gamma} \Box^2  h_{\alpha\gamma}
-\frac12  h  h^{\alpha \beta} \partial_\alpha \partial_\beta \Box  h
\nonumber\\ && \hspace{20mm}
+\frac18  h  \Box h^{\alpha \beta}  \partial_\alpha \partial_\beta  h 
+\frac14  h^{\mu\alpha}   h_\mu {}^\beta \partial_\alpha \partial_\beta \Box  h 
+ h^{\alpha\mu} \Box  h^{\beta\nu}\partial_\alpha \partial_\beta  h_{\mu\nu}
\nonumber\\ &&\hspace{20mm}
\left.\left.
+\frac12  h^{\alpha\mu}   h^{\beta\nu}\partial_\alpha \partial_\beta \Box  h_{\mu\nu}
-\frac12  h^{\alpha\beta}  h^{\mu\nu}\partial_\alpha \partial_\beta  \Box  h_{\mu\nu}
-\frac14   h^{\mu\nu} \Box  h^{\alpha\beta} \partial_\alpha \partial_\beta  h_{\mu\nu} \right] \right\}.
\label{g3}
\end{eqnarray}
This gives the 3-point vertex function for $ h_{\mu\nu}(p_1) h_{\alpha\beta}(p_2) h_{\gamma\lambda}(p_3)$, 
but, since the expression is lengthy, we show it in Appendix~\ref{SUPPLE}.
Here we show the 3-point vertices for modes with basis elements $\e_{\mu\nu}^{(\sigma)}$ and/or $\theta_{\mu\nu}/ \sqrt{3}$ 
(see Eqs.\eq{2acH} and \eq{2acI} for the decomposition of $ h_{\mu\nu}$ in this basis). 
In the derivation, we use the relations among $p_i^\mu$, $e^{\sigma}_{i,\mu\nu}$ and $\theta_{i,\mu\nu}$ shown in Appendix~\ref{SUPPLE}, 
which is obtained by direct calculations.
\begin{eqnarray}
&& \lambda_3^{\mu\nu,\alpha\beta,\gamma\lambda} \e_{1\mu\nu} \e_{2\alpha\beta} \e_{3\gamma\lambda}
\nonumber \\ && = 
\left[ \frac1{2\kappa^2}\left( p_1^2 +p_2^2+p_3^2 \right) -\frac{\beta}2 \left( p_1^4 +p_2^4+p_3^4 \right)  \right]
\e_{1\nu}^\mu \e_{2\alpha}^\nu \e_{3\mu}^\alpha
\nonumber \\ &&
+\left[ -\frac1{\kappa^2}+\beta \left( p_1^2+p_2^2+p_3^2 \right)
\right]
\left( p_{1\mu} \e^\mu_{2\nu} \e^\nu_{1\alpha} \e^\alpha_{3\beta} p_1^\beta 
+p_{2\mu} \e^\mu_{1\nu} \e^\nu_{2\alpha} \e^\alpha_{3\beta} p_2^\beta
+ p_{3\mu} \e^\mu_{2\nu} \e^\nu_{3\alpha} \e^\alpha_{1\beta} p_3^\beta\right)
\nonumber \\ &&
+ \left[ 
\frac1{4\kappa^2}-\frac\beta 4 \left(p_1^2+p_2^2+p_3^2 \right)
\right]
\left( \e_{1\mu\nu} \e_{2}^{\mu\nu}  p_1^\gamma p_1^\lambda \e_{3\gamma\lambda}
+\e_{1\mu\nu} \e_{3}^{\mu\nu}  p_1^\gamma p_1^\lambda \e_{2\gamma\lambda}
\right. \nonumber \\ && \hspace{5mm}  \left.
+\e_{1\mu\nu} \e_{2}^{\mu\nu}  p_2^\gamma p_2^\lambda \e_{3\gamma\lambda}
+\e_{3\mu\nu} \e_{2}^{\mu\nu}  p_2^\gamma p_2^\lambda \e_{1\gamma\lambda}
+\e_{1\mu\nu} \e_{3}^{\mu\nu}  p_3^\gamma p_3^\lambda \e_{2\gamma\lambda}
+\e_{3\mu\nu} \e_{2}^{\mu\nu}  p_3^\gamma p_3^\lambda \e_{1\gamma\lambda} \right), \label{3-ver1}
\end{eqnarray}
\begin{eqnarray}
&& \lambda_3^{\mu\nu,\alpha\beta,\gamma\lambda} \e_{1\mu\nu}  \e_{2\alpha\beta} \left( \frac1{\sqrt{3}} \theta_{3\gamma\lambda} \right)
\nonumber \\ && = 
\sqrt{3}\left[
-\frac1{4\kappa^2}\left( p_3^2- \frac12 \left(p_1^2+p_2^2\right)+ \frac16 \frac{\left( p_1^2-p_2^2 \right)^2}{p_3^2} \right)
+\left(3\alpha + \beta\right) \left( - \frac12 p_3^4 -\frac16 p_3^2 \left( p_1^2+p_2^2\right)\right) 
\right. \nonumber \\ && \qquad \qquad  \left.
+ \frac{\beta}{24} \left(-5 \left( p_1^4+p_2^4\right) + \frac{\left(p_1^2-p_2^2\right)^2\left(p_1^2+p_2^2\right)}{p_3^2} \right)
\right] \e_{1\mu\nu} \e_2^{\mu\nu}
\nonumber \\ &&  \hspace{5mm}
+\sqrt{3}\left[
\frac{1}{3\kappa^2}  - \frac{1}{6\kappa^2} \frac{\left( p_1^2+p_2^2 \right)}{p_3^2}  +\frac23\left(3\alpha+\beta\right) p_3^2
+\frac\beta6 \frac{p_1^4+p_2^4}{p_3^2}
\right] p_{2\mu} \e^\mu_{1\nu} \e^\nu_{2\alpha} p_1^\alpha , \label{3-ver2}
\end{eqnarray}
\begin{eqnarray}
&& \lambda_3^{\mu\nu,\alpha\beta,\gamma\lambda} \e_{1\mu\nu} \left( \frac1{\sqrt{3}} \theta_{2\alpha\beta} \right)\left( \frac1{\sqrt{3}} \theta_{3\gamma\lambda} \right)
\nonumber \\ && =  
\frac{p_2^\mu \e_{1\mu\nu}p_2^\nu}{24 p_2^2p_3^2} 
\left\{ 
\kappa^{-2}\left( -p_1^4+3 p_1^2(p_2^2+p_3^2) +2 (p_2^4+p_3^4)-12p_2^2p_3^2  \right)
\right.
\nonumber \\ && \qquad \qquad \qquad
+(3\alpha + \beta)
\left(  12 p_1^2(p_2^4+p_3^4)  +4  (p_2^6+p_3^6) -36 (p_2^4p_3^2+p_2^2p_3^4) \right) 
\nonumber \\ && \qquad \qquad \qquad \left.
+\beta
\left( p_1^6 +3p_1^4(p_2^2+p_3^2) 
 \right)
\right\},\label{3-ver3}
\end{eqnarray}
\begin{eqnarray}
&& \lambda_3^{\mu\nu,\alpha\beta,\gamma\lambda} \left( \frac1{\sqrt{3}} \theta_{1\mu\nu} \right)\left( \frac1{\sqrt{3}} \theta_{2\alpha\beta} \right)\left( \frac1{\sqrt{3}} \theta_{3\gamma\lambda} \right)
\nonumber \\ && 
=\frac{\sqrt{3}}{144 p_1^2p_2^2p_3^2}
\Biggl[
\frac1{\kappa^2}\left(-(p_1^8+p_2^8+p_3^8) +8 (p_1^2p_2^6+p_1^2p_3^6+p_1^6p_2^2+p_1^6p_3^2+p_2^2p_3^6+p_2^6p_3^2)
\right. \nonumber \\ && \qquad \qquad \qquad\left.
-14(p_1^4p_2^4+p_1^4p_3^4+p_2^4p_3^4)-28p_1^2p_2^2p_3^2(p_1^2+p_2^2+p_3^2)
\right)
\nonumber \\ && \qquad 
+2(3\alpha+\beta) \left(-(p_1^{10}+p_2^{10}+p_3^{10}) 
+11(p_1^2p_2^8+p_1^2p_3^8+p_1^8p_2^2+p_1^8p_3^2+p_2^2p_3^8+p_2^8p_3^2)
\right. \nonumber \\ && \qquad \qquad \qquad \qquad 
-10(p_1^4p_2^6+p_1^4p_3^6+p_1^6p_2^4+p_1^6p_3^4+p_2^4p_3^6+p_2^6p_3^4)
\nonumber \\ && \qquad \qquad \qquad\qquad \left.
-34p_1^2p_2^2p_3^2(p_1^4+p_2^4+p_3^4)
-42p_1^2p_2^2p_3^2(p_1^2p_2^2+p_1^2p_3^2+p_2^2p_3^2)
\right)
\Biggr], \label{3-ver4}
\end{eqnarray}
where we omit the suffix $(\sigma)$ denoting the helicites in $\e_{\mu\nu}^{(\sigma)}$, and 
$p_i^\mu$'s are the momenta of  modes associated with  basis $\e_{i,\mu\nu}$ or $\theta_{i,\mu\nu}/\sqrt{3}$.



\section{Amplitudes for matter-graviton scattering}\label{SA}
The simplest scattering involving gravitons with positive norm and/or negative norm is 
a matter-graviton two-body scattering, 
which is the concern of this paper. 
The graviton field $h_{\mu\nu}$ is decomposed into the massless part  $H_{\mu\nu}$ and the massive  part $I_{\mu\nu}$ (see Eq.\eq{h=H+I}). 
$H_{\mu\nu}$ has two massless transverse-traceless (TT) degrees of freedom $e_{\mu\nu}^{(\sigma)}$ (see Eq.\eq{2acH}) with positive norm, 
while $I_{\mu\nu}$ is composed of five negative-norm massive transverse-traceless degrees of freedom $e_{\mu\nu}^{(\sigma)}$ and one positive-norm massive TT degree of freedom $\theta_{\mu\nu}$ (see Eq.\eq{2acI}). 
We expand $H_{\mu\nu}$ and $I_{\mu\nu}$ as 
\begin{eqnarray}
&& H_{\mu\nu} = \sum_{\sigma=1}^2 \int d^3p H^{(\sigma)} ({\bm p}) e_{\mu\nu}^{(\sigma)} ({\bm p}) , \label{Hint}\\
&& I_{\mu\nu} = \sum_{\sigma=1}^5 \int d^3p I^{(\sigma)} ({\bm p}) e_{\mu\nu}^{(\sigma)} ({\bm p})  
+ \int d^3p I^{(s)}  ({\bm p}) \frac1{\sqrt{3}}\theta_{\mu\nu} ({\bm p}) .\label{Iint}
\end{eqnarray}
$H^{(\sigma)} ({\bm p})$, $I^{(\sigma)} ({\bm p})$ are the operators for  TT components,  and  $I^{(s)}  ({\bm p})$ is that for the trace component. 
The two $e^{(\sigma)}_{\mu\nu}$ appearing in Eqs.~\eqref{Hint} and \eqref{Iint} are the same except that the mass is set to zero in the former. 
The operators $H^{(\sigma)} ({\bm p})$, $I^{(\sigma)} ({\bm p})$ are written with the annihilation and the creation operators (see Eqs.\eq{2acH} and \eq{2acI}) as
\begin{eqnarray}
&&H^{(\sigma)} ({\bm p}) = \frac{\kappa}{\sqrt{(2\pi)^3 p_0}} \left\{ a_H^{(\sigma)}({\bm p}) e^{-ipx}+a_H^{(\sigma)}{}^\dagger({\bm p}) e^{ipx} \right\} , \label{Hs}\\
&&I^{(\sigma)} ({\bm p}) =\frac{\kappa}{\sqrt{(2\pi)^3p_0}} \left\{ a_T^{(\sigma)} ({\bm p}) e^{-ipx}+a_T^{(\sigma)}{}^\dagger ({\bm p}) e^{ipx} \right\} , \\
&&I^{(s)}  ({\bm p}) = \frac{\kappa}{\sqrt{(2\pi)^3p_0}} \left\{ a_S ({\bm p}) e^{-ipx}+a_S^\dagger ({\bm p}) e^{ipx} \right\}.
\label{Is}
\end{eqnarray} 
We introduce $h^{(\sigma)} ({\bm p})$ denoting either of TT modes $H^{(\sigma)} ({\bm p})$ or $I^{(\sigma)} ({\bm p})$, 
since the scattering amplitudes become the same except for the value of mass.
 
\begin{figure}[tb]
  \begin{center}
  \begin{tabular}{c}
        \begin{minipage}{0.23\hsize}
        \begin{center}  
    \includegraphics[clip,width=2.8cm]{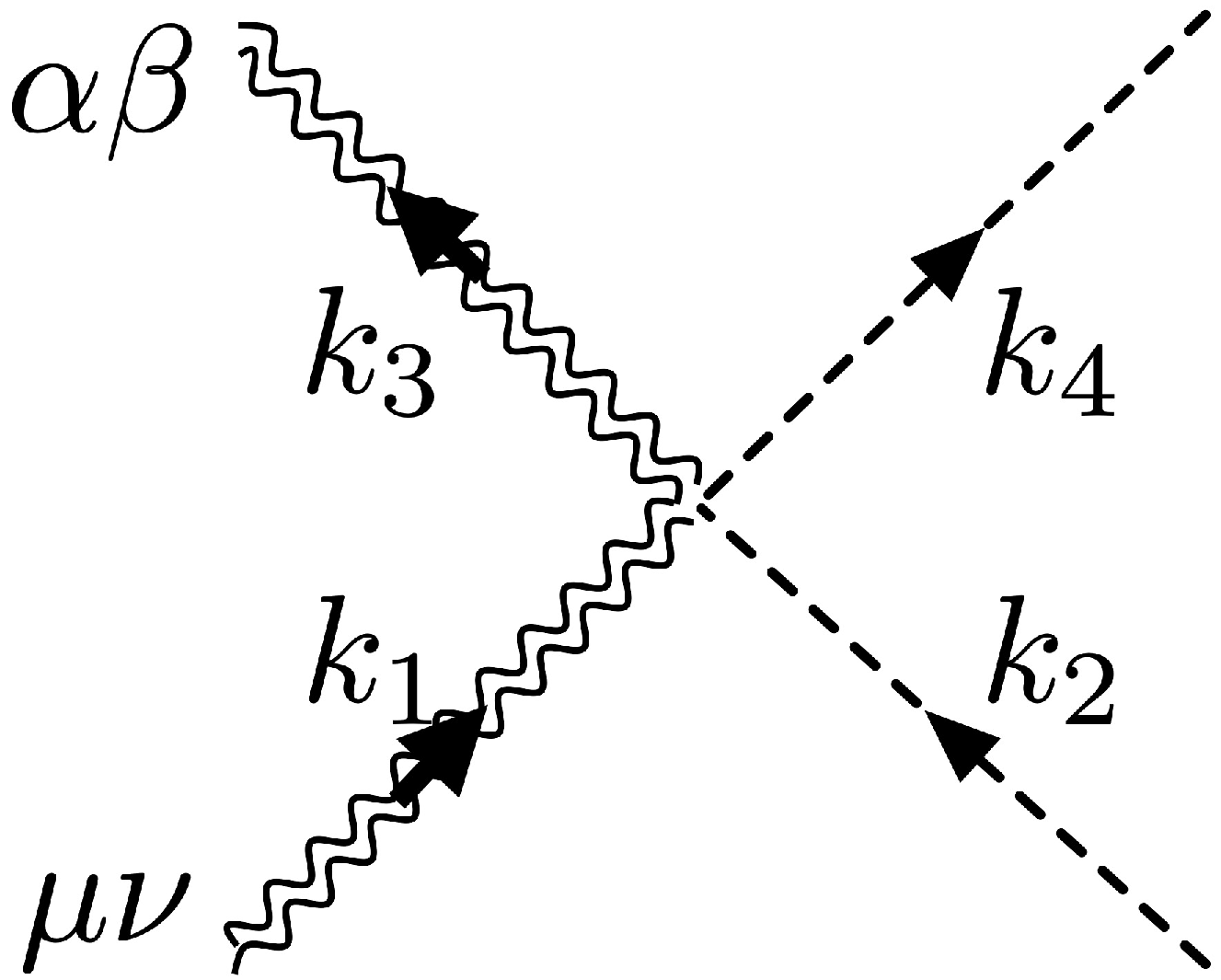}
        \end{center}
      \end{minipage}
      \begin{minipage}{0.20\hsize}
        \begin{center}
    \includegraphics[clip,width=2.5cm]{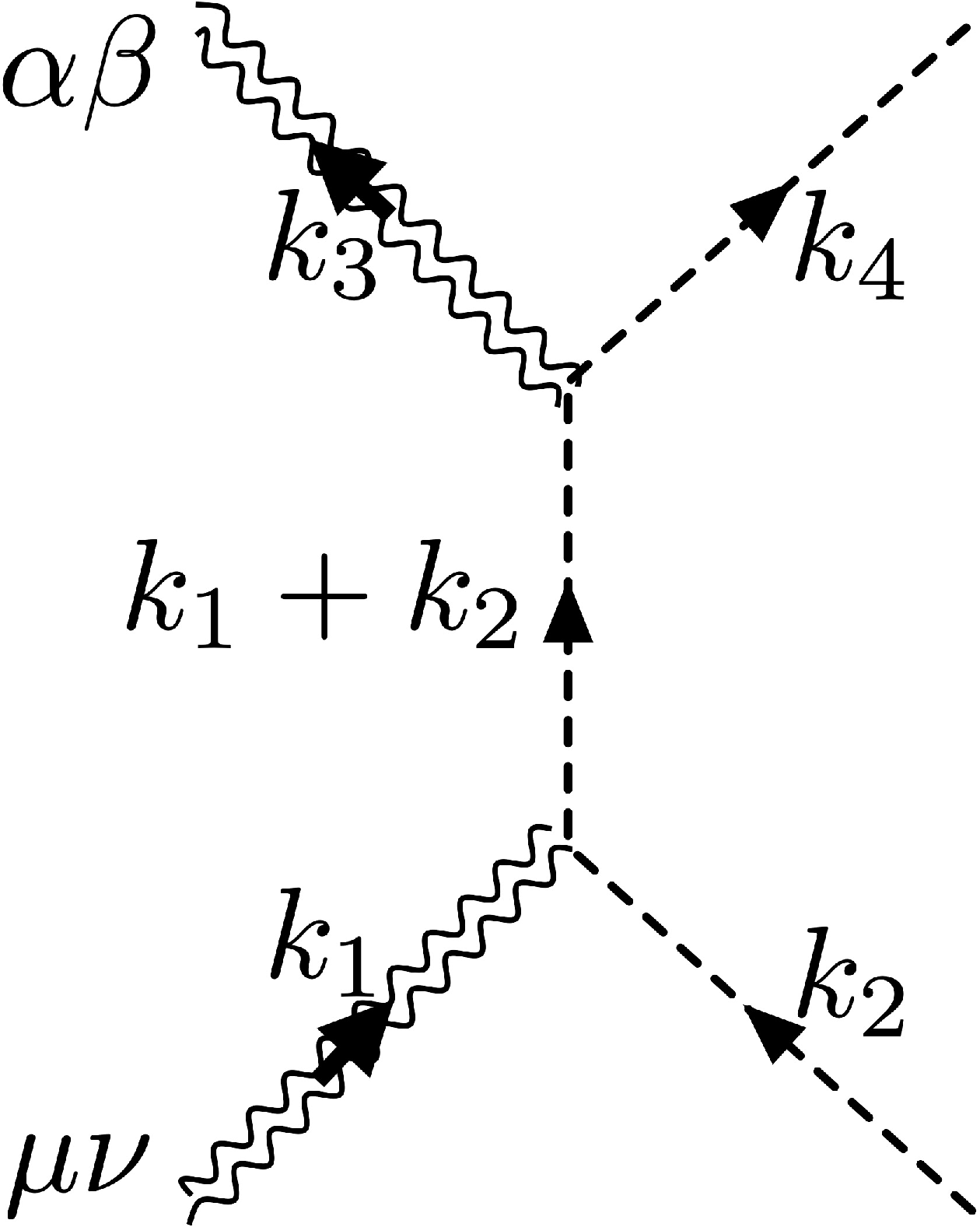}
        \end{center}
      \end{minipage}
      \begin{minipage}{0.20\hsize}
        \begin{center}  
    \includegraphics[clip,width=2.5cm]{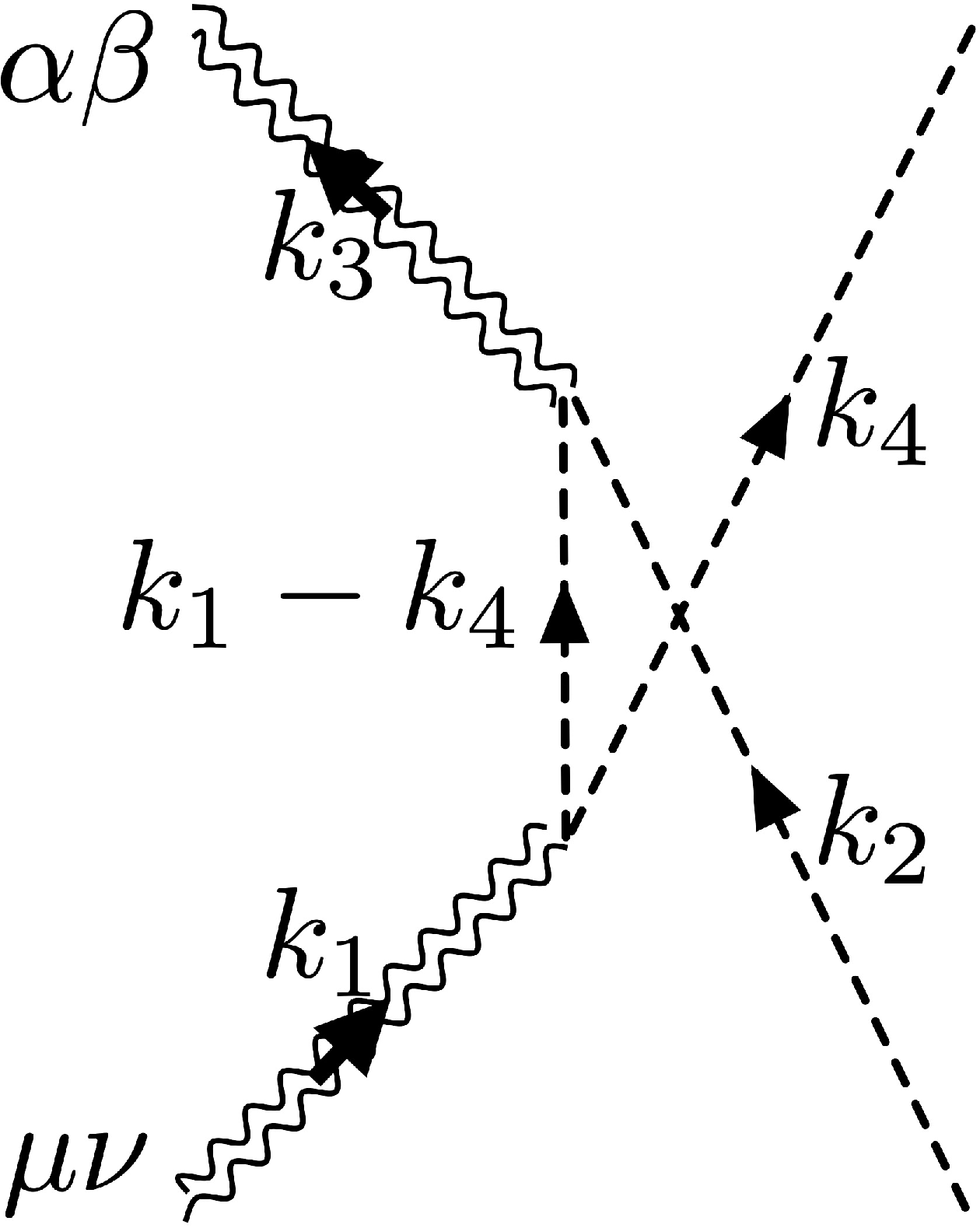}
        \end{center}
      \end{minipage}
      \begin{minipage}{0.25\hsize}
        \begin{center}  
    \includegraphics[clip,width=3.8cm]{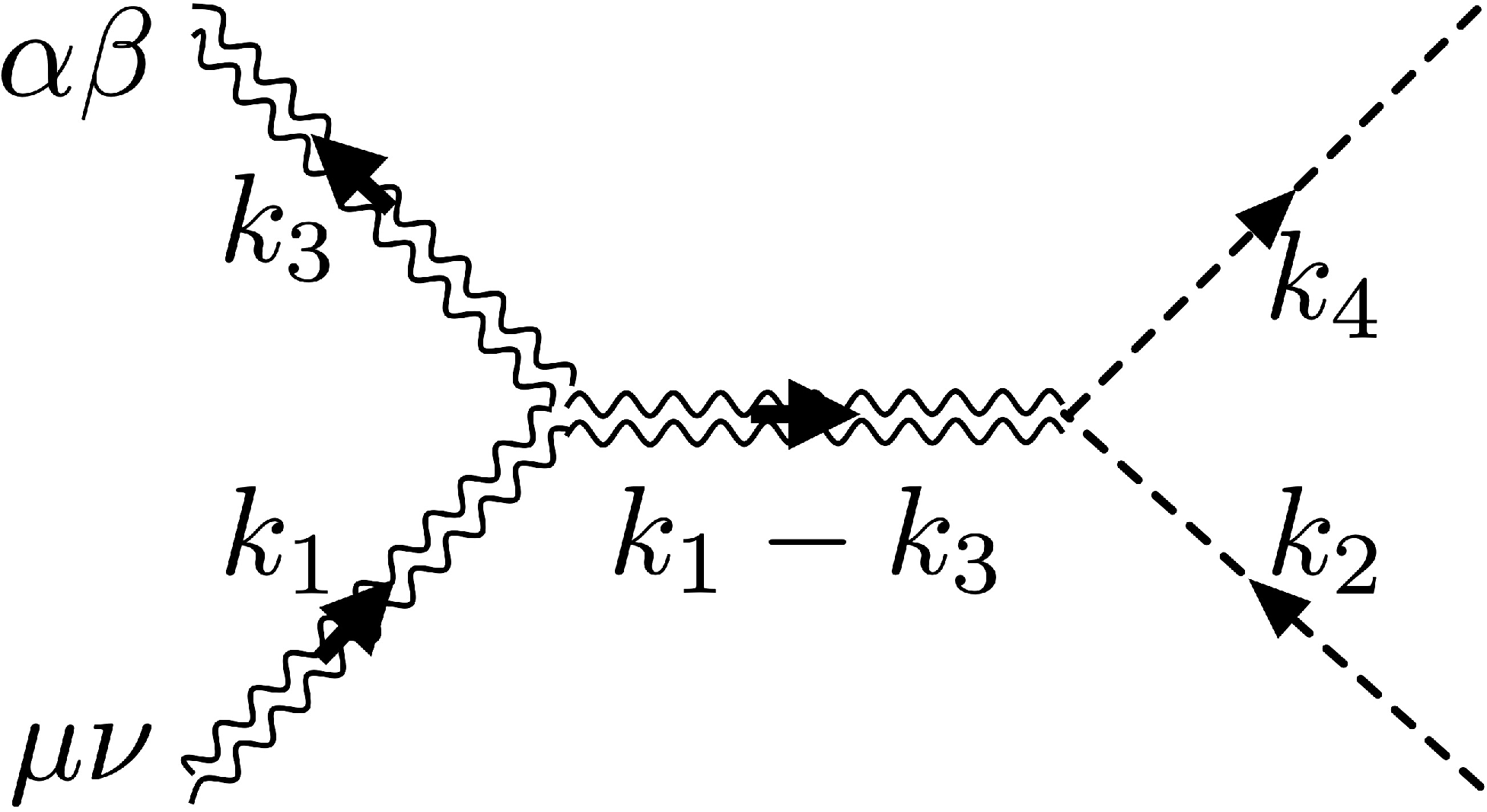}
        \end{center}
      \end{minipage}
\end{tabular}
    \caption{Contact, $s$-, $u$- and $t$-channel exchange diagrams (from left to right): 
    The double wiggly lines correspond to graviton $h_{\mu\nu}$, while the scalar field $\phi$ is described by dotted lines. 
    The initial (final) state is composed of a graviton $h_{\mu\nu}$ ($h_{\alpha\beta}$) with momentum $k_1$ ($k_3$) and a scalar field $\phi$ with momentum $k_2$ ($k_4$).
    }
    \label{Fig:cha}
  \end{center}
\end{figure}

We are now ready to discuss our main issue, the $h_{\mu\nu}$-$\phi$ scattering amplitudes.
We fix the kinematics of $h_{\mu\nu}$-$\phi$ scattering by 
\begin{eqnarray}
h_{\mu\nu}({\bm k}_1, e_{1,\mu\nu} \ \mbox{or} \ \theta_{1,\mu\nu}) + \phi({\bm k}_2) \  \to \ 
h_{\alpha\beta}({\bm k}_3, e_{3,\alpha\beta} \ \mbox{or} \ \theta_{3,\alpha\beta}) + \phi({\bm k}_4) ,
\end{eqnarray}
as shown in figure \ref{Fig:cha}. 
In our calculation at the tree level, there are four types of graphs, 
contact term due to $\lambda_4^{\mu\nu,\alpha\beta}$, the $s$- and $u$-channel exchanges of $\phi$ propagator and the $t$-channel exchange of  $h_{\mu\nu}$ propagator, denoted by $\A_c$, $\A_s$, $\A_u$, and $\A_t$, respectively.
Figure \ref{Fig:cha} shows the corresponding Feynman diagrams of them.

Three cases of $h_{\mu\nu}$-$\phi$ scattering are considered separately: 
\begin{eqnarray}
\begin{matrix}
\mbox{{\it \bf A}}&:& h^{(\sigma)} \phi \to h^{(\sigma') }\phi, && \\
\mbox{{\it \bf B}}&:& h^{(\sigma)} \phi \to I^{(s) }\phi; &I^{(s) } \phi \to h^{(\sigma) }\phi, & \qquad \mbox{(two are related)} \\
\mbox{{\it \bf C}}&:& I^{(s)}  \phi \to I^{(s) }\phi. &&
\end{matrix}
\end{eqnarray}
Since the expression of the amplitudes turns out to be the same regardless of the positivity or negativity of the graviton norm, 
we do not need to distinguish them here.

The field operators $H^{(\sigma)} ({\bm p})$, $I^{(\sigma)} ({\bm p})$ and $I^{(s)}  ({\bm p})$ of Eqs.\eq{Hs}-\eq{Is} are not 
canonically normalized. 
Taking the normalization into account, we should add a factor $\kappa$ per an external graviton. 
Since we now have two external gravitons, we should multiply the usual calculation from the Feynmann diagram by $\kappa^2$.
We will show each contribution and then give the scattering amplitude by summing it up. 
Since the calculations of the $t$-channel exchange and summation are complicated, we show the detailed derivation  in Appendix~\ref{AppSA}.

\subsection{$h^{(\sigma)}\phi\, \to \, h^{(\sigma')}\phi$}

Contributions from the contact interaction ${\A}_c$, the $s$-channel exchange ${\A}_s$ and the $u$-channel exchange ${\A}_u$  are 
obtained as follows,
\begin{eqnarray}
{\A}_c 
\!\!\!\! &   = & \!\!\!\!  
\kappa^2 \left[
 \dfrac12 \left( k_2 \cdot k_4 +m^2 \right)  \mbox{Tr} [e_1\cdot e_3] - \left(k_2\cdot e_1 \cdot  e_3 \cdot k_4+ k_2 \cdot  e_3 \cdot e_1 \cdot k_4 \right) \right]
 \nonumber\\
&&\qquad - 
\xi \kappa^2 \left[ \left(2k_1^2+2k_3^2 -3 k_{1}\cdot k_3 \right) 
\Tr\left[ \e_{1}\cdot  \e_3 \right]
+ 2  k_{3}\cdot  \e_{1} \cdot  \e_{3} \cdot  k_1  \right]
  ,\\
{\A}_s 
\!\!\!\! &   = & \!\!\!\!  
\kappa^2   \cfrac{  k_2 \cdot e_1 \cdot k_2 \  k_4\cdot e_3 \cdot k_4 }{(k_1+k_2)^2+m^2} ,  
\end{eqnarray}
and ${\A}_u $ is obtained from ${\A}_s$ by crossing relation $ k_2 \, \leftrightarrow \, - k_4$, as seen from figure \ref{Fig:cha}.
The calculation of the $t$-channel exchange ${\A}_t$ is more involved because of the three point vertex of gravitons, and is given in Appendix~\ref{AppSA},
\begin{eqnarray}
{\A}_t 
\!\!\!\! &  = & \!\!\!\! \kappa^2 \left(\frac{1}{( k_1 - k_3)^2} +\frac{\beta(k_1^2+k_3^2)}{\beta (k_1-k_3)^4 - \kappa^{-2} (k_1-k_3)^2}\right)
\nonumber \\ && 
\Bigl[
-(k_1\cdot k_2+k_1\cdot k_4)
\left(k_2\cdot \e_1 \cdot \e_3 \cdot k_4 -k_4\cdot \e_1 \cdot \e_3 \cdot k_2\right)
+\left(k_2\cdot \e_1\cdot k_2\right)\left(k_4\cdot \e_3\cdot k_4\right)
\nonumber \\ && 
+\left(k_4\cdot \e_1\cdot k_4\right)\left(k_2\cdot \e_3\cdot k_2\right)
-2\left(k_2\cdot \e_1\cdot k_4\right)\left(k_2\cdot \e_3\cdot k_4\right)
+ \frac14  (k_1\cdot k_2+k_1\cdot k_4)^2
\Tr[\e_1 \cdot \e_3] 
\Bigr]
\nonumber \\ && \qquad
+\frac{ \kappa^2 }{2} 
\left(k_2\cdot \e_1 \cdot \e_3 \cdot k_4 +k_4\cdot \e_1 \cdot \e_3 \cdot k_2\right) 
+ \kappa^2  \left[ \frac3{16} (k_1-k_3)^2 + \frac{(k_1^2+k_3^2)}{16}\right] \Tr[\e_1\cdot\e_3]
\nonumber \\ && \qquad
+ \frac{ \kappa^2 \beta k_1^2k_3^2((k_1-k_3)^2+4m^2)}{8\left(\beta (k_1-k_3)^4 - \kappa^{-2} (k_1-k_3)^2\right)} \Tr[\e_1\cdot\e_3]
 \nonumber\\
&&\qquad +
\xi \kappa^2 \left[ \left(2k_1^2+2k_3^2 -3 k_{1}\cdot k_3 \right) 
\Tr\left[ \e_{1}\cdot  \e_3 \right]
+ 2  k_{3}\cdot  \e_{1} \cdot  \e_{3} \cdot  k_1  \right]
\end{eqnarray}
The $t$-channel exchange contribution $\A_t$ by itself satisfies the crossing relation. 
The sum of four terms is (calculation is shown in Appendix~\ref{AppSA})
\begin{eqnarray}
&&{\cal A}={\A}_s+{\A}_t+{\A}_u+{\A}_c
\nonumber \\ &&  \quad 
=\frac{\kappa^2}{( k_1 - k_3)^2 \left[ \beta (k_1-k_3)^2 - \kappa^{-2} \right] }
\Biggl[ \left( 2\beta (k_1^2+k_3^2 ) - 2 \beta (k_1\cdot k_3 ) - \kappa^{-2} \right)
\nonumber \\ && \qquad
\times\left( -\frac14 (k_1^2-2k_1\cdot k_4 ) (k_1^2+2k_1\cdot k_2) \mbox{Tr} [e_1\cdot e_3]
+(k_1^2-2k_1\cdot k_4 ) k_2 \cdot e_1 \cdot e_3 \cdot k_4
\right.\nonumber \\ && \qquad \qquad\qquad 
+ (k_1^2+2k_1\cdot k_2) k_4 \cdot e_1 \cdot e_3 \cdot k_2
-2 (k_2 \cdot e_1\cdot k_4 )( k_2 \cdot e_3 \cdot k_4)
\nonumber \\ && \qquad \qquad\qquad
-\frac{ k_1^2-2k_1\cdot k_4 }{k_1^2+2k_1\cdot k_2} (k_2 \cdot e_1\cdot k_2 )( k_4 \cdot e_3 \cdot k_4)
\nonumber \\ && \qquad \qquad\qquad \left.
-\frac{k_1^2+2k_1\cdot k_2}{ k_1^2-2k_1\cdot k_4 } (k_4 \cdot e_1\cdot k_4 )( k_2 \cdot e_3 \cdot k_2)
\right)
\nonumber \\ &&\qquad
+ \beta k_1^2 k_3^2 
\left(
\frac18 (k_1^2+k_3^2+4m^2) \mbox{Tr} [e_1\cdot e_3]
-k_2 \cdot e_1 \cdot e_3 \cdot k_4
- k_4 \cdot e_1 \cdot e_3 \cdot k_2
\right.\nonumber \\ &&  \qquad\qquad\qquad \qquad \qquad\qquad \qquad 
+\frac{ 2 }{k_1^2+2k_1\cdot k_2} (k_2 \cdot e_1\cdot k_2)( k_4 \cdot e_3 \cdot k_4)
\nonumber \\ && \left. \qquad \qquad\qquad \qquad \qquad\qquad\qquad
+\frac{2}{ k_1^2-2k_1\cdot k_4 } (k_4 \cdot e_1\cdot k_4 )( k_2 \cdot e_3 \cdot k_2)
\right) \Biggr],
\label{sum1}
\end{eqnarray}
where dot means contraction by spacetime index $\mu$, for instance,  
$k_2 \cdot e_1\cdot k_4 = k_2^\mu e_{1,\mu\nu} k_4^\nu$, 
and  $\mbox{Tr} [e_1\cdot e_3]$ is $e_{1,\mu\nu} e_3^{\mu\nu}$. 
Curiously, the amplitude \eq{sum1} is independent of $\alpha$ and $\xi$. 
The physical meaning of this is yet unclear to us. 

Taking the Einstein limit ($\alpha,\beta,\xi \to 0$), and setting the basis $e_{\mu\nu}$ to be the helicity-2 shown in Eq.\eq{basis2o} and Eq.\eq{basis2e}, we can reproduce the result of Ref.~\cite{Berends:1974gk} from the amplitude \eq{sum1}.%
\footnote{The amplitudes in Ref.~\cite{Berends:1974gk} are calculated in a different basis, labeled by $2$ and $-2$. 
Their amplitudes of $2\to2$ and $-2 \to -2$ become the same, {\it i.e.} ${\cal A}(2\to 2)={\cal A}(-2 \to -2) =: {\cal A}_{dig}$. 
(In their notation, the amplitude ${\cal A}(2\to 2)$ is denoted as $- i \phi_{2;2}$, and the signature of spacetime metric is flipped.) 
In addition, ${\cal A}(2\to -2)={\cal A}(-2 \to 2) =: {\cal A}_{off}$ holds. 
Hence, the scattering amplitude is expressed as a matrix
\begin{eqnarray}
{\cal A} = 
\begin{pmatrix}
{\cal A}_{dig} & {\cal A}_{off}  \\
{\cal A}_{off}  & {\cal A}_{dig} \\
\end{pmatrix}
. \label{ampmat}
\end{eqnarray}
Our basis elements $(2,o)$ and $(2,e)$, defined in Eq.\eqref{basisele}, are different from those in Ref.~\cite{Berends:1974gk}, 
and give the diagonal matrix of the scattering amplitude \eqref{ampmat}. 
Since the difference between the matrix of Ref.~\cite{Berends:1974gk} and ours is caused by the different choices of basis, 
their eigenvalues should coincide with each other. 
We confirm the coincidence and see that ${\cal A}$ in Eq.\eqref{sum1} satisfies ${\cal A}(2,o\to 2,o)|_{\beta=0} ={\cal A}_{dig} - {\cal A}_{off}$
and ${\cal A}(2,e\to 2,e)|_{\beta=0} ={\cal A}_{dig} + {\cal A}_{off}$.
 }

\subsection{$h^{(\sigma)} \phi \, \to\, I^{(s)} \phi$}

The scattering amplitude is calculated in the similar manner to $h^{(\sigma)} \phi \, \to\, h^{(\sigma)} \phi$.
The contributions from $\A_c$, $\A_s$, $\A_u$ and $\A_t$, and their sum are obtained as follows.
\begin{eqnarray}
{\A}_c 
\!\!\!\! &  = & \!\!\!\!
- \frac{\sqrt{3 }\kappa^2}{ 6m_S^2}
\left[(k_1 \cdot k_4 +k_3 \cdot k_4) k_2 \cdot e_1 \cdot k_2
-(k_1 \cdot k_2+k_2 \cdot k_3) k_4 \cdot e_1 \cdot k_4
-k_1^2\  k_2  \cdot e_1  \cdot k_4 \right]
\nonumber \\ && \qquad
-\frac{ \xi \kappa^2}{\sqrt{3}}  \frac{\left(k_3\cdot \e_1 \cdot k_3\right)} {m_S^2}
\left( 2k_1^2 + 4 m_S^2 - k_1\cdot k_3 \right) 
 ,\\
{\A}_s 
\!\!\!\! &  = & \!\!\!\!-\dfrac{ \sqrt{3} \kappa^2 k_2 \cdot e_1 \cdot k_2 \left(3 m_S^2 k_3 \cdot k_4  +2 m^2 m_S^2-  2(k_3 \cdot k_4)^2\right)}{6m_S^2\left[(k_1+k_2)^2+m^2\right]}  
-2\sqrt{3} \xi \kappa^2 m_S^2 \dfrac{ k_2\cdot \e_1 \cdot k_2 } {\left[(k_1+k_2)^2+m^2\right]},
\nonumber \\
%
\end{eqnarray}
The $u$-channel exchange contribution $\A_u$ is related to $\A_s$ by crossing relation $ k_2 \, \leftrightarrow \, - k_4$. 
\begin{eqnarray}
{\A}_t 
\!\!\!\! &  = & \!\!\!\! 
\frac{\sqrt{3} \kappa^2 }{24 m_S^2} \left((k_1+k_2)^2-(k_1-k_4)^2\right) \left( k_2 \cdot \e_1 \cdot k_2- k_4 \cdot \e_1 \cdot k_4\right) 
\nonumber \\ && \ 
+\frac{\sqrt{3} \kappa^2 }{24 m_S^2}\left((k_1-k_3)^2+k_1^2 +5 m_S^2\right)\left( k_2 \cdot \e_1 \cdot k_2+ k_4 \cdot \e_1 \cdot k_4 \right) 
-\frac{\sqrt{3} \kappa^2 }{6 m_S^2}k_1^2\  k_2 \cdot \e_1 \cdot k_4
\nonumber \\ && \ 
- \frac{\sqrt{3} \kappa^2 }6\left( 2\frac{m^2}{(k_1-k_3)^2} + \frac{2(3\alpha+\beta)}{\left(2(3\alpha+\beta)(k_1-k_3)^2+ \kappa^{-2}\right)}
\frac{m_S^2}{(k_1-k_3)^2} \left( (k_1-k_3)^2-2m^2\right)
\right)
\nonumber \\ && \qquad \qquad \qquad \qquad \qquad
\times
k_3 \cdot \e_1 \cdot k_3 
\nonumber \\ && \ 
-
\frac{\xi\kappa^2}{\sqrt{3}m_S^2}\Biggl\{ 
 \left(k_{1\alpha}k_3^\alpha \right) -2 k_1^2 - 4 m_S^2
+ \frac{12(3\alpha+\beta)m_S^4}{2(3\alpha+\beta)(k_1-k_3)^2 + \kappa^{-2} }\Biggr\}
k_3 \cdot \e_1 \cdot k_3 
,
\end{eqnarray}
\begin{eqnarray}
{\cal A}&=&\A_s+\A_u+\A_t+\A_c
\nonumber \\
&=&
- \frac{\sqrt{3} \kappa^2 }{6} \left(m_S^2+ 2m^2\right)
\left(
\frac{k_2 \cdot \e_1 \cdot k_2}{(k_1+k_2)^2+m^2} 
+\frac{k_4 \cdot \e_1 \cdot k_4}{(k_1-k_4)^2+m^2} 
\right)
\nonumber \\ &&
- \frac{\sqrt{3} \kappa^2 }6\left( 2\frac{m^2}{(k_1-k_3)^2} + \frac{2(3\alpha+\beta)}{\left(2(3\alpha+\beta)(k_1-k_3)^2+ \kappa^{-2}\right)}
\frac{m_S^2}{(k_1-k_3)^2} \left( (k_1-k_3)^2-2m^2\right)
\right)
\nonumber \\ &&  \qquad \qquad \qquad \qquad
\times
k_3 \cdot \e_1 \cdot k_3
\nonumber \\ && 
-2\sqrt{3} \xi \kappa^2 m_S^2 \left( 
\frac{2(3\alpha+\beta)\left( k_3 \cdot e_1 \cdot k_3\right)}{2(3\alpha+\beta)(k_1-k_3)^2 + \kappa^{-2} }
+\dfrac{ k_2\cdot \e_1 \cdot k_2 } {\left[(k_1+k_2)^2+m^2\right]}
+\dfrac{ k_4\cdot \e_1 \cdot k_4 } {\left[(k_1-k_4)^2+m^2\right]} \right).
\nonumber \\
.\label{sum2}
\end{eqnarray}
We note that $\alpha$ and $\beta$ appear only in combination $(3\alpha + \beta)$. 
The dependence of the total amplitude on momentum is 4-th order lower than that of each contribution: $\A_s$, $\A_u$, $\A_t$ and $\A_c$. 
Hence, cancelation occurs in the UV limit.

\subsection{$I^{(s)} \phi\,\to\, I^{(s)} \phi$}

The scatteing amplitude in this case is also obtained in the similar manner,  
\begin{eqnarray}
{\A}_c 
\!\!\!\! &  = & \!\!\!\!
-\dfrac{\kappa^2}{12}\left[k_2  \cdot k_4 + 5m^2 \right] - \dfrac{\kappa^2}6 \dfrac{k_1 \cdot k_2\  k_1 \cdot k_4}{m_S^2} 
- \dfrac{\kappa^2}6 \dfrac{k_2 \cdot k_3 \ k_3 \cdot k_4 }{m_S^2} 
\nonumber \\ \qquad &&
+ \kappa^2  \dfrac{k_1 \cdot k_3 }{6m_S^4} \left(
(k_2  \cdot k_4+m^2) k_1 \cdot k_3    
-2 k_1 \cdot k_2 \    k_3 \cdot k_4   
-2 k_1 \cdot k_4 \   k_2 \cdot k_3   
\right) 
\nonumber \\ \qquad &&
+
\frac{\xi \kappa^2 }{3}\left[ \frac{\left(k_1\cdot k_3 \right)^3}{m_S^2} - 8 \frac{\left(k_1\cdot k_3 \right)^2}{m_S^2}
- \left(k_1\cdot k_3 \right) + 2 m_S^2 \right] ,\\
{\A}_s 
\!\!\!\! &  = & \!\!\!\!
\dfrac{ \kappa^2 \left(3 m_S^2 k_1 \cdot k_2  +2 m^2 m_S^2  - 2 (k_1 \cdot k_2)^2\right)
\left(3  m_S^2  k_3 \cdot k_4  +2 m^2 m_S^2 - 2 ( k_3 \cdot k_4 )^2 \right)}{12 m_S^4 \left[(k_1+k_2)^2+m^2\right]}     
\nonumber \\ \qquad &&
- \xi \kappa^2\left( 2 (k_1 \cdot k_2 ) -2m_S^2 - \frac{2m_S^4+4 m^2 m_S^2}{\left[ 2 (k_1 \cdot k_2 )- m_S^2 \right]}   \right)   + \cfrac{ 12 \xi^2 \kappa^2  m_S^4   }{(k_1+k_2)^2+m^2} \ .
%
\end{eqnarray}
The $u$-channel exchange contribution $\A_u$ is obtained from ${\A}_s$ by crossing relation $ k_2 \, \leftrightarrow \, - k_4$. 
\begin{eqnarray}
{\A}_t 
\!\!\!\! &  = & \!\!\!\!\frac{\kappa^2}{48 m_S^4} 
\Biggl[
(k_1+k_2)^2(k_1-k_4)^2(k_1-k_3)^2 - 6 m_S^2(k_1+k_2)^2(k_1-k_4)^2
+(4m_S^2-m^2)(k_1-k_3)^4
\nonumber \\ &&\hspace{20mm}
+(11m_S^4  - 4m^2 m_S^2 -m^4)(k_1-k_3)^2-10m_S^6-12m^2 m_S^4 +6m^4 m_S^2\Biggr]
\nonumber \\ &&
-
\kappa^2 \frac{(k_1-k_3)^4 -6 (k_1+k_2)^2(k_1-k_4)^2 
+4(m_S^2+m^2)(k_1-k_3)^2+6(m_S^2+m^2)^2-8m^2 m_S^2
}{12 \left(\kappa^2 \beta (k_1-k_3)^2 -1 \right)(k_1-k_3)^2}
\nonumber \\ &&
+\frac{\kappa^2}{6}\frac{3m_S^4(k_1-k_3)^2+4m_S^2 m^2 (k_1-k_3)^2-2m_S^4m^2}{(k_1-k_3)^2\left((k_1-k_3)^2+m_S^2\right)}\nonumber \\ && 
+\frac{\xi \kappa^2}{3}  
\left[
-\frac{\left( k_1\cdot k_3 \right)^3}{m_S^4} +8\frac{\left( k_1\cdot k_3 \right)^2}{m_S^2} +7 \left( k_1\cdot k_3 \right) +10 m_S^2
+\frac{12m_S^4}{\left((k_1-k_3)^2 +m_S^2 \right)}
\right],
\end{eqnarray}
\begin{eqnarray}
&&{\cal A}=\A_s+\A_u+\A_t+\A_c
\nonumber \\ && \quad
=-\frac{\kappa^2}{6} \left( m_S^2 +m^2 \right)
+\frac{\kappa^2}{6}\frac{3m_S^4(k_1-k_3)^2+4m_S^2 m^2 (k_1-k_3)^2-2m_S^4m^2}{(k_1-k_3)^2\left((k_1-k_3)^2+m_S^2\right)}
\nonumber \\ &&\qquad
-
\kappa^2 \frac{(k_1-k_3)^4 -6 (k_1+k_2)^2(k_1-k_4)^2 
+4(m_S^2+m^2)(k_1-k_3)^2+6(m_S^2+m^2)^2- 8m_S^2m^2
}{12 \left(\kappa^2 \beta (k_1-k_3)^2 -1 \right)(k_1-k_3)^2}
\nonumber \\ &&\qquad
+\kappa^2\frac{m_S^4+4m^2m_S^2+4m^4}{12}\left( \frac{1}{(k_1+k_2)^2+m^2}+\frac{1}{(k_1-k_4)^2+m^2} \right) 
\nonumber \\ && \qquad
+
2\xi \kappa^2  
\left(
4 m_S^2
+\frac{2m_S^4}{\left((k_1-k_3)^2 +m_S^2 \right)}
+ \cfrac{  m_S^4 +2m^2m_S^2   }{(k_1+k_2)^2+m^2}
+ \cfrac{  m_S^4 +2m^2m_S^2  }{(k_1-k_4)^2+m^2}
\right).
\nonumber \\
\label{sum3}
\end{eqnarray}
Here, we use $k_1^{-2}=k_3^{-2}=-m_S^{-2} =-2 \kappa^2 (3 \alpha + \beta)$. 
Thus, the amplitude \eq{sum3} depends on $(3 \alpha + \beta)$ through $m_S^{2}$, 
but does not depend on $\alpha$ and $\beta$ separately, the same as for $h^{(\sigma)} \phi \, \to\, I^{(s)} \phi$.


\section{Scattering amplitude in the UV limit} \label{SAUV}

We are interested in the high energy behavior of the scattering amplitude to investigate the relation between renormalizability and UV unitarity. 
Two kinds of high energy limits are often investigated, namely, the Regge limit and the hard scattering limit. 
Whether these two different ways of analyzing the high-energy 2-2 amplitudes are related and how if so has not been studied. 
As discussed in Ref.~\cite{Cornwall:1973tb,LlewellynSmith:1973yud,Cornwall:1974km,Fujimori:2015mea}, we take the hard scattering limit here due to the following reason. 
The basis of two-particle states can be independently labeled in the total energy $E$, the total momentum ${\bf P}$ and the scattering angle $\theta$. 
The hard scattering limit, that is, $E \to \infty$ with ${\bf P}=0$ and $\theta$ fixed, does not change the measure for ${\bf P}$ and $\theta$ in the inner product. 
Since the optical theorem, providing the tree unitarity, involves the inner product, the analysis in the hard scattering limit straightforwardly gives results.%
\footnote{
The Regge limit would provide a different condition for the tree unitarity. 
It is known that the Regge limit gives constraints on the amplitude of the forward scattering~\cite{Froissart:1961ux,Martin:1962rt}.
}

Although the scattering amplitudes have been derived in the previous section, 
their dependences on the total energy are hard to be seen. 
One of the reasons is that each element of the basis has a nontrivial energy dependence. 
The situation is similar to that in gauge theories~\cite{Lee:1977eg}. 
Another reason is that nontrivial cancelation occurs. 
In section \ref{MD}, the exact forms of the basis elements will be shown. 
We will see there that they have the total energy dependence. 
With the exact forms of the elements, 
all amplitudes of matter-graviton scattering will be derived in section \ref{SAeach}. 
There exist thirteen non-zero amplitudes in all.

\subsection{Mode Decomposition}\label{MD}

Without loss of generality, we always take the center of mass (CoM) frame due to Lorentz symmetry.
In CoM frame, for massive particle, their component can be expressed as
\begin{eqnarray}
&&k_{1\mu} = \left( \sqrt{k^2+m_1^2}, k, 0,0 \right) \\
&&k_{2\mu} = \left( \sqrt{k^2+m_2^2}, -k, 0,0 \right) \\
&&k_{3\mu} = \left( \sqrt{q^2+m_3^2}, q \cos \theta, q \sin \theta,0 \right) \\
&&k_{4\mu} = \left( \sqrt{q^2+m_4^2}, -q \cos \theta, -q \sin \theta,0 \right) , 
\end{eqnarray}
where $k$ and $q$ are the amplitude of 3-momenta of initial and final particles. 
Moreover, $m_2$ and $m_4$ are equal to the mass of the scalar field $m$. 
The energy conservation law gives,
\begin{eqnarray}
E=\sqrt{k^2+m_1^2}+\sqrt{k^2+m^2}=\sqrt{q^2+m_3^2}+\sqrt{q^2+m^2}.\label{Etot}
\end{eqnarray}
In the UV limit, $q$ is described by $k$ and masses, 
\begin{eqnarray}
q=k \left(
1 + \frac{m_1^2-m_3^2}{4k^2}- 
\frac{\left(m_1^2-m_3^2\right)\left(m_1^2-m^2 \right)} {16k^4} + \cO \left(k^{-6}\right)
\right) .
\end{eqnarray}
We define basis elements for vector to construct bases for gravitons.
Transverse elements are denoted as $t_\mu$ and $u_\mu$, while $l_\mu$ is the element for longitudinal. 
Their components is explicitly written in
\begin{eqnarray}
&&t_{1\mu} = \left( 0,0,1,0 \right) ,\qquad
l_{1\mu} = \left( k, \sqrt{k^2+m_1^2}, 0,0 \right) / m_1 , \nonumber  \\
&&t_{3\mu} = \left( 0 , - \sin \theta,  \cos \theta,0 \right), \quad
l_{3\mu} = \left( q, \sqrt{q^2+m_3^2} \cos \theta, \sqrt{q^2+m_3^2} \sin \theta,0 \right)/m_3 ,\\
&&u_\mu \left ( =u_{1\mu}  =u_{3\mu} \right)  = \left( 0,0,0,1 \right).
\nonumber
\end{eqnarray}
The basis elements for graviton is constructed from the vector elements~\cite{Abe:2020ikj,Aubert:2003je}, 
\begin{eqnarray}
&&\e_{i\mu\nu}^{(0)} = \frac2{\sqrt {6}} l_{i\mu}l_{i\nu}-\frac1{\sqrt{6}} t_{i\mu}t_{i\nu}-\frac1{\sqrt{6}} u_{\mu}u_{\nu} ,\quad
\e_{i\mu\nu}^{(1,e)} = \frac1{\sqrt {2}} \left( l_{i\mu}t_{i\nu}+ t_{i\mu}l_{i\nu} \right), \nonumber \\
&&\e_{i\mu\nu}^{(1,o)} = \frac1{\sqrt {2}} \left( l_{i\mu}u_{\nu}+ u_{\mu}l_{i\nu} \right),\ 
\e_{i\mu\nu}^{(2,e)} = \frac1{\sqrt {2}} \left( t_{i\mu}t_{i\nu}- u_{\mu}u_{\nu} \right), \label{basisele}\\
&&\e_{i\mu\nu}^{(2,o)} = \frac1{\sqrt {2}} \left( t_{i\mu}u_{\nu}+ u_{\mu}t_{i\nu} \right), \quad
e_{i\mu\nu}^{(S)} = \frac1{\sqrt {3}} \left( l_{i\mu}l_{i\nu}+t_{i\mu}t_{i\nu}+u_{\mu}u_{\nu} \right) = \frac1{\sqrt{3}} \theta_{i\mu\nu} . \nonumber
\end{eqnarray}
Here, $\e_{i\mu\nu}^{(0)}$, $\{\e_{i\mu\nu}^{(1,e)},\e_{i\mu\nu}^{(1,o)}\}$, $\{\e_{i\mu\nu}^{(2,e)},\e_{i\mu\nu}^{(2,o)}\}$
and $e_{i\mu\nu}^{(S)}$ are helicity-0 mode, helicity-1 modes, helicity-2 modes and scalar mode, respectively.
For massless graviton, the basis elements of helicity-2 modes, which are only onshell states,
can be obtained by taking massless limit $m_i \to 0$.
Although in the massless limit the helicity-0 and -1 elements are ill-defined,
it does not matter because the massless graviton does not have helicity-0 and -1 modes.

We can categorize the basis elements into two groups: $\{ \e_{i\mu\nu}^{(0)},\e_{i\mu\nu}^{(1,e)},\e_{i\mu\nu}^{(2,e)} ,e_{i\mu\nu}^{(s)} \}$ and $\{ \e_{i\mu\nu}^{(1,o)},\e_{i\mu\nu}^{(2,o)} \}$. 
The former and the latter are named even and odd modes, respectively, which stem from the number of $u_\mu$. 
Since the inner product with $u_\mu$ and the others ($t_\mu$, $l_{\mu}$, $k_{\mu}$) gives zero, 
any scalar quantity, including scattering amplitude, with the odd number of $u_{\mu}$ vanishs. 
This leads to the fact that the scattering amplitudes involving both even and odd modes, all $\Tr[\e_1 \cdot \e_3]$, 
$(k \cdot \e_1 \cdot \e_3 \cdot k)$ and $(k \cdot \e_1 \cdot k)(k \cdot \e_3 \cdot k)$ become zero 
for any $k$. 
Therefore only the scattering amplitudes with two odd gravitons or two even gravitons give non-zero values.
Let us see each case.

\subsection{Scattering amplitudes in UV limit} \label{SAeach}

The UV behaviors of scattering amplitudes are shown in this subsection. 
Since the calculations are lengthy, we give the derivations in Appendix~\ref{AppUV}.  
Although the scattering amplitudes are shown in the previous section, 
they depend on the basis elements $e_{\mu\nu}$.
Since the different basis elements $e_{\mu\nu}$ have the different momentum dependence, 
the individual analysis for each is required. 
As in Weinberg-Salam theory~\cite{Lee:1977eg}, 
the nontrivial cancelation occurs, and thus 
the naive estimation by separately checking the total-energy dependence of each term gives the false result. 
The careful calculations are required.

\subsubsection{$h^{(2,o)}\phi \, \to h^{(2,o)}\phi$} \label{odd22}

The basis elements $e_{1\mu\nu}^{(2,o)}$ and $e_{3\mu\nu}^{(2,o)}$ become
\begin{eqnarray}
e_{1\mu\nu}^{(2,o)} = \frac1{\sqrt {2}} \left( t_{1\mu}u_{\nu}+ u_{\mu}t_{1\nu} \right),\qquad
e_{3\mu\nu}^{(2,o)} = \frac1{\sqrt {2}} \left( t_{3\mu}u_{\nu}+ u_{\mu}t_{3\nu} \right).
\label{basis2o}
\end{eqnarray}
Substituting the aboves into Eq.\eq{sum1} we have
\begin{eqnarray}
{\A}_s +{\A}_t +{\A}_u +{\A}_d 
=
-\kappa^2 k^2\frac{1+\cos \theta }{1-\cos \theta}
+\cO\left( k^0 \right)
\end{eqnarray}
The UV behavior of scattering amplitude has $k^2$ dependence. 
This dominant contribution  does not depend on the initial nor final graviton mass, 
which means that the UV behavior of scattering amplitude is independent of 
whether the initial and final graviton is massless or negative-norm massive state.

\subsubsection{$h^{(2,o)}\phi \, \to I^{(1,o)}\phi$}\label{odd21}

The basis elements for them are described as
\begin{eqnarray}
\e_{1\mu\nu}^{(2,o)} = \frac1{\sqrt {2}} \left( t_{1\mu}u_{\nu}+ u_{\mu}t_{1\nu} \right),\qquad
\e_{3\mu\nu}^{(1,o)} = \frac1{\sqrt {2}} \left( l_{3\mu}u_{\nu}+ u_{\mu}l_{3\nu} \right),
\end{eqnarray}
The scattering amplitude becomes
\begin{eqnarray}
{\A}_s +{\A}_t +{\A}_u +{\A}_d 
=
\kappa^2\frac{m_I k  \sin \theta }{2(1-\cos \theta)}
+\cO\left( k^{-1} \right)
\end{eqnarray}
Note that all terms in $k^3$ order are canceled. 
The dominant contribution does not depend on the initial graviton (helicity-2 graviton) mass but depends on the final graviton (helicity-1 graviton) mass.

\subsubsection{$I^{(1,o)}\phi \, \to I^{(1,o)}\phi$}\label{odd11}

The basis elements for them become
\begin{eqnarray}
\e_{1\mu\nu}^{(1,o)} = \frac1{\sqrt {2}} \left( l_{1\mu}u_{\nu}+ u_{\mu}l_{1\nu} \right),\qquad
\e_{3\mu\nu}^{(1,o)} = \frac1{\sqrt {2}} \left( l_{3\mu}u_{\nu}+ u_{\mu}l_{3\nu} \right).
\end{eqnarray}
The  scattering amplitude behaves
\begin{eqnarray}
{\A}_s +{\A}_t +{\A}_u +{\A}_d 
=
-\frac{\kappa^2 m_I^2}{8} \frac{2^2 +(1+\cos\theta)^2 +(1-\cos\theta)^2 }{(1-\cos \theta)^2}+ 
\cO\left( k^{-2} \right),
\end{eqnarray}
where the cancelation occurs in  $k^4$ and $k^2$ orders. 

\subsubsection{$h^{(2,e)}\phi \, \to h^{(2,e)}\phi$} \label{22}

The basis elements for them are given as
\begin{eqnarray}
\e_{1\mu\nu}^{(2,e)} = \frac1{\sqrt {2}} \left( t_{1\mu}t_{1\nu}- u_{\mu}u_{\nu} \right),\qquad
\e_{3\mu\nu}^{(2,e)} = \frac1{\sqrt {2}} \left( t_{3\mu}t_{3\nu}- u_{\mu}u_{\nu} \right).
\label{basis2e}
\end{eqnarray}
Substituting the above into Eq.\eq{sum1}, we have 
\begin{eqnarray}
&&{\A}_s +{\A}_t +{\A}_u +{\A}_d 
=\kappa^2 k^2  \frac{1+\cos \theta}{1-\cos \theta}
+
\cO\left( k^0 \right) .
\end{eqnarray}
As is the case of \ref{odd22} the dominant  $k^2$-order contribution, does not depends on the initial and final graviton masses. 

\subsubsection{$h^{(2,e)}\phi \, \to I^{(1,e)}\phi$}

The  basis elements are written as
\begin{eqnarray}
\e_{1\mu\nu}^{(2,e)} = \frac1{\sqrt {2}} \left( t_{1\mu}t_{1\nu}- u_{\mu}u_{\nu} \right),\qquad
\e_{3\mu\nu}^{(1,e)} = \frac1{\sqrt {2}} \left( l_{3\mu}t_{3\nu}+ t_{3\mu}l_{3\nu} \right).
\end{eqnarray}
They give
\begin{eqnarray}
&&{\A}_s +{\A}_t +{\A}_u +{\A}_d 
=
-\kappa^2\frac{m_I k  \sin \theta }{2(1-\cos \theta)}
+\cO\left( k^{-1} \right) .
\end{eqnarray}
The cancelation occurs in $k^3$ order and  
the dominant contribution does not depend on the initial graviton mass 
as is the case of \ref{odd21}.

\subsubsection{$I^{(1,e)}\phi \, \to I^{(1,e)}\phi$}
Substituting the basis elements for them
\begin{eqnarray}
\e_{1\mu\nu}^{(1,e)} =\frac1{\sqrt {2}} \left( l_{1\mu}t_{1\nu}+ t_{1\mu}l_{1\nu} \right), \qquad
\e_{3\mu\nu}^{(1,e)} =\frac1{\sqrt {2}} \left( l_{3\mu}t_{3\nu}+ t_{3\mu}l_{3\nu} \right)
\end{eqnarray}
into Eq.\eq{sum1}, we have 
\begin{eqnarray}
&&{\A}_s +{\A}_t +{\A}_u +{\A}_d 
=
-\frac{\kappa^2 m_I^2}{8} \frac{2^2 +(1+\cos\theta)^2 +(1-\cos\theta)^2 }{(1-\cos \theta)^2}+ 
\cO\left( k^{-2} \right) ,
\end{eqnarray}
where the cancelation occurs 
as is the case of \ref{odd11}.

\subsubsection{$h^{(2,e)}\phi \, \to I^{(0)}\phi$}
In this case, we have
\begin{eqnarray}
\e_{1\mu\nu}^{(2,e)}  \!\!\!\! &  = & \!\!\!\! \frac1{\sqrt {2}} \left( t_{1\mu}t_{1\nu}- u_{\mu}u_{\nu} \right), \nonumber \\
\e_{3\mu\nu}^{(0)} \!\!\!\! &  = & \!\!\!\! \frac2{\sqrt {6}} l_{3\mu}l_{3\nu} - \frac1{\sqrt{6}} \left( t_{3\mu}t_{3\nu}+ u_{\mu}u_{\nu} \right) 
\nonumber \\ 
\!\!\!\! &  = & \!\!\!\! 
-\frac{2}{\sqrt{6}k_3^2} k_{3\mu}k_{3\nu} + \frac{1}{\sqrt{6}} \left( 2\eta_{\mu\nu} -3 t_{3\mu}t_{3\nu}-3 u_{\mu}u_{\nu} \right).
\end{eqnarray}
The cancelation occurs in $k^4$ and $k^2$ orders in the scattering amplitude and finally it becomes
\begin{eqnarray}
&&\A_s+\A_t+\A_u+\A_d 
= \cO(k^0).
\end{eqnarray}

\subsubsection{$I^{(1,e)}\phi \, \to I^{(0)}\phi$}

Substituting
\begin{eqnarray}
&&\e_{1\mu\nu}^{(1,e)} =\frac1{\sqrt {2}} \left( l_{1\mu}t_{1\nu}+ t_{1\mu}l_{1\nu} \right), \nonumber \\
&&\e_{3\mu\nu}^{(0)} =
-\frac{2}{\sqrt{6}k_3^2} k_{3\mu}k_{3\nu} + \frac{1}{\sqrt{6}} \left( 2\eta_{\mu\nu} -3 t_{3\mu}t_{3\nu}-3 u_{\mu}u_{\nu} \right), 
\end{eqnarray}
into Eq.\eq{sum1}, we have
\begin{eqnarray}
&&\A_s+\A_t+\A_u+\A_d 
= 
 \cO(k^{-1}),
\end{eqnarray}
where the cancelation occurs in $k^5$, $k^3$ and $k^1$ orders.

\subsubsection{$I^{(0)}\phi \, \to I^{(0)}\phi$}

The basis elements are written in 
\begin{eqnarray}
&&\e_{1\mu\nu}^{(0)} =
-\frac{2}{\sqrt{6}k_1^2} k_{1\mu}k_{1\nu} + \frac{1}{\sqrt{6}} \left( 2\eta_{\mu\nu} -3 t_{1\mu}t_{1\nu}-3 u_{\mu}u_{\nu} \right), \nonumber \\
&&\e_{3\mu\nu}^{(0)} =
-\frac{2}{\sqrt{6}k_3^2} k_{3\mu}k_{3\nu} + \frac{1}{\sqrt{6}} \left( 2\eta_{\mu\nu} -3 t_{3\mu}t_{3\nu}-3 u_{\mu}u_{\nu} \right).
\end{eqnarray}
The scattering amplitude becomes
\begin{eqnarray}
\A_s + \A_t+ \A_u + \A_c = 
-\frac{\kappa^2}{2} \frac{m_I^2(1+\cos \theta)}{(1-\cos \theta)^2}  -\frac{\kappa^2}{8} (m_I^2 +2m^2)   + \cO(k^{-2}),
\end{eqnarray}
where the cancelation occurs in $k^6$, $k^4$ and $k^2$ orders.

\subsubsection{$h^{(2,e )}\phi \, \to I^{(s)}\phi$}
Substituting the basis element for the initial graviton
\begin{eqnarray}
\e_{1\mu\nu}^{(2,e)}  \!\!\!\! &  = & \!\!\!\! \frac1{\sqrt {2}} \left( t_{1\mu}t_{1\nu}- u_{\mu}u_{\nu} \right),
\end{eqnarray}
into Eq.\eq{sum2}, we have
\begin{eqnarray}
\A_s + \A_t+ \A_u + \A_c 
=  \cO(k^{0}). 
\end{eqnarray}

\subsubsection{$I^{(1,e )}\phi \, \to I^{(s)}\phi$}
The basis element for the initial graviton is expressed as
\begin{eqnarray}
\e_{1\mu\nu}^{(1,e)} =\frac1{\sqrt {2}} \left( l_{1\mu}t_{1\nu}+ t_{1\mu}l_{1\nu} \right), 
\end{eqnarray}
and the amplitude behaves as
\begin{eqnarray}
\A_s + \A_t+ \A_u + \A_c =  \cO(k^{-1}).
\end{eqnarray}
Note that the cancelation occurs in $k$ order, in addition to that in the derivation of Eq.\eq{sum2}.

\subsubsection{$I^{(0)}\phi \, \to I^{(s)}\phi$}
The basis element for the initial graviton is
\begin{eqnarray}
&&\e_{1\mu\nu}^{(0)} =
-\frac{2}{\sqrt{6}k_1^2} k_{1\mu}k_{1\nu} + \frac{1}{\sqrt{6}} \left( 2\eta_{\mu\nu} -3 t_{1\mu}t_{1\nu}-3 u_{\mu}u_{\nu} \right).
\end{eqnarray}
Then, the amplitude behaves as
\begin{eqnarray}
\A_s + \A_t+ \A_u + \A_c =  \cO(k^{0}).
\end{eqnarray}
The cancelation occurs in $k^2$ order.

\subsubsection{$I^{(s)}\phi \, \to I^{(s)}\phi$}
The amplitude is given in Eq.\eq{sum3}, which indicates the scattering amplitude behaves as
\begin{eqnarray}
&&\A_s + \A_t+ \A_u + \A_c= 
-\frac{\kappa^2}{6} \left(m_S^2+m^2\right) - \frac{1}{12\beta}+8 \xi \kappa^2 m_S^2 + \cO\left(k^{-2}\right). 
\label{IsIs}
\end{eqnarray}
Each contribution, $\A_s$, $\A_t$, $\A_u$ and $\A_c$, indeed has $k^6$ terms, and hence 
the cancelation occurs in the summation.

\section{$S$-matrix unitarity} \label{SMU}

The $S$-matrix unitarity 
\begin{eqnarray}
S S^\dagger = I \label{SS}
\end{eqnarray}
should be satisfied in any healthy theories, 
where $I$ is the identity matrix. 
In theories without negative norm states, 
to see the perturbative unitarity of a theory in the UV limit,
we usually check the unitarity bound for amplitude of two-body elastic scattering $\A(2 \to 2)$ at tree level.
The unitarity bound is written in 
\begin{eqnarray}
|\A | \le E^\alpha, \qquad \alpha \le 0 \qquad (E\to \infty)
\label{UB}
\end{eqnarray}
where $E$ is the total energy in the center-of-mass frame. 
Since in our 2-2 scattering problem the total energy $E$ is given by $E \simeq 2k$  at high energy (see Eq.\eq{Etot}),  
the UV limit means the limit $E\to\infty$ or $k\to\infty$. 
In theories without ghost fields, the bound \eq{UB} is a necessary consequence from \eq{SS}.
Since $R_{\mu\nu}^2$ gravity has negative norm states, however
the unitarity bound is not necessarily satisfied. 
Instead of the unitarity bound, we will check its analog for the $S$-matirix unitarity which is valid in the theories with negative norm states. 
To construct the analog, we begin with the optical theorem, which is the starting point of the derivation of  the unitarity bound \eq{UB}.

The matrix element of $T$ is obtained from the scattering amplitude $\A$ as
\begin{eqnarray}
\langle \Phi |T| \Psi \rangle = \delta^{4} \left( {p}_{\Psi} - {p}_{\Phi} \right) \A \left( \Psi \to \Phi \right).
\label{defM}
\end{eqnarray}
The optical theorem can be written by the scattering amplitude $\A$ as
\begin{eqnarray}
2{\rm{Im}}\, \A \left(\Psi \to \Psi\right)=\sum_{\Phi} \epsilon_\Phi \delta^{4}  \left( {p} - {p}_\Phi \right)|\A\left(\Psi\to \Phi\right)|^2,
\label{OT}
\end{eqnarray}
where the summation is over all possible intermediate on-shell states, and 
$\epsilon_\Phi$ is $1$ or $-1$, if the state $\Phi$ is normalized, depending on the norm of $\Phi$ being positive or negative.
We consider the following inequality,
\begin{eqnarray}
|\A(\Psi \to \Psi)|\ge \mbox{Im} \, \A(\Psi \to \Psi) = \sum_{\Phi} \epsilon_\Phi \delta^{4}  \left( {p} - {p}_\Phi \right)|\A\left(\Psi\to \Phi\right)|^2, \label{op}
\end{eqnarray}
where we ignore the unimportant numerical factors. 
If theory does not have negative norm states, all $\epsilon_\Phi$ are positive, and thus we have%
\footnote{In the four dimensional spacetime, the dependence of $\sum_{\Phi} \epsilon_\Phi \delta^{4} \left( {p}_{\Psi} - {p}_{\Phi} \right)$ on total energy in the UV region is estimated as $E^0$. It leads to the inequality \eq{UB}. In other dimension, the bound for $\alpha$ in \eq{UB} is different.} 
\begin{eqnarray}
|\A(\Psi \to \Psi)|\ge \mbox{Im} \, \A(\Psi \to \Psi) = \sum_{\Phi} \delta^{4}  \left( {p} - {p}_\Phi \right)|\A\left(\Psi\to \Phi\right)|^2 \ge | \A\left(\Psi\to \Psi\right)|^2. 
\label{op1}
\end{eqnarray}
This inequality is simplified into
\begin{eqnarray}
1 \ge | \A\left(\Psi\to \Psi\right)|,
\end{eqnarray}
which results in the unitarity bound \eq{UB}. 
However, if theory has negative norm states, the right hand side of the optical theorem \eq{OT} (or Eq.\eq{op}) is not bounded as the last inequality in Eq.\eq{op1}. 
Therefore, we should directly analyse the inequality \eq{op}. 
Hence, the inequality \eq{op} is the analog of the unitarity bound for $S$-matrix unitarity.

The tree-level approximation of the unitarity bound \eq{UB}  represents unitarity of theories well, such as gauge theories and Einstein gravity.  
As is the case in the unitarity bound, the tree-level approximation of the analog is expected to give the information of $S$-matrix unitarity, 
which has already been discussed in scalar field theory~\cite{Abe:2018rwb}. 
Therefore, we study the analog  of the unitarity bound for $S$-matrix unitarity, that is the inequality \eq{op}, at tree level.

Let us move to the analysis of the inequality (\ref{op}). 
We have to rewrite the inequality of Eq.(\ref{op}) in the terms of momentum basis, 
which gives some contributions to order of total energy from summation and delta functions for energy-momentum. 
However, in four dimension, such contributions are accidentally canceled.%
\footnote{It is really accidental. In other dimension, such contribution is important and a factor depending on $E$ in Eq.\eq{ineqA} appears.
The detail is discussed in~\cite{Fujimori:2015wda,Fujimori:2015mea}}
Suppose that scattering amplitudes for $(h^{(\sigma)}\phi \to h^{(\sigma')}\phi )$, $(h^{(\sigma)}\phi \to I^{(S)}\phi )$ and $(I^{(S)}\phi \to I^{(S)}\phi )$ behave in UV limit ($E\to\infty$) as
\begin{eqnarray}
&&\A (h^{(\sigma)}\phi \to h^{(\sigma')}\phi ) \to \beta_{\sigma\sigma'} E^{\alpha_{\sigma\sigma'}} , \\
&&\A (h^{(\sigma)}\phi \to I^{(S)}\phi)  \to \beta_{\sigma s} E^{\alpha_{\sigma s}} ,\\
&&\A (I^{(S)}\phi \to I^{(S)}\phi)  \to \beta_{ss} E^{\alpha_{ss} } .
\end{eqnarray}
Then, the inequality of Eq.(\ref{op}) for $\Phi = h^{(\sigma)}\phi$ is written as
\begin{eqnarray}
\left| \beta_{\sigma\sigma} E^{\alpha_{\sigma\sigma}} \right| 
&\simeq& \left| \A (h^{(\sigma)}\phi \to h^{(\sigma)}\phi ) \right|
\nonumber \\
&\ge& \left|  \sum_\Phi \epsilon_\Phi \left| \A (h^{(\sigma)}\phi \to \Phi ) \right|^2 \right|
\nonumber \\
&\simeq& \left|  \sum_{\sigma'} \epsilon_{\sigma'} \left| \A (h^{(\sigma)}\phi \to h^{(\sigma')}\phi ) \right|^2 
- \sum_{s} \left| \A (h^{(\sigma)}\phi \to I^{(s)}\phi ) \right|^2 \right|
\nonumber \\
&\simeq& \left|  \sum_{\sigma'} \epsilon_{\sigma'} \left| \beta_{\sigma\sigma'} \right|^2 E^{2\alpha_{\sigma\sigma'}}
- \sum_{s} \left|\beta_{\sigma s} \right|^2 E^{2\alpha_{\sigma s}} \right| , \label{ineqA}
\end{eqnarray}
where, in the last (approximated) equality, we ignore the states consisting of more than two particles in the summation
which is higher order in the perturbative expansion. 
The right hand side of Eq.\eq{ineqA} has 
$\left| \beta_{\sigma\sigma} \right|^2 E^{2\alpha_{\sigma\sigma}}$ in the summation, 
while the total energy dependence of the left hand side is $E^{\alpha_{\sigma\sigma}}$.
Thus, in the naive estimation, 
if $\alpha_{\sigma\sigma} > 0$ (for helicity-two mode it is true since $\alpha_{\sigma\sigma}$ is two), the right hand side might be larger than the left hand side in the UV limit $E \to \infty$, 
which means the violation of the inequality. 
However, since the right hand side of Eq.\eq{ineqA} has not only positive contribution but also negative ones due to the negative norm states, 
there is a possibility that cancelation occurs and then the inequality is satisfied. 
We will see that it is the case in the gravity theory with quadratic curvatures.

\begin{table}[t]
  \begin{minipage}[t]{.45\textwidth}
    \begin{center}
  \begin{tabular}{|c|c|}  \hline
$\quad h^{(2,o)}\phi \, \to h^{(2,o)}\phi\quad $ & $\quad \alpha = 2\quad $ \\
$\quad h^{(2,o)}\phi \, \to h^{(1,o)}\phi\quad $ & $\quad \alpha = 1\quad $ \\
$\quad h^{(1,o)}\phi \, \to h^{(1,o)}\phi\quad $ & $\quad \alpha =0\quad $ \\ 
 \hline
  \end{tabular}
    \end{center}
    \caption{}
    \label{T1}
  \end{minipage}
   \hfill
  \begin{minipage}[t]{.45\textwidth}
    \begin{center}
  \begin{tabular}{|c|c|}  \hline
$\quad h^{(2,e)}\phi \, \to h^{(2,e)}\phi\quad $ & $\quad \alpha = 2\quad $ \\
$\quad h^{(2,e)}\phi \, \to h^{(1,e)}\phi\quad $ & $\quad \alpha = 1\quad $ \\
$\quad h^{(2,e)}\phi \, \to h^{(0)}\phi\quad $ & $\quad \alpha = 0\quad $ \\
$\quad h^{(2,e)}\phi \, \to I^{(s)}\phi\quad $ & $\quad \alpha = 0\quad $ \\
$\quad h^{(1,e)}\phi \, \to h^{(1,e)}\phi\quad $ & $\quad \alpha =0\quad $ \\
$\quad h^{(1,e)}\phi \, \to h^{(0)}\phi\quad $ & $\quad \alpha \le -1\quad $ \\
$\quad h^{(1,e)}\phi \, \to I^{(s)}\phi\quad $ & $\quad \alpha \le -1\quad $ \\
$\quad h^{(0)}\phi \, \to I^{(0)}\phi\quad $ & $\quad \alpha = 0\quad $ \\
$\quad h^{(0)}\phi \, \to I^{(s)}\phi\quad $ & $\quad \alpha \le 0\quad $ \\
$\quad I^{(s)}\phi \, \to I^{(s)}\phi\quad $ & $\quad \alpha = 0\quad $ \\  \hline
  \end{tabular}
    \end{center}
    \caption{}
    \label{T2}
  \end{minipage}
\end{table}

The high energy behavior of scattering amplitude derived in the previous subsection is summarized in Table \ref{T1} and \ref{T2}. 
We find that some of scattering amplitudes violate the unitarity bound. 
For instance, the scattering amplitude for $h^{(2,e)}\phi \, \to h^{(2,e)}\phi$ diverges as $\propto$ $k^2$,
which does not satisfy the unitarity bound~\eqref{UB}. 
However, as we explained,  it does not mean the violation of the $S$-matrix unitarity due to the negative norm states. 
In order to explore the $S$-matrix unitarity, we should see the inequality \eq{ineqA}.  
Let us see the case for the forward scattering of the scalar field $\phi$ and the massless positive norm graviton $H^{(2,e)}$,
\begin{eqnarray}
&&\left|\A(H^{(2,e)}\phi \, \to H^{(2,e)}\phi)\right| \ge \sum_\sigma \left|\A(H^{(2,e)}\phi \, \to H^{(\sigma)} \phi )\right|^2
\nonumber \\
&&\qquad\qquad
-\sum_\sigma \left|\A(H^{(2,e)}\phi \, \to I^{(\sigma)} \phi )\right|^2
+ \left|\A(H^{(2,e)}\phi \, \to I^{(s)} \phi )\right|^2. \label{op22}
\end{eqnarray}
The left hand side behaves as $\propto$ $k^2$, while one of the terms in the right hand side $\left|\A(H^{(2,e)}\phi \, \to H^{(2,e)}\phi)\right|$ behaves as $\propto$ $k^4$. 
Therefore, in the naive estimate, 
in the high energy limit $k^4$ terms in the right hand side becomes larger than the left hand side, and  
the inequality of Eq.\eq{op22} seems to be violated. 
However, since a cancelation occurs, the naive estimate does not give the correct result. 
In the right hand side, $\left|\A(H^{(2,e)}\phi \, \to I^{(2,e)}\phi)\right|$ appears with the negative sign. 
As we see in section \ref{SAUV}, the leading-order behaviors of $\left|\A(H^{(2,e)}\phi \, \to H^{(2,e)}\phi)\right|$ and $\left|\A(H^{(2,e)}\phi \, \to I^{(2,e)}\phi)\right|$ are the same, 
because the dominant contribution of the amplitude does not depend on the graviton mass.
Therefore, their $k^4$ terms are canceled and the sub-leading parts behave as $\propto$ $k^2$. 
Other terms in the inequality of Eq.\eq{op22}, such as $\left|\A(H^{(2,e)}\phi \, \to H^{(1,e)}\phi)\right|$, 
behave as  $\propto$ $k^2$ or the lower power. 
Thus, the right hand side behaves as $\propto$ $k^2$. 
Then, the leading-order terms in the left and right hand sides have the same power. 
The right hand side is supressed by the coupling constant (which should be small in the perturbative analysis), 
the inequality of Eq.\eq{op22} is satisfied. 

In conclusion, for all the other cases, we can see that the similar inequality obtained from the optical theorem is satisfied. 
It means that $S$-matrix unitarity is satisfied.


\section{Summary}
\label{Summary}

In this paper, we have given the matter-graviton scattering amplitude in the quadratic gravity. 
We have seen that the {\it perturbative $S$-matrix unitarity} ($SS^\dagger =1$) is satisfied, although the {\it tree unitarity} fails. 
The tree unitarity was introduced to investigate the UV consistency in full order perturbation~\cite{Bell:1973ex,Cornwall:1973tb,LlewellynSmith:1973yud,Cornwall:1974km,Fujimori:2015wda,Fujimori:2015mea}. 
However, it was pointed out that the derivation of the tree unitarity is valid only if the theory has no negative norm states~\cite{Abe:2018rwb}. 
The {\it perturbative $S$-matrix unitarity} is a generalization of the tree unitarity and it is applicable to the theories with negative norm states as wll. 
Therefore, the analysis of the {\it perturbative $S$-matrix unitarity} is expected to give a new sight on understanding the UV consistency, as the tree unitarity contributed to the progress in the particle physics~\cite{Lee:1977eg}.
Our result that the quadratic gravity satisfies the {\it perturbative $S$-matrix unitarity} 
shows that the quadratic gravity behaves better in the UV than the Einstein gravity which violates the tree unitarity~\cite{Berends:1974gk},
and it should be related to the renormalizability as 
discussed for the tree unitarity\cite{Bell:1973ex,Cornwall:1973tb,LlewellynSmith:1973yud,Cornwall:1974km,Fujimori:2015wda,Fujimori:2015mea}.

One concern about the quadratic gravity is the existence of negative norm excitations. 
Negative norm states violate the interpretation of probability in quantum system. 
However, all the negative norm states in the quadratic gravity are massive such that 
all of them decay into light particles, and thus they do not appear as asymptotic states~\cite{Lee:1969fy,Lee:1970iw}. 
Therefore, the Hilbert space in this theory, which should be defined in the asymptotic region as usual, has no negative norm states.
This idea has been recently applied to the quadratic gravity~\cite{Holdom:2021hlo}. 
The disappearance of negative norm asymptotic states may mean a possibility that
there exists a sector without negative norm states in psuedo-Hilbert space of the quadratic gravity, 
although it would not be written in local operators, similar to dressed states. 
This may imply that, for the description of quantum gravity with local operators, an infinite number of operators are required. 

The information paradox problem of black holes can be said to be the problem of unitarity. 
This problem appears to stem from the fact that Einstein gravity is not unitary. 
The problem roughly states that the correspondence between initial and final states is not a one-to-one relation. 
The {\it $S$-matrix unitarity}, that is shown to be satisfied for the quadratic gravity in this paper, quarantees
the one-to-one correspondence between initial and final states, even if negative norm states appear.
Naively it seems to us that the solution to the information paradox problem can be seen in the quadratic gravity. 
This is an interesting future direction.

The exact form of scattering amplitudes, that we have derived, helps us with revealing the other properties of the quadratic gravity. 
For instance, the quadratic curvature correction of gravity enables us to predict the stability of  the Higgs potential with quantum gravitational corrections near Planck scale ~\cite{Smolin:1979ca,Bhattacharjee:2012my,Abe:2016irv,Salvio:2014soa,Abe:2016hdj}. 
Our results of the amplitude of matter-graviton scattering would provide other insight into the effect of UV physics from the viewpoint of particle phenomenology. 
Furthermore,  the gravitational positivity bound is recently discussed in Einstein gravity~\cite{Tokuda:2020mlf,Alberte:2020jsk,Noumi:2021uuv}, 
aiming at deriving constraints in the inflation cosmology. 
It is interesting to investigate whether different constraints may arise in the quadratic gravity. 
The difficult problem of the negative-norm graviton is recently discussed in Ref.~\cite{Donoghue:2019fcb}. 
Together with these, namely, the gravitational positivity bound, treatment with the negative-norm graviton and our results of 
the scattering amplitude, the quadratic gravity appears as a UV complete quantum gravity theory. 
These topics are left for the future works.

\section*{Acknowledgements}

Y. A. is supported by JSPS Grants-in-Aid for Early-Career
Scientists (No. 19K14719). 
K. I. is supported by Grant-Aid for Scientific Research from Ministry of Education, Science, Sports and Culture of Japan (Nos. 17H01091, JP20H01902, JP21H05182, JP21H05189) and JSPS Bilateral Joint Research Projects (JSPS-DST
collaboration) (JPJSBP120227705). 


\appendix


\section{Quantization of massive graviton $I_{\mu\nu}$} \label{mq}
In this appendix, we give the canonical quantization for massive graviton $I_{\mu\nu}$. 
The canonical quantization for graviton is given in \cite{Kimura:1976px,Nakanishi:1977gt} for the massless graviton, in \cite{Chang:1966zza}  for the massive graviton, 
in \cite{Kawasaki:1981zw} for the quadratic gravity and in \cite{Kubo:2022lja} for the conformal gravity. 
Some of them involve the  Faddeev-Popov ghosts in addition to the gauge fixing terms. 
Since they do not contribute to the tree amplitudes that we focus on, 
in this appendix we give the minimal requirement for our purpose.

The quadratic action for $I_{\mu\nu}$ is given in \eq{ga5}, 
\begin{eqnarray}
S_{I2}=
-\frac1{4\kappa^2} \int d^4 x \, \left[
 I_{\mu\nu} {\cal L}^{\mu\nu,\alpha\beta}I_{\alpha\beta}
+ \frac1{\beta \kappa^2} \left( I_{\mu\nu}^2- \frac{\alpha}{4\alpha+\beta}I^2\right)  
\right]. \label{ie1}
\end{eqnarray}
We take the Cartesian coordinate 
\begin{eqnarray}
ds^2=-dt^2+d{\bm x}^2,  
\end{eqnarray}
which gives the $(1+3)$-decomposition of spacetime based on constant $t$ surface. 
The action in this decomposition becomes
\begin{eqnarray}
S_{I2}\!\!\!\! &  = & \!\!\!\!
-\frac1{4\kappa^2} \int d^4 x \, \Biggl[
2I_{00}\left( \Delta \delta_{ij} - \p_i \p_j \right) I_{ij}
-2I_{0i}\left( \Delta \delta_{ij} - \p_i \p_j \right) I_{0j}
\nonumber  \\
&& \qquad \qquad
+ 2 \dot I_{ij} \left( 2 \p_k \delta_{ij} - \p_i \delta_{jk} - \p_j \delta_{ik} \right) I_{0k} 
+ \dot I_{ij}  {\cal G}_3^{ij,kl} \dot I_{kl}
+  I_{ij}  {\cal L}_3^{ij,kl} I_{kl}
\nonumber  \\
&& \qquad \qquad
+ \frac1{\beta \kappa^2} \left( I_{00}^2-2I_{0i}^2+I_{ij}^2- \frac{\alpha}{4\alpha+\beta}(-I_{00}+I_{ii})^2\right)  
\Biggr].
\label{2ActionHI}
\end{eqnarray}
where dot means time derivative and
\begin{eqnarray}
&&{\cal G}_3^{ij,kl}:= 
\delta_{i(k}\delta_{l)j}-\delta_{ij}\delta_{kl}, \\
&&{\cal L}_3^{ij,kl}:= 
\Delta \delta^{i(k}\delta^{l)j} 
- \p^i \p^{(k}\delta^{l)j} 
- \p^j \p^{(k}\delta^{l)i} 
-\Delta \delta^{ij}\delta^{kl}
+ \p^i \p^j\delta^{kl} 
+ \delta^{ij}\p^k \p^l. \label{ia1}
\end{eqnarray}
Here, Latin letters $\{i,j,\dots\}$ are the indices for the 3-dimensional space.

The canonical variables are obtained by the variation of action \eq{2ActionHI} with respect to $\dot I_{\mu\nu}$,
\begin{eqnarray}
&&\pi^{00}=0, \qquad  \pi^{0i}= 0, \label{PrimCons} \\
&&\pi^{ij}= -\frac1{2\kappa^2} \left[ \left( 2 \p_k \delta_{ij} - \p_i \delta_{jk} - \p_j \delta_{ik} \right) I_{0k} 
+   {\cal G}_3^{ij,kl} \dot I_{kl} \right] . \label{defpi}
\end{eqnarray}
The first two are interpreted as the constraints, while the last equation can be solved for $\dot I_{ij}$ as
\begin{eqnarray}
\dot I_{ij} = -{\cal G}_3^{-1}{}_{ij,kl} \left( 2 \kappa^2 \pi_{kl} +\left( 2 \p_m \delta_{kl} - \p_k \delta_{lm} - \p_l \delta_{km} \right) I_{0m} \right),
\end{eqnarray}
where ${\cal G}_3^{-1}{}_{ij,kl}$ is the inverse of  ${\cal G}_3^{ij,kl}$
\begin{eqnarray}
{\cal G}_3^{-1}{}_{ij,kl}:= 
\delta_{i(k}\delta_{l)j}-\frac12\delta_{ij}\delta_{kl}, 
\end{eqnarray}
satisfying 
\begin{eqnarray}
{\cal G}_3^{ij,kl}{\cal G}_3^{-1}{}_{kl,mn}:= \delta_{i(m}\delta_{n)j}.
\end{eqnarray}

The Hamiltonian density ${\cal H}$ is expressed in terms of the canonical momenta as 
\begin{eqnarray}
{\cal H} \!\!\!\! &  := & \!\!\!\! \pi^{ij}\dot I_{ij} - {\cal L} + \lambda_\mu \pi^{0\mu} \nonumber\\
\!\!\!\! &  = & \!\!\!\!
\frac1{4\kappa^2} 
\Biggl[ - 4 \kappa^4 \pi^{ij} {\cal G}_3^{-1}{}_{ij,kl} \pi^{kl} +8 \kappa^2 I_{0i} \p_j \pi^{ij}
+2I_{00}\left( \Delta \delta_{ij} - \p_i \p_j \right) I_{ij}
+  I_{ij}  {\cal L}_3^{ij,kl} I_{kl}
\nonumber \\
&& \qquad \qquad
+ \frac1{\beta \kappa^2} \left( I_{00}^2-2I_{0i}^2+I_{ij}^2- \frac{\alpha}{4\alpha+\beta}(-I_{00}+I_{ii})^2\right)  
\Biggr]
 + \lambda_\mu \pi^{0\mu},
 \label{calH}
\end{eqnarray}
where $\lambda_\mu$ is a multiplier.
The time evolutions of the primary constraints \eqref{PrimCons} should be set to zero. This requirement reads as follows. Define
\begin{eqnarray}
&&\frac1{2\kappa^2}{\cal C}_0:= \dot \pi^{00} = [ {\cal H}, \pi_{00} ] = \frac {\delta {\cal H}}{\delta I_{00}} 
\nonumber \\
&&\hspace{6.5mm}
= \frac1{2\kappa^2} \left( \Delta \delta_{ij} - \p_i \p_j \right) I_{ij} + \frac1{2\beta \kappa^4} \left( I_{00}- \frac{\alpha}{4\alpha+\beta}(I_{00}+I_{ii})\right)  , \label{c0}
\\
&&2 {\cal C}_i:= \dot \pi^{0i} = [ {\cal H}, \pi_{0i} ] = \frac {\delta {\cal H}}{\delta I_{0i}} 
= 2 \p_j \pi^{ij}- \frac1{\beta \kappa^4} I_{0i}. \label{ci}
\end{eqnarray}
The secondary constraints amount to setting  
\begin{eqnarray}
&&{\cal C}_\mu =0.
\label{SecondCons}
\end{eqnarray}
The time evolution of the  secondary  constraints \eqref{SecondCons} are required to be zero, that is, $\dot {\cal C}_\mu =0$.
This is achieved by adjustment of the Lagrange multipliers $\lambda_\mu$ in Eq.~\eqref{calH}, 
and generates no more constraints.

According to Dirac's method of quantization, we put together all constraints in an $8$-component vector 
${\cal C}_A:=(\pi^{0\mu},{\cal C}_\nu )$, 
and calculate a matrix $M_{AB}:=[{\cal C}_A, {\cal C}_B]$. 
Here, $A$, $B$ run from $0$ to $7$; 
$0$ to $3$ corresponding to $0$ to $3$ of $\pi^{0\mu}$~'s $\mu$ 
and $4$ to $7$ to $0$ to $3$ of ${\cal C}_\mu$'s $\mu$, respectively. 
We note that $\mbox{det}M \neq 0$ is satisfied, and hence all $C_A=0$ are the second class constraints. 
The Dirac bracket should be applied for the quantization of the $I_{ij}$ and $\pi^{kl}$.
The $4\times 4$ submatrix appearing the upper left in $M_{AB}$ is zero, 
and thus, in the inverse submatrix, the lower  right $4\times 4$ submatrix is zero. 
The components with spacelike indices, $I_{ij}$ and $\pi^{ij}$, are commutative (in Poisson bracket) 
with the first four components of the constraints, {\it i.e.} $\pi^{0\mu}$. 
Therefore, for $I_{ij}$ and $\pi^{ij}$, Dirac bracket becomes the same as Poisson bracket,
\begin{eqnarray}
&&[I_{ij} ({\bm x}) ,\pi^{kl}({\bm y})]_D=[I_{ij}({\bm x}),\pi^{kl}({\bm y})]_P= i \delta_{i(k} \delta_{l)j} \delta^3 ({\bm x}-{\bm y}). 
\label{Ipi} \\
&&[I_{ij} ({\bm x}) ,I^{kl}({\bm y})]_D=[I_{ij}({\bm x}),I^{kl}({\bm y})]_P= 0, \\
&&[\pi_{ij} ({\bm x}) ,\pi^{kl}({\bm y})]_D=[\pi_{ij}({\bm x}),\pi^{kl}({\bm y})]_P= 0, \label{pipi}
\end{eqnarray}
where $[\cdot,\cdot]_D$ and $[\cdot,\cdot]_P$ are Dirac and Poisson brackets, respectively.

The variation of the action \eq{ie1} with respect to $I_{\mu\nu}$ gives the equation of motion (EoM) for $I_{\mu\nu}$:
\begin{eqnarray}
2{\cal L}^{\mu\nu,\alpha\beta}I_{\alpha\beta}
+ \frac2{\beta \kappa^2} \left( I_{\mu\nu}- \frac{\alpha}{4\alpha+\beta}I\eta_{\mu\nu}\right)  =0,\label{ei}
\end{eqnarray}
where we omit the overall constant factor. 
Operating $\p_\mu$ to \eq{ei} and noting $\p_\mu {\cal L}^{\mu\nu,\alpha\beta}I_{\alpha\beta}=0$, we have 
\begin{eqnarray}
\p_\mu I^{\mu\nu}= \frac{\alpha}{4\alpha+\beta} \p^\nu I. \label{ci}
\end{eqnarray}
Then, the trace of  \eq{ei} can be expressed as 
\begin{eqnarray}
0\!\!\!\! &  = & \!\!\!\! 2\left(\p_\alpha \p_\beta I^{\alpha\beta} - \Box I \right) + \frac1{\kappa^2} I  \nonumber \\
\!\!\!\! &  = & \!\!\!\!-2 (3\alpha + \beta) \Box I + \frac1{\kappa^2} I. \label{ti}
\end{eqnarray}
This equation shows that the trace part of $I_{\mu\nu}$ behaves as a scalar field with mass 
\begin{eqnarray}
m_S^2=\dfrac1{2 (3\alpha + \beta)\kappa^2}.
\end{eqnarray} 
The general solution is given as 
\begin{eqnarray}
I = \int d^3p \left\{ A_S ({\bm p}) e^{-ipx}+A_S^\ast ({\bm p}) e^{ipx} \right\} \hspace{10mm} \left(\mbox{where}\quad p_0=\sqrt{{\bm p}^2+m_S^2}\right) ,
\end{eqnarray} 
where $A_S$ is an integration constant depending on ${\bm p}$. 

We separate the degrees of freedom in $I_{\mu\nu}$ into that of $I$ and the rest by expressing the latter by $\tilde I_{\mu\nu}$. This can be made by writing
\begin{eqnarray}
\tilde I_{\mu\nu} = I_{\mu\nu}+ \frac{2 \kappa^2 \beta (3\alpha +\beta) } {3 (4\alpha + \beta)} \p_\mu \p_\nu I - \frac{3\alpha+\beta}{3(4\alpha+\beta)} I \eta_{\mu\nu}.
\end{eqnarray}
Then, $\tilde I_{\mu\nu}$ satisfies the transverse-traceless condition
\begin{eqnarray}
\p^\mu \tilde I_{\mu\nu} = 0, \qquad \tilde I =0,
\end{eqnarray}
where we use the EoM for trace mode $I$, Eq.\eq{ti}. 
Equation \eq{ei} becomes 
\begin{eqnarray}
\Box \tilde I_{\mu\nu} + \frac{1}{\beta \kappa^2 } \tilde I_{\mu\nu} =0 .
\end{eqnarray}
Hence, $\tilde I_{\mu\nu}$ is a transverse-traceless mode with mass $m_I^2 = -(\beta \kappa^2)^{-1}$, 
that is,  interpreted as a massive spin-2 field. 
The general solution for $\tilde I_{\mu\nu}$ is given as
\begin{eqnarray}
\tilde I_{\mu\nu} = \sum_{\sigma} \int  d^3p \left\{ A_T^{(\sigma)} ({\bm p}) e^{-ipx}+A_T^{(\sigma)}{}^\ast ({\bm p}) e^{ipx} \right\} e^{(\sigma)}_{\mu\nu}  \hspace{5mm} \left(p_0=\sqrt{{\bm p}^2+m_I^2}\right) ,
\end{eqnarray} 
where  $A_T^{(\sigma)}$ is an integration constant depending on ${\bm p}$ and $e^{(\sigma)}_{\mu\nu}$ shows the five elements of the transverse traceless modes. 
In section \ref{MD}, five degrees of freedom of $(\sigma)$ are expressed as $(0)$, $(1,o)$, $(1,e)$, $(2,o)$ and  $(2,e)$.
The solution for $I_{\mu\nu}$ is given as
\begin{eqnarray}
I_{\mu\nu} =  \tilde I_{\mu\nu}- \frac{2 \kappa^2 \beta (3\alpha +\beta) } {3 (4\alpha + \beta)} \p_\mu \p_\nu I + \frac{3\alpha+\beta}{3(4\alpha+\beta)} I \eta_{\mu\nu}.
\end{eqnarray}

Now, we quantize $I$ and $\tilde I_{\mu\nu}$. 
It can be done by replacing $\{A_S,A_T\}$ and $\{A_S^\ast,A_T^\ast\}$ by annihilation and creation operators, respectively.  
The normalization factors are fixed such that the commutation relations \eq{Ipi}-\eq{pipi} hold.
To this end, we need  the explicit form of the basis $e_{\mu\nu}^{(\sigma)}$.
In preparation, we introduce a basis of vector as
\begin{eqnarray}
l_{\mu} (p^\nu)= \left( p, \sqrt{p^2+m_I^2}, 0,0 \right) / m , \quad
t_{\mu} (p^\nu)= \left( 0,0,1,0 \right) , \quad
u_\mu (p^\nu)=\left( 0,0,0,1 \right). 
\end{eqnarray}
Here, $l_\mu$, $t_{\mu}$ and $u_{\mu}$ are normalized spacelike vectors normal to momentum $p^\nu$. 
\begin{eqnarray}
p^\nu= \left( \sqrt{p^2+m_I^2},p, 0,0 \right).
\end{eqnarray}
Five elements for transverse-traceless modes are written with these vectors as~\cite{Abe:2020ikj,Aubert:2003je}
\begin{eqnarray}
&&\e_{\mu\nu}^{(0)} (p^\alpha) = \frac2{\sqrt {6}} l_{\mu}l_{\nu}-\frac1{\sqrt{6}} t_{\mu}t_{\nu}-\frac1{\sqrt{6}} u_{\mu}u_{\nu} , \nonumber\\
&&\e_{\mu\nu}^{(1,e)} (p^\alpha) = \frac1{\sqrt {2}} \left( l_{\mu}t_{\nu}+ t_{\mu}l_{\nu} \right),  \qquad
\e_{\mu\nu}^{(1,o)} (p^\alpha) = \frac1{\sqrt {2}} \left( l_{\mu}u_{\nu}+ u_{\mu}l_{\nu} \right), \label{tildee} \\
&&\e_{\mu\nu}^{(2,e)} (p^\alpha) = \frac1{\sqrt {2}} \left( t_{\mu}t_{\nu}- u_{\mu}u_{\nu} \right),  \qquad 
\e_{\mu\nu}^{(2,o)} (p^\alpha) = \frac1{\sqrt {2}} \left( t_{\mu}u_{\nu}+ u_{\mu}t_{\nu} \right). \nonumber
\end{eqnarray}

Since the commutation relations \eq{Ipi}-\eq{pipi}, it is useful to define the three dimensional elements for tensors. 
We use the elements for three-dimensional parts of $t_\mu$, $u_\mu$ and 
\begin{eqnarray}
\hat p_i = (1,0,0).
\end{eqnarray}
The $(i,j)$-components of $e^{(\sigma)}_{\mu\nu}$ can be written as
\begin{eqnarray}
&&\e_{ij}^{(0)} = \frac2{\sqrt {6}} \left(\frac{p^2}{m_I^2}+1\right) \hat p_i\hat p_j-\frac1{\sqrt{6}} t_it_j-\frac1{\sqrt{6}} u_iu_j , \nonumber\\
&&\e_{ij}^{(1,e)} = \frac{\sqrt{p^2+m_I^2}}{\sqrt {2}m_I} \left( \hat p_it_j+ t_i\hat p_j \right),  \qquad
\e_{ij}^{(1,o)} = \frac{\sqrt{p^2+m_I^2}}{\sqrt {2}m_I} \left( \hat p_iu_j+ u_i\hat p_j \right), \\
&&\e_{ij}^{(2,e)} = \frac1{\sqrt {2}} \left( t_it_j- u_iu_j \right),  \qquad 
\e_{ij}^{(2,o)} = \frac1{\sqrt {2}} \left( t_iu_j+ u_it_j \right). \nonumber
\end{eqnarray}
The space components of $I_{\mu\nu}$ can be written as
\begin{eqnarray}
&&I_{ij} = \sum_{\sigma} \int  d^3p \left\{ A_T^{(\sigma)} ({\bm p}) e^{-ipx}+A_T^{(\sigma)}{}^\dagger ({\bm p}) e^{ipx} \right\} e^{(\sigma)}_{ij} \nonumber \\
&&\qquad\qquad +
\frac{3\alpha+ \beta}{\sqrt{3}(4\alpha+\beta)}  \int  d^3p \left\{ A_S ({\bm p}) e^{-ipx}+A_S^\dagger ({\bm p}) e^{ipx} \right\} \e^{(S)}_{ij},
\end{eqnarray} 
where
\begin{eqnarray}
\e^{(S)}_{ij}:= \frac1{\sqrt{3}}\left(1 - 2\frac{p^2} {m_I^2}   \right) \hat p_i \hat p_j +\frac1{\sqrt{3}}\left(t_it_j+u_iu_j\right).
\end{eqnarray} 

The spatial components of the canonical momentum $\pi^{ij}$, defined in Eq.\eq{defpi}, are expressed as 
\begin{eqnarray}
&&\pi^{ij} = \frac{i}{2\kappa^2} \Biggl[ \sum_{\sigma} \int  d^3p \sqrt{p^2+m_I^2} \left\{ A_T^{(\sigma)} ({\bm p}) e^{-ipx}-A_T^{(\sigma)}{}^\dagger ({\bm p}) e^{ipx} \right\}  \bar e^{(\sigma)}_{ij} \nonumber \\
&&\qquad\qquad  
-\frac{2(3\alpha+ \beta)}{\sqrt{3}(4\alpha+\beta)}  \int  d^3p \sqrt{p^2+m_S^2} \left\{ A_S ({\bm p}) e^{-ipx} - A_S^\dagger ({\bm p}) e^{ipx} \right\} \bar e_{ij}^{(S)} \Biggr] .
\end{eqnarray} 
where
\begin{eqnarray}
&&\bar e_{ij}^{(0)} = \frac2{\sqrt {6}}  \hat p_i\hat p_j+\frac1{\sqrt{6}} \left( 2\frac{p^2}{m_I^2} -1 \right)\left( t_it_j+ u_iu_j  \right), \qquad
\bar e_{ij}^{(1,q)} = \frac{m_I^2 }{p^2+m_I^2} \e_{ij}^{(1,q)}, \nonumber \\
&&
\bar e_{ij}^{(2,q)} = \e_{ij}^{(2,q)} , \qquad
\bar e_{ij}^{(S)} = \frac1{\sqrt{3}} \hat p_i \hat p_j + \frac1{\sqrt{3}}\left( \frac{p^2}{m_I^2}+1\right) \left( t_it_j+u_i u_j \right) ,
\end{eqnarray}
and  $q$ represents $e$ or $o$.
The basis elements $\bar e_{ij}^{(\sigma)}$ are the dual of  the basis elements $\e_{ij}^{(\sigma)}$,
\begin{eqnarray}
\e^{(\sigma)}_{ij} \bar e^{(\tau)}_{ij} = \delta_{\sigma\tau}.
\end{eqnarray}

Substituting $I_{ij}$ and $\pi^{ij}$ into the commutation relations \eq{Ipi}-\eq{pipi} fixes the normalization factors, and finally we have
\begin{eqnarray}
&&I_{\mu\nu} = \sum_{\sigma} \int d^3p \frac{\kappa}{\sqrt{(2\pi^3)p_0}} \left\{ a_T^{(\sigma)} ({\bm p}) e^{-ipx}+a_T^{(\sigma)}{}^\dagger ({\bm p}) e^{ipx} \right\} e^{(\sigma)}_{\mu\nu} \nonumber \\
&&\hspace{15mm} 
+
 \int d^3p \frac{\kappa}{\sqrt{(2\pi^3)p_0}} \left\{ a_S ({\bm p}) e^{-ipx}+a_S^\dagger ({\bm p}) e^{ipx} \right\} \frac1{\sqrt{3}}\left(\eta_{\mu\nu}- 2\frac{p_\mu p_\nu} {m_I^2}  \right),
\end{eqnarray} 
where
\begin{eqnarray}
[a_T^{(\sigma)} ({\bm k}) , a_T^{(\tau)}{}^\dagger ({\bm p})] = - \delta_{\sigma\tau} \delta^3 \bigl({\bm k}-{\bm p}\bigr) , \qquad
[a_S ({\bm k}) , a_S^\dagger ({\bm p})] = \delta^3 ({\bm k}-{\bm p}),
\end{eqnarray} 
and other commutations are zero.

Field $H_{\mu\nu}$ has the gauge symmetry. In the free-theory limit, it is written as 
\begin{eqnarray}
H_{\mu\nu} \to H_{\mu\nu} + \p_{(\mu} \xi_{\nu)}. 
\end{eqnarray} 
Therefore, if we shift the field $H_{\mu\nu}$ as 
\begin{eqnarray}
H_{\mu\nu} \to H_{\mu\nu} + p_{\mu} p_{\nu} \Phi ,
\end{eqnarray} 
where $\Phi$ is an arbitrary operator, the obtained theory becomes the same. 
The fundamental field  appearing the theory is $h_{\mu\nu}$ and it is shifted in the above transformation as
\begin{eqnarray}
h_{\mu\nu}=H_{\mu\nu} + I_{\mu\nu} \to H_{\mu\nu} + I_{\mu\nu}+ p_{\mu} p_{\nu} \Phi .
\end{eqnarray} 
This can be interpreted  that the field $I_{\mu\nu}$ can be shifted by $p_{\mu} p_{\nu} \Phi $. 
Using this gauge degree of freedom, we can eliminate the component propotional to $p_{\mu} p_{\nu}$, 
and then $I_{\mu\nu}$ can be
\begin{eqnarray}
&&I_{\mu\nu} = \sum_{\sigma} \int d^3p \frac{\kappa}{\sqrt{(2\pi^3)p_0}} \left\{ a_T^{(\sigma)} ({\bm p}) e^{-ipx}+a_T^{(\sigma)}{}^\dagger ({\bm p}) e^{ipx} \right\} e^{(\sigma)}_{\mu\nu} \nonumber \\
&&\hspace{20mm}
 +
 \int d^3p \frac{\kappa}{\sqrt{(2\pi^3)p_0}} \left\{ a_S ({\bm p}) e^{-ipx}+a_S^\dagger ({\bm p}) e^{ipx} \right\} \frac1{\sqrt{3}} \theta_{\mu\nu}.
\end{eqnarray} 
%
%


\section{Perturvative expansion by variation}\label{PEV}

One way to expand a function $F(g)$ of the  metric $g_{\mu\nu}$ with respect to the perturbation 
\begin{eqnarray}
h_{\mu\nu}:=g_{\mu\nu}-g^{(0)}_{\mu\nu}, 
\end{eqnarray} 
where $g^{(0)}_{\mu\nu}$ is a background metric (not need to be Minkowski metric $\eta_{\mu\nu}$), 
is substituting  $g_{\mu\nu}=g^{(0)}_{\mu\nu}+h_{\mu\nu}$ into the function $F(g)$. 
The derivation has not been explicitly given in the past literature, as far as we can find. 
Here we rely on another way to derive the perturbative expansion by use of Taylor series. 
Taylor series of function $F(g)$ is obtained (where the indices for metric $g_{\mu\nu}$ are omitted for simplicity) as
\begin{eqnarray}
&&F(g) = \delta^0 F (g^{(0)}) +   \delta^1 F (g^{(0)}; h) + \delta^2 F (g^{(0)}; h,h) +\dots \nonumber \\
&&\hspace{9mm}
= \sum_{k=0} \frac{1}{k!} \delta^k F (g^{(0)}; \overbrace{h,\cdots, h}^{\displaystyle k}).
\end{eqnarray} 
%
The $k$-th order variation $\delta^k F (g^{(0)}; h, \cdots, h)$ relates to the $(k-1)$-th order variation  $\delta^{k-1}F (g^{(0)}; h, \cdots, h)$ as follows. 
Suppose we have an operator $\delta^{k-1}F (g; h, \cdots, h)$ on an arbitrary background $g$. 
$F^{(k-1)} (g^{(0)}; h, \cdots, h)$ is described as $\delta^{k-1}F (g; h, \cdots, h)|_{g=g^{(0)}}$.
Then, $\delta^kF (g^{(0)}; h, \cdots, h)$ is the first-order part of the expansion of $\delta^{k-1}F (g; h, \cdots, h)$
with respect to $h_{\mu\nu}$ divided by $k$.
This relation gives $\delta^k F (g^{(0)}; h, \cdots, h)$  successively.
Taylor series can give the perturbative expansion more easily than the direct substitution, 
especially if we go to higher orders.

Let us calculate the expansion of gravitational action with respect to $h_{\mu\nu}$, by Taylor expansion. 
In this paper, the background metric is taken to be $\eta_{\mu\nu}$. 
Hence, we introduce the symbol ``$\sim$'' denoting that  $\eta_{\mu\nu}$ is taken as the background metric  {\it i.e.} 
\begin{eqnarray}
\delta^k F(g; h, \cdots, h) \sim \delta^k F(\eta; h, \cdots, h) ,
\end{eqnarray} 
while, in this appendix, the symbol of equality ``$=$'' means that the background metric is an arbitrary metric $g$.
Be mindful that in the form of $\delta^k F(g; h, \cdots, h)$, 
raising and/or lowering indices should be done by $g$, not by $g^{(0)}$ nor by $\eta$. 
Moreover, in this appendix, functions without arguments show those of  an arbitrary background metric $g_{\mu\nu}$ and $h_{\mu\nu}$, that is, 
\begin{eqnarray}
\delta^k F:=\delta^k F(g; h, \cdots, h) .
\end{eqnarray} 

\subsection{Christoffel symbols}
The variations of the Christoffel symbols can be calculated as 
\begin{eqnarray}
&&\delta \Gamma^\alpha_{\beta\gamma}= \frac12  g^{\alpha\omega}\left( \nabla_\beta h_{\omega\gamma}+\nabla_\gamma h_{\omega\beta}-\nabla_\omega h_{\beta\gamma} \right) \\
&&\delta^2 \Gamma^\alpha_{\beta\gamma} =- 2 g^{\alpha\omega} h_{\omega\lambda} \delta \Gamma^\lambda_{\beta\gamma} \label{gamma2}
\end{eqnarray}
The $n$-th order variation is obtained by using Eq.\eq{gamma2} itereatively,
\begin{eqnarray}
\delta^k \Gamma^\alpha_{\beta\gamma} = (-1)^{k-1} k! \left( h^{k-1} \right){}^\alpha{}_\lambda \delta \Gamma^{\lambda}_{\beta\gamma},
\end{eqnarray}
where
\begin{eqnarray}
&&\left( h^k \right){}^\alpha{}_\lambda := g^{\alpha\mu_1}h_{\mu_1\nu_1}g^{\nu_1\mu_2}h_{\mu_2\nu_2}g^{\nu_2\mu_3}\cdots
g^{\nu_{k-2}\mu_{k-1}}h_{\mu_{k-1}\nu_{k-1}}g^{\nu_{k-1}\mu_k}h_{\mu_k\lambda} ,
\end{eqnarray}
that is, the Christoffel symbols are expanded as 
\begin{eqnarray}
\Gamma^\alpha_{\beta\gamma}= \Gamma^\alpha_{\beta\gamma}|_{g=g^{(0)}} 
+ \sum _{k=0} (-1)^k \left( h^k \right){}^\alpha{}_\lambda \delta \Gamma^\lambda_{\beta\gamma}|_{g=g^{(0)}},
\end{eqnarray}
where
\begin{eqnarray}
\left( h^0 \right){}^\alpha{}_\lambda := \delta^\alpha{}_\lambda .
\end{eqnarray}

\subsection{Riemann curvature}
The variations of the Riemann curvature can be calculated as 
\begin{eqnarray}
&&\delta R_{\alpha \mu\beta\nu}=
-\frac12 \left( 
\nabla_{(\alpha} \nabla_{\beta)} h_{\mu\nu}-\nabla_{(\alpha} \nabla_{\nu)} h_{\mu\beta
}-\nabla_{(\mu} \nabla_{\beta)} h_{\alpha\nu}+\nabla_{(\mu} \nabla_{\nu)} h_{\alpha\beta}
\right. \nonumber\\
&&\hspace{50mm}
\left.
+R_{\alpha\mu[\beta}{}^{\lambda} h_{\nu]\lambda}+R_{\beta\nu[\alpha}{}^{\lambda} h_{\mu]\lambda}
 \right),
\\
&&\delta^2 R_{\alpha \mu\beta\nu}=
-2g_{\lambda\epsilon} 
\delta^2 S^{\epsilon}{}_{\alpha\beta}{}^\lambda{}_{\mu\nu},
\end{eqnarray}
where
\begin{eqnarray}
\delta^2 S^{\epsilon}{}_{\alpha\gamma}{}^\lambda{}_{\mu\nu} :=   \delta \Gamma^{(\epsilon}_{\alpha\gamma}\delta \Gamma^{\lambda)}_{\mu\nu}-\delta \Gamma^{(\epsilon}_{\mu\gamma}\delta \Gamma^{\lambda)}_{\alpha\nu} .
\end{eqnarray}
The variations of $\delta^2 S^{\epsilon}{}_{\alpha\gamma}{}^\lambda{}_{\mu\nu}$ becomes
\begin{eqnarray}
\delta \left( \delta^2 S^{\epsilon}{}_{\alpha\gamma}{}^\lambda{}_{\mu\nu} \right) =
-2g^{\epsilon\kappa}h_{\kappa\omega} \delta^2 S^{\omega}{}_{\alpha\gamma}{}^\lambda{}_{\mu\nu}
-2g^{\lambda\kappa}h_{\kappa\omega} \delta^2 S^{\epsilon}{}_{\alpha\gamma}{}^\omega{}_{\mu\nu} .
\end{eqnarray}
The higher-order variation of the Riemann curvature can be obtained by induction, 
\begin{eqnarray}
\delta^k R_{\alpha \mu\beta\nu}=
(-1)^{k-1}k! g_{\lambda\kappa} \left( h^{k-2} \right)^\kappa{}_\epsilon  
\delta^2 S^{\epsilon}{}_{\alpha\beta}{}^\lambda{}_{\mu\nu},
\end{eqnarray}
Therefore, we can compile the Taylor series of the Rieman curvature into the following simple form,  
\begin{eqnarray}
R_{\alpha \mu\beta\nu} =
R_{\alpha \mu\beta\nu}|_{g=g^{(0)}} + \delta R_{\alpha \mu\beta\nu}|_{g=g^{(0)}} + \sum_{k=0}   
(-1)^{k-1} g_{\lambda\kappa} \left( h^{k} \right)^\kappa{}_\epsilon  
\delta^2 S^{\epsilon}{}_{\alpha\beta}{}^\lambda{}_{\mu\nu}|_{g=g^{(0)}}.
\end{eqnarray}

With Leibniz rule, the Ricci curvature and the Ricci scalar are expressed as 
\begin{eqnarray}
&&\delta^n R_{\alpha\beta} = \sum_{k=0}^n \frac{n!}{k!(n-k)!} \delta^k g^{\mu\nu}\,  \delta^{n-k} R_{\alpha\mu\beta\nu} \\
&&\delta^n R_{\alpha\beta} = \sum_{k=0}^n \sum_{l=0}^k \frac{n!}{l!(k-l)!(n-k)!} \delta^l g^{\alpha\beta}\, \delta^k g^{\mu\nu}\,  \delta^{n-k} R_{\alpha\mu\beta\nu}
\end{eqnarray}


\subsection{Einstein-Hilbert action}
Here, we show the perturbative expansion of Einstein-Hilbert action:
\begin{eqnarray}
S_{EH}:= \int \sqrt{-g} R d^4x.
\end{eqnarray}
The first-order variation becomes
\begin{eqnarray}
\delta S_{EH} \simeq   -  \int \sqrt{-g}\left( g^{\mu\alpha}g^{\nu\beta} -\frac12 g^{\mu\nu} g^{\alpha \beta} \right) R_{\alpha \beta} h_{\mu\nu} d^4x \sim 0 .
\end{eqnarray}
where $\simeq$ means that total derivative terms are ignored. 
The quadratic-order variation can be obtained by taking another variation of the above, 
\begin{eqnarray}
\delta^2 S_{EH}
&\simeq& 
-\int \left[ \left\{ \delta \left[  \sqrt{-g}\left( g^{\mu\alpha}g^{\nu\beta} -\frac12 g^{\mu\nu} g^{\alpha \beta} \right) \right] \right\} R_{\alpha \beta} h_{\mu\nu} \right.
\nonumber \\ && \qquad\qquad\qquad \left.
-   \sqrt{-g}\left( g^{\mu\alpha}g^{\nu\beta} -\frac12 g^{\mu\nu} g^{\alpha \beta} \right)\left[ \delta R_{\alpha \beta} \right] h_{\mu\nu} \right] d^4 x \\
&\sim&
-\int \frac12  \left( h^{\alpha\beta} -\frac12 h \eta^{\alpha \beta} \right)
\left ( \partial_\gamma \partial_\alpha h^{\gamma}{}_\beta+\partial_\gamma \partial_\beta h^{\gamma}{}_\alpha 
-\partial_\gamma \partial^\gamma h_{\alpha\beta} -\partial_\alpha \partial_\beta h  \right) d^4 x 
\nonumber \\
 &\simeq&
\frac12\int  \left( h^{\alpha\beta} \partial_\gamma \partial^\gamma h_{\alpha\beta} 
-2h^{\alpha\beta}  \partial_\gamma \partial_\alpha h^{\gamma}{}_\beta 
+2h^{\alpha \beta}\partial_\alpha \partial_\beta h  
- h\partial_\gamma \partial^\gamma h   \right) d^4x .
 \end{eqnarray}
The cubic-order variation becomes 
\begin{eqnarray}
\delta^3 S_{EH}&\simeq& \int \left(
 -\left\{\delta^2  \left[ \sqrt{-g}\left( g^{\mu\alpha}g^{\nu\beta} -\frac12 g^{\mu\nu} g^{\alpha \beta} \right)\right] \right\} R_{\alpha \beta} h_{\mu\nu}
\right.\nonumber \\
&&\qquad
- 2\left\{\delta \left[ \sqrt{-g}\left( g^{\mu\alpha}g^{\nu\beta} -\frac12 g^{\mu\nu} g^{\alpha \beta} \right)\right] \right\}\left[ \delta R_{\alpha \beta} \right] h_{\mu\nu} 
\nonumber \\
&&\qquad \left.
-   \sqrt{-g}\left( g^{\mu\alpha}g^{\nu\beta} -\frac12 g^{\mu\nu} g^{\alpha \beta} \right) \left[ \delta^2 R_{\alpha \beta} \right] h_{\mu\nu} \right) d^4x
\nonumber \\
&\sim&\int\left(
-\frac38 \h^2 \Box \h 
+\frac34  \h \h^{\alpha\beta} \Box \h_{\alpha\beta} 
+\frac98 \h^{\alpha\beta}\h_{\alpha\beta} \Box \h
-\frac32 \h^{\alpha\beta} \h_{\beta}{}^{\gamma} \Box \h_{\alpha\gamma}
+\frac32 \h \h^{\alpha\beta} \partial_\alpha\partial_\beta \h 
\right.\nonumber \\
&&\left. \qquad \qquad
-\frac32 \h^{\alpha\beta} \h_{\alpha}{}^{\gamma} \partial_\beta\partial_ \gamma \h 
+ 3 \h^{\alpha\gamma} \h^{\beta\lambda}  \partial_\alpha \partial_\beta \h_{\gamma \lambda} 
-\frac32 \h^{\alpha\beta} \h^{\gamma\lambda} \partial_\alpha \partial_\beta \h_{\gamma\lambda} \right) d^4x ,
\end{eqnarray}
where, in the last equality ``$\sim$'', we omit the surface terms. 
Then, $S_{EH}$ is expanded as
\begin{eqnarray}
S_{EH}= \delta S_{EH} + \frac12 \delta^2 S_{EH}+ \frac16 \delta^3 S_{EH} + \cO \left(h^4 \right)
\end{eqnarray}


\subsection{$R_{\mu\nu}^2$ term}

Let us expand the quadratic curvature action
\begin{eqnarray}
S_{R_{\mu\nu}^2} :=\int \sqrt{-g} R_{\mu\nu}R^{\mu\nu} d^4 x = \int  \sqrt{-g} g^{\mu\alpha}g^{\nu\beta} R_{\mu\nu}R_{\alpha\beta} d^4 x .
\end{eqnarray}
The linear, qudratic and cubic variations can be calculated as
\begin{eqnarray}
\delta S_{R_{\mu\nu}^2} 
\!\!\!\! &  = & \!\!\!\! \int \biggl( \left[\delta\left( \sqrt{-g} g^{\mu\alpha}g^{\nu\beta}\right)\right] R_{\mu\nu}R_{\alpha\beta} 
+ 2 \sqrt{-g} g^{\mu\alpha}g^{\nu\beta}R_{\mu\nu}  \delta R_{\alpha\beta} \biggr) d^4x \sim 0 , \\
\delta^2 S_{R_{\mu\nu}^2} 
\!\!\!\! &  = & \!\!\!\! \int \biggl( \left[\delta^2\left( \sqrt{-g} g^{\mu\alpha}g^{\nu\beta}\right)\right] R_{\mu\nu}R_{\alpha\beta} 
+4\left[\delta\left( \sqrt{-g} g^{\mu\alpha}g^{\nu\beta}\right)\right]    R_{\mu\nu} \delta R_{\alpha\beta} 
\nonumber \\ && \hspace{30mm}
+ 2\sqrt{-g} g^{\mu\alpha}g^{\nu\beta}  R_{\mu\nu} \delta^2R_{\alpha\beta} 
+ 2\sqrt{-g} g^{\mu\alpha}g^{\nu\beta} \delta R_{\mu\nu} \delta R_{\alpha\beta} \biggr) d^4 x
\nonumber \\  \hspace{13mm}
\!\!\!\! & \sim & \!\!\!\! \frac 12 \int  \left( \partial_\gamma \partial_\nu h^\gamma{}_\mu -\partial_\gamma \partial^\gamma h_{\mu\nu } 
-\partial_\mu \partial_\nu h+ \partial_\mu \partial^\gamma h_{\gamma\nu } \right) \nonumber \\
&&\hspace{30mm}
\left( \partial_\lambda \partial^\nu h^\lambda{}^\mu -\partial_\lambda \partial^\lambda h^{\mu\nu } 
-\partial^\mu \partial^\nu h+ \partial^\mu \partial^\lambda h_{\lambda}{}^{\nu } \right) d^4 x, \\
\delta^3 S_{R_{\mu\nu}^2} 
\!\!\!\! &  = & \!\!\!\! \int \biggl( \left[\delta^3\left( \sqrt{-g} g^{\mu\alpha}g^{\nu\beta}\right)\right] R_{\mu\nu}R_{\alpha\beta} 
+6\left[\delta^2\left( \sqrt{-g} g^{\mu\alpha}g^{\nu\beta}\right)\right]  R_{\mu\nu} \delta R_{\alpha\beta} 
\nonumber \\ && \hspace{20mm}
+6\left[\delta\left( \sqrt{-g} g^{\mu\alpha}g^{\nu\beta}\right)\right]   R_{\mu\nu} \delta^2R_{\alpha\beta} 
+6\left[\delta\left( \sqrt{-g} g^{\mu\alpha}g^{\nu\beta}\right)\right]  \delta R_{\mu\nu}  \delta R_{\alpha\beta} 
\nonumber \\ && \hspace{20mm}
+2 \sqrt{-g} g^{\mu\alpha}g^{\nu\beta}  R_{\mu\nu}  \delta^3  R_{\alpha\beta} 
+ 6 \sqrt{-g} g^{\mu\alpha}g^{\nu\beta} \delta R_{\mu\nu}  \delta^2  R_{\alpha\beta} \biggr) d^4 x
\nonumber \\ 
&\sim&
6 \int \left[
-\frac1{16} \h  \Box \h \Box \h
+\frac3{32} \h^2 \Box^2 \h
-\frac3{16} \h_{\alpha\beta}\h^{\alpha\beta}\Box^2 \h
+\frac18 \h  \Box \h_{\alpha\beta}\Box \h^{\alpha\beta} \right.
\nonumber\\ && \qquad \qquad
-\frac18\h^{\alpha\beta} \Box \h_{\alpha\beta} \Box \h   
-\frac14 \h^{\alpha\beta}   \h_{\beta}{}^{\gamma} \Box^2 \h_{\alpha\gamma}
-\frac12 \h \h^{\alpha \beta} \partial_\alpha \partial_\beta \Box \h
+\frac18 \h  \Box\h^{\alpha \beta}  \partial_\alpha \partial_\beta \h 
\nonumber\\ && \qquad \qquad
+\frac14 \h^{\mu\alpha}  \h_\mu {}^\beta \partial_\alpha \partial_\beta \Box \h 
+\h^{\alpha\mu} \Box \h^{\beta\nu}\partial_\alpha \partial_\beta \h_{\mu\nu}
+\frac12 \h^{\alpha\mu}  \h^{\beta\nu}\partial_\alpha \partial_\beta \Box \h_{\mu\nu}
\nonumber\\ && \qquad \qquad\left.
-\frac12 \h^{\alpha\beta} \h^{\mu\nu}\partial_\alpha \partial_\beta  \Box \h_{\mu\nu}
-\frac14  \h^{\mu\nu} \Box \h^{\alpha\beta} \partial_\alpha \partial_\beta \h_{\mu\nu} \right] d^4 x .
\end{eqnarray}
Then, $S_{R_{\mu\nu}^2} $ is expanded as
\begin{eqnarray}
S_{R_{\mu\nu}^2} = \delta S_{R_{\mu\nu}^2}  + \frac12 \delta^2 S_{R_{\mu\nu}^2} + \frac16 \delta^3 S_{R_{\mu\nu}^2}  + \cO \left(h^4 \right) .
\end{eqnarray}


\subsection{$R^2$ term}

We have another quadratic curvature action in four dimensional spacetime,
\begin{eqnarray}
S_{R^2}:= \int \sqrt{-g} R^2 d^4 x = \int \sqrt{-g} g^{\mu\nu}g^{\alpha\beta} R_{\mu\nu}R_{\alpha\beta} d^4 x.
\end{eqnarray}
The variations are calculated as
\begin{eqnarray}
&&\delta S_{R^2} 
=\int \biggl( \left[\delta\left( \sqrt{-g} g^{\mu\nu}g^{\alpha\beta}\right)\right] R_{\mu\nu}R_{\alpha\beta} 
+ 2 \sqrt{-g} g^{\mu\nu}g^{\alpha\beta}  R_{\mu\nu} \delta R_{\alpha\beta} \biggr) d^4 x \sim 0 ,
\\
&&\delta^2 S_{R^2}
= \int \biggl( \left[\delta^2\left( \sqrt{-g} g^{\mu\nu}g^{\alpha\beta}\right)\right] R_{\mu\nu}R_{\alpha\beta} 
+4\left[\delta\left( \sqrt{-g} g^{\mu\nu}g^{\alpha\beta}\right)\right]    R_{\mu\nu} \delta R_{\alpha\beta} 
\nonumber \\ && \hspace{30mm}
+ \sqrt{-g} g^{\mu\nu}g^{\alpha\beta}  R_{\mu\nu} \delta^2 R_{\alpha\beta} 
+ \sqrt{-g} g^{\mu\nu}g^{\alpha\beta} \delta R_{\mu\nu} \delta R_{\alpha\beta} \biggr) d^4x
\nonumber \\ && \hspace{11mm}
\sim 2  \int \left( \partial_\mu\partial^\mu h -\partial_\mu \partial_\nu h^{\mu\nu} \right)^2 d^4 x ,
\\
&&\delta^3 S_{R^2}
= \int \biggl( \left[\delta^3\left( \sqrt{-g} g^{\mu\nu}g^{\alpha\beta}\right)\right] R_{\mu\nu}R_{\alpha\beta} 
+6\left[\delta^2\left( \sqrt{-g} g^{\mu\nu}g^{\alpha\beta}\right)\right]   R_{\mu\nu} \delta  R_{\alpha\beta}
\nonumber \\ && \hspace{20mm}
+6\left[\delta\left( \sqrt{-g} g^{\mu\nu}g^{\alpha\beta}\right)\right]  R_{\mu\nu}  \delta^2 R_{\alpha\beta} 
+6\left[\delta\left( \sqrt{-g} g^{\mu\nu}g^{\alpha\beta}\right)\right]  \delta R_{\mu\nu} \delta R_{\alpha\beta} 
\nonumber \\ && \hspace{20mm}
+2 \sqrt{-g} g^{\mu\nu}g^{\alpha\beta}  R_{\mu\nu} \delta^3 R_{\alpha\beta} 
+6 \sqrt{-g} g^{\mu\nu}g^{\alpha\beta} \delta R_{\mu\nu}  \delta^2 R_{\alpha\beta} \biggr)
\nonumber \\ && \hspace{10mm}
\simeq
6\int \biggl[\frac14 \h^2 \Box^2 \h 
-\frac34 \h^{\alpha\beta}\h_{\alpha\beta} \Box^2 \h 
- \frac12 h^{\alpha\beta} \Box \h_{\alpha \beta} \Box \h
\nonumber \\ && \hspace{30mm}
-2  \h \h^{\alpha\beta} \partial_\alpha\partial_\beta  \Box \h
+\h^{\mu\alpha} \h_\mu{}^\beta \partial_\alpha \partial_\beta \Box \h
 \biggr] d^4 x .
\end{eqnarray}
Then, $S_{R^2} $ is expanded as
\begin{eqnarray}
S_{R^2} = \delta S_{R^2}  + \frac12 \delta^2 S_{R^2} + \frac16 \delta^3 S_{R^2}  + \cO \left(h^4 \right) .
\end{eqnarray}

\subsection{$\phi^2 R$ term}
The expansion of $\phi^2 R$ term
\begin{eqnarray}
S_{\phi^2R} := \int \sqrt{-g} \phi^2 R d^4x
\end{eqnarray}
  is written as 
\begin{eqnarray}
S_{\phi^2R}&=& \delta S_{\phi^2R}+ \frac12 \delta S_{\phi^2R} + \cO\left( h^3 \right), \\
\delta S_{\phi^2R}&=&\int \left(\left[ \delta \sqrt{-g}\right] \phi^2 R  + \sqrt{-g} \phi^2  \delta R \right) d^4x \sim - \int \phi^2  \Box \h \, d^4x ,\\
\delta^2 S_{\phi^2R}&=&\int \left(\left[ \delta^2 \sqrt{-g}\right] \phi^2 R + 2\left[ \delta \sqrt{-g}\right] \delta \phi^2 R  + \sqrt{-g} \phi^2  \delta^2 R \right) d^4x
\nonumber \\
&\sim& \int \phi^2 \left(  - \h \Box \h - \frac12 \left( \p_\mu \h \right) \left( \p^\mu \h \right) 
+ 2 \h^{\mu\nu} \p_\mu\p_\nu \h + 2\h^{\mu\nu} \Box \h_{\mu\nu} 
\right.
\nonumber \\
&&\left. \hspace{20mm}
+ \frac32 \left(\p_\alpha \h_{\mu\nu}\right) \left(\p^\alpha \h^{\mu\nu}\right) 
- \left(\p_\alpha \h_{\mu\beta}\right) \left(\p^\beta \h^{\mu\alpha}\right) 
\right) d^4x.
\end{eqnarray}


\section{Supplement for the calculation of 3-point vertex for gravitons}\label{SUPPLE}

We have expanded the gravitational action in Appendix~\ref{PEV}. 
The cubic action for gravitons are written as Eq.\eq{g3}.
After Fourier transform, 3-point vertex function for $h_{\mu\nu}(p_1)h_{\alpha\beta}(p_2)h_{\gamma\lambda}(p_3)$ is obtained as
\begin{eqnarray}
&& \lambda_3^{\mu\nu,\alpha\beta,\gamma\lambda}
\nonumber \\ && = 
\left[\frac1{8\kappa^2} \left( p_1^2+p_2^2+p_3^2 \right)+
\left(\frac\alpha2+\frac{3\beta}{16} \right)\left( p_1^4+p_2^4 +p_3^4 \right)
-\frac\beta8 \left( p_1^2 p_2^2+p_2^2p_3^2+p_3^2 p_1^2 \right)
\right]\eta^{\mu\nu}\eta^{\alpha\beta}\eta^{\gamma\lambda}
\nonumber \\ &&
+\left[ -\frac1{8\kappa^2} \left( p_2^2+p_3^2 \right) - \frac3{8\kappa^2} p_1^2 
-\frac38\left( 4\alpha+\beta \right) p_1^4 
- \frac 18 \left( 4\alpha+\beta \right) \left(p_1^2p_2^2+p_1^2p_3^2 \right)  
+ \frac \beta4 p_2^2p_3^2 \right]
\nonumber \\ && \qquad \qquad  \times
\frac12  \eta^{\mu\nu} \left(\eta^{\alpha\gamma}\eta^{\beta \lambda}+\eta^{\alpha\lambda}\eta^{\beta \gamma}\right) 
+ \left[(p_1,\mu,\nu)\leftrightarrow(p_2,\alpha,\beta)\leftrightarrow(p_3,\gamma,\lambda);(2\ \mbox{terms}) \right]
\nonumber \\ &&
+\left[ \frac1{2\kappa^2}\left( p_1^2 +p_2^2+p_3^2 \right) -\frac{\beta}2 \left( p_1^4 +p_2^4+p_3^4 \right)  \right]
\frac18\left( \eta^{\mu\alpha} \eta^{\beta\gamma}\eta^{\lambda\nu}+\eta^{\nu\alpha} \eta^{\beta\gamma}
\eta^{\lambda\mu}
+\eta^{\mu\beta} \eta^{\alpha\gamma}\eta^{\lambda\nu}
\right.
\nonumber \\ && \qquad \qquad \left.
+\eta^{\nu\beta} \eta^{\alpha\gamma}\eta^{\lambda\mu}
+\eta^{\mu\alpha} \eta^{\beta\lambda}\eta^{\gamma\nu}+\eta^{\nu\alpha} \eta^{\beta\lambda}\eta^{\gamma\mu}
+\eta^{\mu\beta} \eta^{\alpha\lambda}\eta^{\gamma\nu}+\eta^{\nu\beta} \eta^{\alpha\lambda}\eta^{\gamma\mu}\right)
\nonumber \\ &&
+\left[ -\frac1{4\kappa^2} + \frac\beta8  p_3^2 - \frac12\left(4 \alpha+ \beta\right) p_2^2 \right]
\eta^{\mu\nu}\eta^{\alpha\beta} p_2^\gamma p_2^\lambda
\nonumber \\ && \qquad \qquad 
+ \left[(p_1,\mu,\nu)\leftrightarrow(p_2,\alpha,\beta)\leftrightarrow(p_3,\gamma,\lambda);(5\ \mbox{terms}) \right]
\nonumber \\ &&
+\left[ \frac1{2\kappa^2} +\frac12\left(4\alpha+\beta\right) p_1^2 \right]
\frac14\eta^{\mu\nu}\left( p_1^\alpha p_1^\lambda \eta^{\beta\gamma}+p_1^\beta p_1^\gamma \eta^{\alpha\lambda}
+p_1^\alpha p_1^\gamma \eta^{\beta\lambda}+p_1^\beta p_1^\lambda \eta^{\alpha\gamma} \right)
\nonumber \\ && \qquad \qquad 
+ \left[(p_1,\mu,\nu)\leftrightarrow(p_2,\alpha,\beta)\leftrightarrow(p_3,\gamma,\lambda);(2\ \mbox{terms}) \right]
\nonumber \\ &&
+\left[- \frac1{\kappa^2}+\beta \left( p_1^2+p_2^2+p_3^2 \right)
\right]
\frac14
\left( \eta^{\mu\alpha}\eta^{\nu\gamma}p_1^\beta p_1^\lambda+\eta^{\mu\beta}\eta^{\nu\gamma}p_1^\alpha p_1^\lambda
+\eta^{\mu\alpha}\eta^{\nu\lambda}p_1^\beta p_1^\gamma+\eta^{\mu\beta}\eta^{\nu\lambda}p_1^\alpha p_1^\gamma \right)
\nonumber \\ && \qquad \qquad 
+ \left[(p_1,\mu,\nu)\leftrightarrow(p_2,\alpha,\beta)\leftrightarrow(p_3,\gamma,\lambda);(2\ \mbox{terms}) \right]
\nonumber \\ &&
+ \left[ 
\frac1{4\kappa^2}-\frac\beta 4 \left(p_1^2+p_2^2+p_3^2 \right)
\right]
\frac12\left( \eta^{\mu\alpha} \eta^{\nu\beta}+\eta^{\mu\beta}\eta^{\nu\alpha}\right) p_1^\gamma p_1^\lambda
\nonumber \\ && \qquad \qquad 
+ \left[(p_1,\mu,\nu)\leftrightarrow(p_2,\alpha,\beta)\leftrightarrow(p_3,\gamma,\lambda);(5\ \mbox{terms}) \right] .
\end{eqnarray}
Multiplying the corresponding basis elements $e^{(\sigma)}_{\mu\nu}$ and/or $\theta_{\mu\nu}/\sqrt{3}$ by the above, 
we can obtain the  3-point vertices for them. 
The results are shown in Eqs.\eq{3-ver1}-\eq{3-ver4} .
In the calculations, we use the following equalities,
\begin{eqnarray}
&&
\e^\mu_{1\nu} \e^\nu_{2\alpha} \theta^\alpha_{3\mu}
= \e^\mu_{1\nu} \e^\nu_{2\mu} -  \frac1{p_3^2} p_{2\mu} \e^\mu_{1\nu} \e^\nu_{2\alpha} p_1^\alpha ,
\qquad
p_{3\alpha}  \e^\alpha_{1\mu} \e^\mu_{2\nu}  p_3^\nu = p_{2\mu} \e^\mu_{1\nu} \e^\nu_{2\alpha} p_1^\alpha ,
\\&&
\left( p_{1\mu} \e^\mu_{2\nu} \e^\nu_{1\alpha} \theta^\alpha_{3\beta} p_1^\beta 
+p_{2\mu} \e^\mu_{1\nu} \e^\nu_{2\alpha} \theta^\alpha_{3\beta} p_2^\beta
+ p_{3\mu} \e^\mu_{2\nu} \theta^\nu_{3\alpha} \e^\alpha_{1\beta} p_3^\beta\right)
= - \frac1{p_3^2} p_2^\mu \e_{1\mu\nu} p_2^\nu \ p_1^\alpha \e_{2\alpha\beta} p_1^\beta,
\nonumber \\ 
\\
&&\left( \e_{1\mu\nu} \e_{2}^{\mu\nu}  p_1^\gamma p_1^\lambda \theta_{3\gamma\lambda}
+\e_{1\mu\nu} \theta_{3}^{\mu\nu}  p_1^\gamma p_1^\lambda \e_{2\gamma\lambda}
+\e_{1\mu\nu} \e_{2}^{\mu\nu}  p_2^\gamma p_2^\lambda \theta_{3\gamma\lambda}
\right. \nonumber \\ && \qquad \qquad  \left.
+\theta_{3\mu\nu} \e_{2}^{\mu\nu}  p_2^\gamma p_2^\lambda \e_{1\gamma\lambda}
+\e_{1\mu\nu} \theta_{3}^{\mu\nu}  p_3^\gamma p_3^\lambda \e_{2\gamma\lambda}
+\theta_{3\mu\nu} \e_{2}^{\mu\nu}  p_3^\gamma p_3^\lambda \e_{1\gamma\lambda} \right)
\nonumber \\&&\quad
= \left(-\frac12 p_3^2 + \left( p_1^2+p_2^2 \right) -\frac1{2p_3^2} \left( p_1^2-p_2^2 \right)^2 \right) \e^\mu_{1\nu} \e^\nu_{2\mu} 
-\frac4{p_3^2} p_2^\mu \e_{1\mu\nu} p_2^\nu \ p_1^\alpha \e_{2\alpha\beta} p_1^\beta,
\\
&&p_2^\mu \e_{1\mu\nu} p_2^\nu =p_3^\mu \e_{1\mu\nu} p_3^\nu, \quad
\e_{1\mu\nu} \theta_{2}^{\mu\nu}= - \frac{p_2^\mu \e_{1\mu\nu} p_2^\nu}{p_2^2}, \quad 
\e_{1\mu\nu} \theta_{3}^{\mu\nu}= - \frac{p_2^\mu \e_{1\mu\nu} p_2^\nu}{p_3^2},
\\
&&\e_{1\mu}{}^{\nu} \theta_{2\nu}{}^{\lambda}\theta_{3\lambda}{}^\mu 
=- \frac{p_1^2 + p_2^2 +p_3^2}{2p_2^2 p_3^2} \left( p_2^\mu \e_{1\mu\nu} p_2^\nu \right) ,
\\
&&p_3^\mu \e_{1\mu}{}^{\nu} \theta_{2\nu}{}^{\lambda} p_{3\lambda} 
=\frac{p_1^2 + p_2^2 -p_3^2}{2p_2^2} \left( p_2^\mu \e_{1\mu\nu} p_2^\nu \right) ,
\\
&& p_1^\mu \theta_{2\mu\nu} \e_1^{\nu\alpha} \theta_{3\alpha\beta} p_1^\beta
= \frac{1}{4p_2^2 p_3^2} \left[
-p_1^4+p_2^4+p_3^4  -2p_2^2p_3^2
\right]\left( p_2^\mu \e_{1\mu\nu} p_2^\nu \right) ,
\\
&& p_2^\mu \e_{1\mu\nu} \theta_2^{\nu\alpha} \theta_{3\alpha\beta} p_2^\beta
= \frac{1}{4p_2^2 p_3^2} \left[
p_1^4-p_2^4+p_3^4  -2p_1^2p_3^2
\right]\left( p_2^\mu \e_{1\mu\nu} p_2^\nu \right) ,
\\
&& p_3^\mu \theta_{2\mu\nu} \theta_3^{\nu\alpha} \e_{1\alpha\beta} p_3^\beta
= \frac{1}{4p_2^2 p_3^2} \left[
p_1^4+p_2^4-p_3^4  -2p_1^2p_2^2
\right]\left( p_2^\mu \e_{1\mu\nu} p_2^\nu \right) ,
\\
&&\theta_{1\mu\nu} \theta_2^{\mu\nu} 
= \frac{1}{4p_1^2 p_2^2} \left[
p_1^4+p_2^4+p_3^4 + 10 p_1^2 p_2^2 -2p_1^2p_3^2 -2p_2^2p_3^2
\right] ,
\\
&&p_1^\mu  \theta_{2\mu\nu} p_1^\nu 
= \frac{1}{4 p_2^2} \left[
-\left(p_1^4+p_2^4+p_3^4\right) + 2 \left( p_1^2 p_2^2 +p_1^2p_3^2 +p_2^2p_3^2 \right)
\right] ,
\\
&&\theta_{1\mu}{}^{\nu} \theta_{2\nu}{}^{\lambda}\theta_{3\lambda}{}^\mu 
= \frac{1}{8p_1^2 p_2^2 p_3^2} \left[
p_1^6+p_2^6+p_3^6 
\right. \nonumber \\ && \left. \qquad \qquad \qquad \qquad
- \left( p_1^2 p_2^4+p_1^4 p_2^2+p_1^2 p_3^4+p_1^4 p_3^2+p_2^2 p_3^4+p_2^4 p_3^2 \right)
+18 p_1^2 p_2^2p_3^2
\right] ,
\\
&&
p_1^\mu\theta_{2\mu\nu} \theta_3^{\nu\lambda}p_{1\lambda} 
= \frac{1}{8p_1^2 p_2^2 p_3^2} \left[
p_1^8-3p_1^6(p_2^2+p_3^2)+3p_1^4(p_2^4+p_3^4) 
\right.\nonumber \\ && \qquad \qquad \qquad \qquad \qquad  \qquad \qquad\left.
+2 p_1^4 p_2^2 p_3^2
-p_1^2(p_3^6+p_2^6)+ p_1^2p_2^2p_3^2(p_2^2+p_3^2)\right] ,
\\
&&p_1^\mu\theta_{2\mu\nu} \theta_1^{\nu\alpha} \theta_{3\alpha\beta} p_1^\beta 
= \frac{1}{16 p_1^2 p_2^2 p_3^2} \left[
p_1^8-(p_2^8+p_3^8) +2p_1^2(p_2^6+p_3^6)-2p_1^2(p_2^4 p_3^2 +p_2^2 p_3^4) 
\right. \nonumber \\ && \left. \qquad \qquad \qquad \qquad \qquad \qquad  \qquad  \qquad 
 -2p_1^6(p_2^2+p_3^2) - 6p_2^4 p_3^4+4(p_2^6 p_3^2+p_2^2 p_3^6)
\right] ,
\\
&&p_1^\mu\theta_{2\mu\nu} \theta_1^{\nu\alpha} \theta_{3\alpha\beta} p_1^\beta
+p_2^\mu\theta_{1\mu\nu} \theta_2^{\nu\alpha} \theta_{3\alpha\beta} p_2^\beta
+p_3^\mu\theta_{2\mu\nu} \theta_3^{\nu\alpha} \theta_{1\alpha\beta} p_3^\beta
\nonumber \\ &&\quad 
= \frac{1}{16 p_1^2 p_2^2 p_3^2} \left[
-(p_1^8+p_2^8+p_3^8)+4(p_1^6p_2^2+p_1^6p_3^2+p_1^2p_2^6+p_2^6p_3^2+p_1^2p_3^6+p_2^2p_3^6)
\right. \nonumber \\ && \qquad \qquad \qquad \qquad \qquad  \left.
-6(p_1^4p_2^4+p_1^4p_3^4+p_2^4p_3^4) -4 p_1^2p_2^2p_3^2(p_1^2+p_2^2+p_3^2)
\right] ,
\\
&&\theta_{1\mu\nu}\theta_2^{\mu\nu} p_1^\alpha \theta_{3\alpha\beta} p_1^\beta
\nonumber \\ &&\quad 
=\frac{1}{16p_1^2 p_2^2p_3^2} \left[
-(p_1^8+p_2^8+p_3^8)+4(p_1^6+p_2^6)p_3^2 +4(p_1^2+p_2^2)p_3^6 
-8(p_1^6p_2^2+p_1^2p_2^6)
\right. \nonumber \\ && \qquad \qquad  \qquad \qquad \left.
+ 18 p_1^4p_2^4-6(p_1^4+p_2^4)p_3^4 -16p_1^2 p_2^2p_3^4+20(p_1^4p_2^2p_3^2+p_1^2p_2^4p_3^2)
\right] ,
\\
&&
\theta_{1\mu\nu}\theta_2^{\mu\nu} p_1^\alpha \theta_{3\alpha\beta} p_1^\beta
+\theta_{1\mu\nu}\theta_3^{\mu\nu} p_1^\alpha \theta_{2\alpha\beta} p_2^\beta
+\theta_{2\mu\nu}\theta_3^{\mu\nu} p_2^\alpha \theta_{1\alpha\beta} p_2^\beta 
\nonumber \\ &&\ 
=\frac{1}{16p_1^2 p_2^2p_3^2} \left[
-3(p_1^8+p_2^8+p_3^8)
+6(p_1^4p_2^4+p_1^4p_3^4+p_2^4p_3^4) +24 p_1^2p_2^2p_3^2(p_1^2+p_2^2+p_3^2)
\right] ,
\nonumber \\
\end{eqnarray}


\section{Calculations of scattering amplitudes}\label{AppSA}

Here we show the calculation of scattering amplitudes in details. 
Since the calculations of the $s$-channel exchange and the contact term are not hard, we show only the result. 
The derivation of the $t$-channel exchange is quite complicated. 
We give some equations which we used in the calculation. 
After showing each of the contributions separately, we sum up all. 
Then, we can see nontrivial cancelations among the $s$-, $t$-, $u$-channel exchanges and the contact term.

Four-momenta of the ingoing graviton, ingoing scalar field, outgoing graviton and outgoing scalar field are denoted by 
$k_1^\mu$, $k_2^\mu$ $k_3^\mu$ $k_4^\mu$, respectively. 
Polarization tensors for ingoing and outgoing gravitons are denoted as $e_{1\mu\nu}$, $e_{3\mu\nu}$, respectively.

\subsection{$s$-channel exchange}
In the $s$-channel exchange, only the scalar field mediates (see the second figure in figure \ref{Fig:cha}) and its contribution is obtained as follows,
\begin{eqnarray}
&&{\A}_s \left(h_{\mu\nu}(k_1) \phi (k_2); h_{\alpha\beta}(k_3) \phi (k_4)\right)  \nonumber \\
&&=\kappa^2 e_{1\mu\nu} \lambda_3^{\mu\nu}\left(p_1=k_2,p_2=-k_1-k_2\right) i G\left( p=k_1+k_2\right)
\lambda_3^{\alpha\beta}\left(p_1=-k_4,p_2=k_3+k_4\right) e_{3\alpha\beta}
\nonumber \\&&
= \frac{\kappa^2}4\left[ \left(-k_{1\mu}k_2^\mu -k_2^2-m^2  \right) e_{1\mu}^\mu +2k_2^\mu k_2^\nu e_{1\mu\nu}+4\xi k_1^2 e^{\mu}_{1\mu}\right]
\left[ \frac{1}{(k_1+k_2)^2+m^2} \right]
\nonumber \\&& \qquad\qquad
\times
\left[ \left(-k_{3\alpha}k_4^\alpha -k_4^2-m^2\right) e_{3\alpha}^\alpha +2k_4^\alpha k_4^\beta e_{3\alpha\beta}+ 4\xi k_3^2 e^{\alpha}_{3\alpha}\right]
\nonumber \\&& 
=  
\begin{cases}
\kappa^2 \cfrac{  k_2 \cdot e_1 \cdot k_2 \,  k_4\cdot e_3 \cdot k_4 }{(k_1+k_2)^2+m^2}  
\hspace{42mm} \left(\mbox{for  $(h^{(\sigma)}\phi \to h^{(\sigma')}\phi )$} \right)\\
-\kappa^2\dfrac{ \sqrt{3} k_2\cdot e_1 \cdot  k_2\left(3 k_{3}\cdot k_4  +2 m^2+ 2\frac{ (k_3\cdot  k_{4})^2} {k_3^2}\right)}{6\left[(k_1+k_2)^2+m^2\right]}   + 2\sqrt{3} \xi \kappa^2 k_3^2 \dfrac{ k_2\cdot \e_1 \cdot k_2 } {\left[(k_1+k_2)^2+m^2\right]}  
\\
\hspace{80mm} \left(\mbox{for $(h^{(\sigma)}\phi \to I^{(S)}\phi )$} \right)   
\\
\kappa^2 \dfrac{ \left(3 k_{1}\cdot k_2  +2 m^2+ 2\frac{ (k_1\cdot  k_{2})^2}{k_1^2}\right)\left(3 k_{3}\cdot k_4  +2 m^2+ 2\frac{ (k_3\cdot  k_{4})^2}{k_3^2}\right)}{12\left[(k_1+k_2)^2+m^2\right]}   
\\
\hspace{10mm}- \xi \kappa^2\left( 2 (k_1\cdot k_{2}) +2k_1^2 +\dfrac{-2k_1^4+4 m^2 k_1^2}{\left[ (k_1+k_2)^2+m^2 \right]}   \right)
+12 \xi^2 \kappa^2  \cfrac{ k_1^2 k_3^2  }{(k_1+k_2)^2+m^2} 
\\
\hspace{80mm} \left(\mbox{for $(I^{(S)}\phi \to I^{(S)}\phi )$} \right)   \, .
\end{cases}
\label{D1}
\end{eqnarray}
where we use the transversality condition $k_1\cdot  e_{1}=0=k_3\cdot  e_{3}$ and on-shell condition for external lines $k_2^2=k_4^2=-m^2$. 
Note that the graviton propagators \eq{G2} and the vertex functions \eq{sgvf3}, \eq{sgvf4} and \eq{g3} are constructed with respect to $h_{\mu\nu}$, not $H_{\mu\nu}$ nor $I_{\mu\nu}$. 
The factor $\kappa^2$  appearing in \eqref{D1} is due t the factor $\kappa$ used in eqs.~\eq{2acH}  and  \eq{2acI}.
The $u$-channel exchange contribution can be obtained by the crossing relation $ k_2 \, \leftrightarrow \, - k_4$.

\subsection{Contact term}
Contact term is calculated as follows,
\begin{eqnarray}
&&{\A}_c \left(h_{\mu\nu}(k_1) \phi (k_2); h_{\alpha\beta}(k_3) \phi (k_4)\right)  \nonumber \\
&&\quad
=\kappa^2 e_{1\mu\nu} \lambda_4^{\mu\nu,\alpha\beta}\left(p_1=k_2, p_2=-k_4 \right) e_{3\alpha\beta}
\nonumber \\&&\quad
= 
\frac{\kappa^2}4 \left[ - \left( k_{2\gamma}k_4^\gamma +m^2 \right)\left( e_{1\mu}^\mu e_{3\nu}^\nu-2 e_{1\mu\nu}e_3^{\mu\nu}\right)\right] 
\nonumber \\ && \qquad
+ \frac{\kappa^2}4 \left[2k_2^\mu k_4^\nu e_{1\mu\nu} e_{3\alpha}^\alpha+2k_2^\mu k_4^\nu e_{3\mu\nu} e_{1\alpha}^\alpha
-4k_2^\alpha k_4^\beta e_{1\alpha\mu}e_{3\beta}^\mu-4k_4^\alpha k_2^\beta e_{1\alpha\mu}e_{3\beta}^\mu \right]
\nonumber \\ && \qquad
+\xi \bigl[
\left( k_1^2+k_3^2- k_{1\mu}k_3^\mu\right) e^\alpha_{1\alpha} e^\beta_{3\beta}
-\left(2 k_1^2+2k_3^2- 3k_{1\mu}k_3^\mu\right) e_{1\alpha\beta} e^{\alpha\beta}_{3}
\nonumber \\ && \qquad
-k_{3}^\mu k_3^\nu e_{1\mu\nu} e_{3\alpha}^\alpha-k_{1}^\mu k_1^\nu e_{3\mu\nu} e_{1\alpha}^\alpha -2k_{1}^\mu k_3^\nu e_{3\mu\alpha} e_{1\nu}^\alpha
\bigr]
\nonumber \\&& \quad
= \left\{
\begin{array}{l}
 \dfrac{\kappa^2}2 \left( k_{2}\cdot k_4 +m^2 \right) \Tr \left[e_{1}\cdot e_3\right] - \kappa^2\left(k_2\cdot   e_{1}\cdot e_{3}\cdot  k_4+ k_2\cdot  e_{3}\cdot e_{1}\cdot k_4 \right)
 \\ \hspace{10mm}
+ \xi \kappa^2  \left[-\left( 2k_1^2 +2 k_3^2 -3 k_1\cdot k_3 \right) \mbox{Tr} \left(\e_1 \cdot \e_3 \right) -
2k_1 \cdot \e_3 \cdot \e_1 \cdot k_3 \right]
\\
\hspace{88mm} \left(\mbox{for  $(h^{(\sigma)}\phi \to h^{(\sigma')}\phi )$} \right) \\
\dfrac{\sqrt{3}\kappa^2}{6k_3^2}
\left[(k_{1}\cdot  k_4+k_{3}\cdot  k_4) k_{2}\cdot   \e_1\cdot k_2
-(k_{1}\cdot  k_2+k_{3}\cdot  k_2) k_{4}\cdot  \e_1 \cdot  k_4
-k_1^2 \, k_{2}\cdot  \e_1 \cdot k_4 \right]
 \\ \hspace{5mm}
+ \dfrac{ \sqrt{3}\xi \kappa^2}{3}  \left[\dfrac{\left(k_3\cdot \e_1 \cdot k_3\right)} {k_3^2}
\left( 2k_1^2 -4 k_3^2 - k_1\cdot k_3 \right) \right]
\hspace{8mm} 
\left(\mbox{for $(h^{(\sigma)}\phi \to I^{(S)}\phi )$} \right)   
 \\
\dfrac{\kappa^2}{12}\left[-k_{2}\cdot  k_4 -5m^2 \right] +\dfrac{\kappa^2}6 \dfrac{k_{1}\cdot  k_2\,  k_{1}\cdot  k_4}{k_1^2} 
+ \dfrac{\kappa^2}6 \dfrac{k_{2}\cdot  k_3\,  k_{3}\cdot  k_4}{k_3^2} 
\\ \quad
+ \dfrac{\kappa^2k_{1}\cdot  k_3 }{6k_1^2 k_3^2} \left(
(k_{2}\cdot  k_4+m^2) k_{1}\cdot  k_3   
-2 k_{1}\cdot  k_2\,   k_{3}\cdot  k_4  
-2 k_{1}\cdot  k_4\,  k_{2}\cdot  k_3  
\right)
 \\ \hspace{3mm}
+\dfrac{\xi \kappa^2 }{3}\left[ \dfrac{\left(k_1\cdot k_3 \right)^3}{m_S^2} - 8 \dfrac{\left(k_1\cdot k_3 \right)^2}{m_s^2}
- \left(k_1\cdot k_3 \right) + 2 m_S^2 \right]
\hspace{3mm} 
 \left(\mbox{for $(I^{(S)}\phi \to I^{(S)}\phi )$} \right)   \, .
 \end{array}
\right.
\end{eqnarray}

\subsection{$t$-channel exchange and total amplitude}

In the $t$-channel exchange, the gravitational field mediates (see the last figure in figure \ref{Fig:cha}). 
It involve the 3-point graviton vertex and thus the calculation becomes quite messy. 
We give the detail in each case: ($h^{(\sigma)}\phi \to h^{(\sigma')}\phi$ ), ($h^{(\sigma)}\phi \to I^{(s)}\phi$ ) 
and ($I^{(s)}\phi \to I^{(s)}\phi $). 
The graviton propagator appearing in the $t$-channel exchange is expressed in eq.\eq{gpro}. 
The projection operators \eq{propa2} and \eq{propa0} in the graviton propagator are written as 
\begin{eqnarray}
&&P^{(2)}_{\alpha\beta,\mu\nu}= \sum_{\sigma} e^{(\sigma)}_{\alpha\beta} e^{(\sigma)}_{\mu\nu}\\
&&P^{(0)}_{\alpha\beta,\mu\nu}= \left( \frac{1}{\sqrt{3}}\theta_{\alpha\beta} \right) \left( \frac{1}{\sqrt{3}}\theta_{\mu\nu} \right)
\end{eqnarray}
With this expression, we decompose the $t$-channel exchange contribution into two parts: 
spin-2 propagator and spin-0 propagator parts. 
We also give the total amplitude, which is the sum of all contribution ${\A}_c+{\A}_s+{\A}_t+{\A}_u$.

\subsubsection{$(h^{(\sigma)}\phi \to h^{(\sigma')}\phi )$}$\ $

The contribution from the $t$-channel exchange is written as
\begin{eqnarray}
&&{\A}_t \left(h^{(\sigma)}_{\mu\nu}(k_1) \phi (k_2); h^{(\sigma')}_{\alpha\beta}(k_3) \phi (k_4)\right)  \nonumber \\
&& = \kappa^2 \e_{1\mu\nu} \e_{3\alpha\beta} \lambda_3^{\mu\nu,\alpha\beta,\gamma\delta} \left( p_1= k_1, p_2=-k_3, p_3=-k_1+k_3\right) 
\nonumber \\ && \qquad\qquad \times
i G_{\gamma\delta,\lambda\omega} \left( p= k_1-k_3 \right) \lambda_3^{\lambda\omega} \left(p_1=k_2,p_2=-k_4 \right)
\nonumber \\&&
 = 
\kappa^2 \left\{ \left[ \frac1{2\kappa^2}\left( k_1^2 +k_3^2 + (k_1-k_3)^2  \right) -\frac{\beta}2 \left( k_1^4 +k_3^4 + (k_1-k_3)^4 \right)  \right]
\e^\alpha_{1\gamma} \e_{3}^{\gamma\beta} 
\right.
\nonumber \\ &&
+\left[ -\frac1{\kappa^2}+\beta \left( k_1^2 +k_3^2 + (k_1-k_3)^2 \right)
\right]
\left( k_{1\gamma} \e^\gamma_{3\lambda} \e^{\lambda\alpha}_{1} k_1^\beta 
+k_{3\gamma} \e^\gamma_{1\lambda} \e^{\lambda\alpha}_{3}k_3^\beta
- k_{1\gamma} \e^{\gamma\alpha}_{3}  \e^{\beta\lambda}_1 k_{3\lambda}\right)
\nonumber \\ &&
+ \left[ 
\frac1{4\kappa^2}-\frac\beta 4 \left(k_1^2 +k_3^2 + (k_1-k_3)^2 \right)
\right]
\left( 2\e_{1\gamma\lambda} \e_{3}^{\gamma\lambda}  k_1^\alpha k_1^\beta
+2\e_{1}^{\alpha\beta}   k_1^\gamma k_1^\lambda \e_{3\gamma\lambda}
 \left.
+2\e_{3}^{\alpha\beta}  k_3^\gamma k_3^\lambda \e_{1\gamma\lambda}
\right) \right\}
\nonumber \\ &&
\times \frac{-2}{\beta (k_1-k_3)^4 - \kappa^{-2} (k_1-k_3)^2}
\left\{
 \frac12\left( \theta_{\alpha \mu}\theta_{\beta\nu}+\theta_{\alpha \nu}\theta_{\beta\mu}\right)-\frac13 \theta_{\alpha \beta}\theta_{\mu\nu}
\right\} 
\frac12 \left[ 
k_2^\mu k_4^\nu +k_2^\nu k_4^\mu \right]
\nonumber \\ &&
+\kappa^2  \left\{
\left[
-\frac1{4\kappa^2}\left( (k_1-k_3)^2-\frac12 \left(k_1^2+k_3^2\right)+ \frac16 \frac{\left( k_1^2-k_3^2 \right)^2}{(k_1-k_3)^2} \right)
\right. \right. \nonumber \\ &&  \qquad \qquad
+\left(3\alpha + \beta\right) \left( - \frac12 (k_1-k_3)^4 -\frac16 (k_1-k_3)^2 \left( k_1^2+k_3^2\right)\right) 
\nonumber \\ && \qquad \qquad  \left.
+ \frac{\beta}{24} \left(-5 \left( k_1^4+k_3^4\right) + \frac{\left(k_1^2-k_3^2\right)^2\left(k_1^2+k_3^2\right)}{(k_1-k_3)^2} \right)
\right] \e_{1\mu\nu} \e_3^{\mu\nu}
\nonumber \\ &&  \left.
-\left[
\frac{1}{3\kappa^2}  -\frac{1}{6\kappa^2} \frac{\left( k_1^2+k_3^2 \right)}{(k_1-k_3)^2}  +\frac23\left(3\alpha+\beta\right) (k_1-k_3)^2
+\frac\beta6 \frac{k_1^4+k_3^4}{(k_1-k_3)^2}
\right] k_{3\mu} \e^\mu_{1\nu} \e^\nu_{3\alpha} k_1^\alpha  \right\}
\nonumber \\ &&
\times \frac{-1}{2(3\alpha+\beta)(k_1-k_3)^4 + \kappa^{-2}(k_1-k_3)^2}
\left(- k_{2\gamma}k_4^\gamma -2m^2+6 \xi \left(k_2-k_4\right)^2\right),
\end{eqnarray}
where we use $\theta_{\mu\nu}$, defined in eq.~\eq{deftheta}, with $p=(-k_1+k_3)$. 
We calculate each part of ${\A}_t$ separately,
\begin{eqnarray}
&&\left\{
 \frac12\left( \theta_{\alpha \mu}\theta_{\beta\nu}+\theta_{\alpha \nu}\theta_{\beta\mu}\right)-\frac13 \theta_{\alpha \beta}\theta_{\mu\nu}
\right\} 
\frac12 \left[ 
k_2^\mu k_4^\nu +k_2^\nu k_4^\mu \right]
\nonumber \\ && \qquad\qquad\qquad\qquad
= \frac14 \left( k_{2\alpha} + k_{4\alpha}\right)\left( k_{2\beta} + k_{4\beta}\right) 
- \frac16 \theta_{\alpha\beta} \left( k_{2\gamma}k_4^\gamma -m^2\right) ,
\\
&&\e^\alpha_{1\gamma} \e_{3}^{\gamma\beta} \left\{ 
\frac14 \left( k_{2\alpha} + k_{4\alpha}\right)\left( k_{2\beta} + k_{4\beta}\right) 
- \frac16 \theta_{\alpha\beta} \left( k_{2\gamma}k_4^\gamma -m^2\right)
\right\}
\nonumber \\ && \qquad\qquad\qquad\qquad
=\left(\frac16 -\frac{m^2}{3 (k_2-k_4)^2} \right)
\left(k_2\cdot \e_1 \cdot \e_3 \cdot k_2  +k_4\cdot \e_1 \cdot \e_3 \cdot k_4 \right)
\nonumber \\ && \qquad\qquad\qquad\qquad\qquad
+
\left(\frac13 +\frac{m^2}{3 (k_2-k_4)^2} \right)
\left(k_2\cdot \e_1 \cdot \e_3 \cdot k_4 +k_4\cdot \e_1 \cdot \e_3 \cdot k_2\right)
\nonumber \\ && \qquad\qquad\qquad\qquad\qquad
-\frac{\left( k_{2\gamma}k_4^\gamma -m^2\right)}{6 }  \Tr[\e_1\cdot \e_3] ,
\end{eqnarray}
where 
we use the on-shell condition for matter, {\i.e.} $k_2^2=k_4^2=-m^2$,
\begin{eqnarray}
&&\left( k_{1\gamma} \e^\gamma_{3\lambda} \e^{\lambda\alpha}_{1} k_1^\beta 
+k_{3\gamma} \e^\gamma_{1\lambda} \e^{\lambda\alpha}_{3}k_3^\beta
- k_{1\gamma} \e^{\gamma\alpha}_{3}  \e^{\beta\lambda}_1 k_{3\lambda}\right)
\nonumber \\ && \qquad\qquad\qquad\qquad\times
\left\{ 
\frac14 \left( k_{2\alpha} + k_{4\alpha}\right)\left( k_{2\beta} + k_{4\beta}\right) 
- \frac16 \theta_{\alpha\beta} \left( k_{2\gamma}k_4^\gamma -m^2\right)
\right\}
\nonumber \\ && \quad
=
\frac14\left(k_{2\gamma}+k_{4\gamma}\right) \left(k_1^\gamma+k_3^\gamma\right)
\left(k_2\cdot \e_1 \cdot \e_3 \cdot k_4 -k_4\cdot \e_1 \cdot \e_3 \cdot k_2\right)
\nonumber \\ && \qquad\qquad\qquad\qquad
+ \frac14
\left( k_2\cdot \e_1 \cdot k_2-k_4\cdot \e_1 \cdot k_4 \right)
\left( k_2\cdot \e_3 \cdot k_2-k_4\cdot \e_3 \cdot k_4 \right) 
\nonumber \\ && \qquad\qquad\qquad\qquad
+\frac{\left( k_{2\gamma}k_4^\gamma -m^2\right)}{6(k_1-k_3)^2}
\left( k_2\cdot \e_1 \cdot k_2+k_4\cdot \e_1 \cdot k_4 - 2 k_2\cdot \e_1 \cdot k_4 \right)
\nonumber \\ && \qquad\qquad\qquad\qquad\qquad\qquad\qquad\qquad \times
\left( k_2\cdot \e_3 \cdot k_2+k_4\cdot \e_3 \cdot k_4 - 2 k_2\cdot \e_3 \cdot k_4 \right) ,
\\
&&
\left( 2\e_{1\gamma\lambda} \e_{3}^{\gamma\lambda}  k_1^\alpha k_1^\beta
+2\e_{1}^{\alpha\beta}   k_1^\gamma k_1^\lambda \e_{3\gamma\lambda}
+2\e_{3}^{\alpha\beta}  k_3^\gamma k_3^\lambda \e_{1\gamma\lambda}
\right)
\nonumber \\ && \qquad\qquad\qquad\qquad\times
\left\{ 
\frac14 \left( k_{2\alpha} + k_{4\alpha}\right)\left( k_{2\beta} + k_{4\beta}\right) 
- \frac16 \theta_{\alpha\beta} \left( k_{2\gamma}k_4^\gamma -m^2\right)
\right\}
\nonumber \\ && \quad
= \left\{
\left(
\frac{\left( k_1-k_3\right)^2}{12}
-\frac{k_1^2+k_3^2}6
+\frac{\left( k_1^2-k_3^2 \right)^2}{12 \left(k_1-k_3\right)^2}
\right)
\left( k_{2\gamma}k_4^\gamma -m^2 \right)
\right.
\nonumber \\ && \qquad\qquad\qquad\qquad\qquad\qquad\qquad\qquad\qquad\qquad
\left.
+\frac12 \left[ \left(k_{2\gamma}+k_{4\gamma}\right)k_1^\gamma \right]^2\right\}
\Tr[e_1 \cdot e_3] 
\nonumber \\ && \qquad
+\left( k_2\cdot \e_1 \cdot k_2+k_4\cdot \e_1 \cdot k_4 \right)
\left( k_2\cdot \e_3 \cdot k_2+k_4\cdot \e_3 \cdot k_4 \right)
-4k_2\cdot \e_1 \cdot k_4\ k_2\cdot \e_3 \cdot k_4
\nonumber \\ && \qquad
+\frac{ 2\left( k_{2\gamma}k_4^\gamma -m^2\right)}{3 (k_1-k_3)^2}
\left( k_2\cdot \e_1 \cdot k_2+k_4\cdot \e_1 \cdot k_4 - 2 k_2\cdot \e_1 \cdot k_4 \right)
\nonumber \\ && \qquad\qquad\qquad\qquad\qquad \times
\left( k_2\cdot \e_3 \cdot k_2+k_4\cdot \e_3 \cdot k_4 - 2 k_2\cdot \e_3 \cdot k_4 \right), \\
&&
k_{3\mu} \e^\mu_{1\nu} \e^\nu_{3\alpha} k_1^\alpha 
= -\left(k_2\cdot \e_1 \cdot \e_3 \cdot k_2  +k_4\cdot \e_1 \cdot \e_3 \cdot k_4
-k_2\cdot \e_1 \cdot \e_3 \cdot k_4 -k_4\cdot \e_1 \cdot \e_3 \cdot k_2 \right).
\nonumber \\ 
\end{eqnarray}

The external momenta squared of gravitons $k_1^2$ and $k_3^2$ can be written by the graviton masses squared, that is $m_g^2=0$ (for positive norm modes) and $m_g^2=-1/ (\beta \kappa^2)$ (for negative norm modes) 
where $m_g$ is mass of spin-2 graviton. 
Thus, we have four cases,
\begin{eqnarray}
\left(k_1^2, k_3^2\right)=\left\{ \Bigl(0,0\Bigr),\Bigl(0, \frac1{\beta \kappa^2}\Bigr),\Bigl(\frac1{\beta \kappa^2},0\Bigr), \Bigl(\frac1{\beta \kappa^2},\frac1{\beta \kappa^2}\Bigr) \right\}.
\label{gmass}
\end{eqnarray}
In all cases, equalities
\begin{eqnarray}
\frac{1}{\kappa^2}\left (k_1^2+k_3^2\right) - \beta \left( k_1^4+k_3^4\right) =0 ,\qquad
\left(k_1^2-k_3^2\right)^2\left(\frac{1}{\kappa^2} - \beta \left( k_1^2+k_3^2\right) \right) =0 
\end{eqnarray}
hold true. 
Then the $t$-channel exchange contribution ${\A}_t$ can be simplified as follows,
\begin{eqnarray}
&&{\A}_t = \kappa^2\left(\frac{1}{( k_1 - k_3)^2} +\frac{\beta(k_1^2+k_3^2)}{\beta (k_1-k_3)^4 - \kappa^{-2} (k_1-k_3)^2}\right)
\nonumber \\ && \qquad\qquad\times
\Bigl[
-(k_1\cdot k_2+k_1\cdot k_4)
\left(k_2\cdot \e_1 \cdot \e_3 \cdot k_4 -k_4\cdot \e_1 \cdot \e_3 \cdot k_2\right)
\nonumber \\ && \qquad\qquad\qquad
+\left(k_2\cdot \e_1\cdot k_2\right)\left(k_4\cdot \e_3\cdot k_4\right)
+\left(k_4\cdot \e_1\cdot k_4\right)\left(k_2\cdot \e_3\cdot k_2\right)
\nonumber \\ && \qquad\qquad\qquad
-2\left(k_2\cdot \e_1\cdot k_4\right)\left(k_2\cdot \e_3\cdot k_4\right)
+ \frac14  (k_1\cdot k_2+k_1\cdot k_4)^2
\Tr[\e_1 \cdot \e_3] 
\Bigr]
\nonumber \\ && \qquad
+\frac{ \kappa^2}2 
\left(k_2\cdot \e_1 \cdot \e_3 \cdot k_4 +k_4\cdot \e_1 \cdot \e_3 \cdot k_2\right) 
+ \kappa^2\left[ \frac3{16} (k_1-k_3)^2 + \frac{(k_1^2+k_3^2)}{16}\right] \Tr[\e_1\cdot\e_3]
\nonumber \\ && \qquad
+ \frac{ \kappa^2 \beta k_1^2k_3^2((k_1-k_3)^2+4m^2)}{8\left(\beta (k_1-k_3)^4 - \kappa^{-2} (k_1-k_3)^2\right)} \Tr[\e_1\cdot\e_3]
 \nonumber\\
&&\qquad + 
\xi \kappa^2 \left[ \left(2k_1^2+2k_3^2 -3 k_{1}\cdot k_3 \right) 
\Tr\left[ \e_{1}\cdot  \e_3 \right]
+ 2  k_{3}\cdot  \e_{1} \cdot  \e^\nu_{3} \cdot  k_1  \right]
\end{eqnarray}
Taking the sum ${\A}_s+{\A}_t+{\A}_u+{\A}_d$, we have the total scattering amplitude shown in eq.~\eq{sum1}. 
There we use 
\begin{eqnarray}
(k_1^2+k_3^2) \left( k_1^2+k_3^2 - \frac1{\kappa^2\beta} \right) 
= 
2 k_1^2k_3^2,  
\end{eqnarray}
which is satisfied in any case of Eq.(\ref{gmass}).
Note that the contributions from $\alpha$ $\xi$ are canceled.

\subsubsection{$(h^{(\sigma)}\phi \to I^{(s)}\phi )$}$\ $

Contribution from the $t$-channel exchange is obtained as
\begin{eqnarray}
&&{\A}_t \left(h^{(\sigma)}(k_1) \phi (k_2); I^{(s)}(k_3) \phi (k_4)\right)  \nonumber \\
&&=\kappa^2  \e_{1\mu\nu} \left(\frac{\sqrt{3}}3  \theta_{3\alpha\beta} \right)\lambda_3^{\mu\nu,\alpha\beta,\gamma\delta} \left( p_1= k_1, p_2=-k_3, p_3=-k_1+k_3\right) 
\nonumber \\ && \qquad\qquad \times
iG_{\gamma\delta,\lambda\omega} \left( p= k_1-k_3 \right) \lambda_3^{\lambda\omega} \left(p_1=k_2,p_2=-k_4 \right)
\nonumber \\&&
=\sqrt{3}\kappa^2 \left\{
\left[
-\frac1{4\kappa^2}\left( k_3^2- \frac12 \left(k_1^2+(k_1-k_3)^2\right)+ \frac16 \frac{\left( k_1^2-(k_1-k_3)^2 \right)^2}{k_3^2} 
\right)
\right. \right. \nonumber \\ && \qquad \qquad  
+\left(3\alpha + \beta\right) \left( - \frac12 k_3^4 -\frac16 k_3^2 \left( k_1^2+(k_1-k_3)^2\right)\right) 
 \nonumber \\ && \qquad \qquad  \left.
+ \frac{\beta}{24} \left(-5 \left( k_1^4+(k_1-k_3)^4\right) + \frac{\left(k_1^2-(k_1-k_3)^2\right)^2\left(k_1^2+(k_1-k_3)^2\right)}{k_3^2} \right)
\right] \e_{1\alpha\beta} 
\nonumber \\ &&  \left.
+\left[
-\frac{1}{6\kappa^2} \left( -2 +\frac{\left( k_1^2+(k_1-k_3)^2 \right)}{k_3^2} \right) +\frac23\left(3\alpha+\beta\right) k_3^2
+\frac\beta6 \frac{k_1^4+(k_1-k_3)^4}{k_3^2}
\right] k_{3\gamma} \e^\gamma_{1\alpha} k_{1\beta} \right\}
\nonumber \\ && \qquad\qquad \times
\frac{-2}{\beta (k_1-k_3)^4 - \kappa^{-2} (k_1-k_3)^2}
\left\{
 \frac12\left( \theta_{\alpha \mu}\theta_{\beta\nu}+\theta_{\alpha \nu}\theta_{\beta\mu}\right)-\frac13 \theta_{\alpha \beta}\theta_{\mu\nu}
\right\} 
\frac12 \left[ 
k_2^\mu k_4^\nu +k_2^\nu k_4^\mu \right]
\nonumber \\ &&
+
\sqrt{3}\kappa^2 \Biggl\{ \frac1{\kappa^2}
\left[ -k_1^4 +2\left((k_1-k_3)^4+k_3^4\right)-
12(k_1-k_3)^2k_3^2 +3 k_1^2((k_1-k_3)^2+k_3^2) \right]
\nonumber \\ && \quad
+4(3\alpha + \beta) 
\left[ \left((k_1-k_3)^6+k_3^6\right) -9 \left((k_1-k_3)^4k_3^2+(k_1-k_3)_2^2k_3^4\right)
 +3 k_1^2\left((k_1-k_3)^4+k_3^4\right) \right]
\nonumber \\ && \quad 
+\beta 
\left[ k_1^6 
+3 k_1^4\left((k_1-k_3)^2+k_3^2\right) \right]
\Biggr\}
\nonumber \\ &&\qquad \times
\frac1{72 (k_1-k_3)^2k_3^2}
\frac{-k_3^\mu \e_{1\mu\nu}k_3^\nu}{2(3\alpha+\beta)(k_1-k_3)^4 + \kappa^{-2}(k_1-k_3)^2}
\left(- k_{2\gamma}k_4^\gamma -2m^2+6 \xi \left(k_2-k_4\right)^2\right)
\nonumber\\
\end{eqnarray}
Now, the graviton in the final state is scalar one, mass of which is $(-k_3^2 =)m_S^2= \dfrac1{2(3\alpha+\beta) \kappa^2}$, 
while the graviton mass of the initial state $(-k_1^2=) m_1^2 = 0$ or $-\dfrac1{\kappa^2 \beta}$. 
The latter gives 
\begin{eqnarray}
k_1^2 ( \kappa^{-2}  -\beta k_1^2) =0.
\end{eqnarray}
Using these equations, we have
\begin{eqnarray}
&&-\frac1{4\kappa^2}\left( k_3^2- \frac12 \left(k_1^2+(k_1-k_3)^2\right)+ \frac16 \frac{\left( k_1^2-(k_1-k_3)^2 \right)^2}{k_3^2} 
\right)
 \nonumber \\ && \qquad \qquad  
+\left(3\alpha + \beta\right) \left( - \frac12 k_3^4 -\frac16 k_3^2 \left( k_1^2+(k_1-k_3)^2\right)\right) 
 \nonumber \\ && \qquad \qquad  
+ \frac{\beta}{24} \left(-5 \left( k_1^4+(k_1-k_3)^4\right) + \frac{\left(k_1^2-(k_1-k_3)^2\right)^2\left(k_1^2+(k_1-k_3)^2\right)}{k_3^2} \right)
\nonumber \\ 
&&\qquad  
=\frac1{24} \left( \beta (k_1-k_3)^2 - \frac1{\kappa^2} \right)
\left( \frac{(k_1-k_3)^4}{k_3^2}-5(k_1-k_3)^2-\frac{k_1^2}{k_3^2}(k_1-k_3)^2 \right),
\\
&& 
-\frac{1}{6\kappa^2} \left( -2 +\frac{\left( k_1^2+(k_1-k_3)^2 \right)}{k_3^2} \right) +\frac23\left(3\alpha+\beta\right) k_3^2
+\frac\beta6 \frac{k_1^4+(k_1-k_3)^4}{k_3^2}
\nonumber \\ &&\qquad  
=\frac16\left(\beta (k_1-k_3)^2-\frac1{\kappa^2}\right)  \frac{(k_1-k_3)^2}{k_3^2} ,
\\
&&
 \frac1{\kappa^2}
\left[ -k_1^4 +2\left((k_1-k_3)^4+k_3^4\right)-
12(k_1-k_3)^2k_3^2 +3 k_1^2((k_1-k_3)^2+k_3^2) \right]
\nonumber \\ && \quad 
+4(3\alpha + \beta) 
\left[ \left((k_1-k_3)^6+k_3^6\right) -9 \left((k_1-k_3)^4k_3^2+(k_1-k_3)^2k_3^4\right)
 +3 k_1^2\left((k_1-k_3)^4+k_3^4\right) \right]
\nonumber \\ && \quad 
+\beta 
\left[ k_1^6 
+3 k_1^4\left((k_1-k_3)^2+k_3^2\right)\right]
\nonumber \\
&&\qquad  
=
\left( 2(3\alpha+\beta)(k_1-k_3)^2 + \kappa^{-2} \right)
\left( 2(k_1-k_3)^4 +6k_1^2 (k_1-k_3)^2 -18k_3^2 (k_1-k_3)^2 \right)
\nonumber \\
&&\qquad\qquad\qquad\qquad
-48(3\alpha+\beta)k_3^4 (k_1-k_3)^2 ,
\\
&&\e_{1\alpha\beta}
\left\{
 \frac12\left( \theta_{\alpha \mu}\theta_{\beta\nu}+\theta_{\alpha \nu}\theta_{\beta\mu}\right)-\frac13 \theta_{\alpha \beta}\theta_{\mu\nu}
\right\} 
\frac12 \left[ 
k_2^\mu k_4^\nu +k_2^\nu k_4^\mu \right]
\nonumber \\ &&\quad
= \frac14 \left( k_{2}\cdot \e_1 \cdot k_2+2 k_{2}\cdot \e_1 \cdot k_4+k_4\cdot \e_1 \cdot k_4\right)
+ \frac1{6(k_1-k_3)^2} \left( k_{2\gamma}k_4^\gamma-m^2 \right) k_3\cdot \e_1 \cdot k_3
\nonumber \\ &&\quad
= \frac14 \left( k_{2}\cdot \e_1 \cdot k_2+2 k_{2}\cdot \e_1 \cdot k_4+k_4\cdot \e_1 \cdot k_4\right)
- \left(\frac1{12} + \frac{m^2}{3(k_1-k_3)^2}  \right) k_3\cdot \e_1 \cdot k_3,
\\
&&k_{3\gamma} \e^\gamma_{1\alpha} k_{1\beta}
\left\{
 \frac12\left( \theta_{\alpha \mu}\theta_{\beta\nu}+\theta_{\alpha \nu}\theta_{\beta\mu}\right)-\frac13 \theta_{\alpha \beta}\theta_{\mu\nu}
\right\} 
\frac12 \left[ 
k_2^\mu k_4^\nu +k_2^\nu k_4^\mu \right]
\nonumber \\ &&\quad
= \frac14 \left( k_{2\gamma}+k_{4\gamma} \right) k_1^\gamma \left( k_{2}\cdot \e_1 \cdot k_2- k_4\cdot \e_1 \cdot k_4\right)
\nonumber \\ &&\qquad\qquad
+\left( \frac1{24} + \frac16 \frac{m^2}{(k_1-k_3)^2} \right) \left( (k_1-k_3)^2 +k_1^2 -k_3^2   \right)  k_3\cdot \e_1 \cdot k_3
\nonumber \\ &&\quad
= \frac18 \left[\left(k_1+k_2 \right)^2-\left(k_1-k_4 \right)^2\right]\left( k_{2}\cdot \e_1 \cdot k_2- k_4\cdot \e_1 \cdot k_4\right)
\nonumber \\ &&\qquad\qquad
+\left( \frac1{24} + \frac16 \frac{m^2}{(k_1-k_3)^2} \right) \left( (k_1-k_3)^2 +k_1^2 -k_3^2   \right)  k_3\cdot \e_1 \cdot k_3 .
\end{eqnarray}
These equation simplify the form of $\A_t$ and finally we have
\begin{eqnarray}
&&\A_t= -\frac{\sqrt{3} \kappa^2 }{24 k_3^2} \left((k_1+k_2)^2-(k_1-k_4)^2\right) \left( k_2 \cdot \e_1 \cdot k_2- k_4 \cdot \e_1 \cdot k_4\right) 
\nonumber \\ && \qquad
-\frac{\sqrt{3} \kappa^2 }{24 k_3^2}\left((k_1-k_3)^2+k_1^2 -5 k_3^2\right)\left( k_2 \cdot \e_1 \cdot k_2+ k_4 \cdot \e_1 \cdot k_4 \right) 
+\frac{\sqrt{3} \kappa^2 }{6 k_3^2}k_1^2 k_2 \cdot \e_1 \cdot k_4
\nonumber \\ && \qquad
- \frac{\sqrt{3} \kappa^2 }6\left( 2\frac{m^2}{p^2} - \frac{2(3\alpha+\beta)}{\left(2(3\alpha+\beta)p^2+ \kappa^{-2}\right)}
\frac{k_3^2}{p^2} \left( p^2-2m^2\right)
\right)k_3 \cdot \e_1 \cdot k_3
\nonumber \\ && \qquad
+
\frac{\xi\kappa^2}{\sqrt{3}k_3^2}\Biggl\{ 
 \left(k_{1}\cdot k_3 \right) -2 k_1^2 + 4k_3^2
+ \frac{12(3\alpha+\beta)k_3^4}{2(3\alpha+\beta)(k_1-k_3)^2 + \kappa^{-2} }\Biggr\}
k_3\cdot  \e_{1} \cdot  k_3 .
\end{eqnarray}
The sum $\A_s+\A_u+\A_t+\A_d$ is shown in eq. \eq{sum2}.

\subsubsection{$(I^{(s)}\phi \to I^{(s)}\phi )$}$\ $

Using the following equation
\begin{eqnarray}
&&k_1^\alpha k_1^\beta
\left[\frac14 \left( k_{2\alpha} + k_{4\alpha}\right)\left( k_{2\beta} + k_{4\beta}\right) 
- \frac16 \theta_{\alpha\beta} \left( k_{2\gamma}k_4^\gamma -m^2\right) \right]
\nonumber \\ 
&&\qquad
=\frac1{16} \left((k_1+k_2)^2-(k_1-k_4)^2\right)^2-\frac1{48}\left((k_1-k_3)^2 +4m^2\right)\left( (k_1-k_3)^2-4k_1^2\right)
\nonumber \\ 
&&\qquad
=\frac1{24} \left((k_1-k_3)^4 -6 (k_1+k_2)^2(k_1-k_4)^2 
\right.
\nonumber \\ 
&&\qquad\qquad\qquad
\left.
-4(k_1^2-m^2)(k_1-k_3)^2+6(k_1^2-m^2)^2+8k_1^2m^2\right), 
\end{eqnarray}
the $t$-channel exchange contribution is written as
\begin{eqnarray}
&&{\A}_t \left(I^{(s)}(k_1) \phi (k_2); I^{(s)}(k_3) \phi (k_4)\right)  \nonumber \\
&&=  \kappa^2 \left(\frac{\sqrt{3}}3  \theta_{1\mu\nu} \right)\left(\frac{\sqrt{3}}3  \theta_{3\alpha\beta} \right)\lambda_3^{\mu\nu,\alpha\beta,\gamma\delta} \left( p_1= k_1, p_2=-k_3, p_3=-k_1+k_3\right) 
\nonumber \\ && \qquad\qquad \times
iG_{\gamma\delta,\lambda\omega} \left( p= k_1-k_3 \right) \lambda_3^{\lambda\omega} \left(p_1=k_2,p_2=-k_4 \right)
\nonumber \\&&
=\kappa^2  \frac{k_1^\alpha k_1^\beta}{24 k_1^2k_3^2} 
\left\{ 
\kappa^{-2}\left( -(k_1-k_3)^4+3 (k_1-k_3)^2(k_1^2+k_3^2) +2 (k_1^4+k_3^4)-12k_1^2k_3^2  \right)
\right. \nonumber \\ && \qquad \qquad \qquad
+(3\alpha + \beta)
\left(  12 (k_1-k_3)^2(k_1^4+k_3^4)  +4  (k_1^6+k_3^6) -36 (k_1^4k_3^2+k_1^2k_3^4) \right)
\nonumber \\ && \qquad \qquad \qquad \left.
+\beta
\left( (k_1-k_3)^6 +3(k_1-k_3)^4(k_1^2+k_3^2) 
 \right)
\right\}
\nonumber \\ && \qquad\qquad \times
\frac{-2}{\beta (k_1-k_3)^4 - \kappa^{-2} (k_1-k_3)^2}
\left\{
 \frac12\left( \theta_{\alpha \mu}\theta_{\beta\nu}+\theta_{\alpha \nu}\theta_{\beta\mu}\right)-\frac13 \theta_{\alpha \beta}\theta_{\mu\nu}
\right\} 
\frac12 \left[ 
k_2^\mu k_4^\nu +k_2^\nu k_4^\mu \right]
\nonumber \\ &&
+
\frac{\kappa^2 }{144 k_1^2k_3^2(k_1-k_3)^2}
\Biggl[
\frac1{\kappa^2}\left(-(k_1^8+(k_1-k_3)^8+k_3^8) 
\right. \nonumber \\ && \qquad\quad \left.
+8 (k_1^2(k_1-k_3)^6+k_1^2k_3^6+k_1^6(k_1-k_3)^2+k_1^6k_3^2+(k_1-k_3)^2k_3^6+(k_1-k_3)^6k_3^2)
\right. \nonumber \\ && \qquad\quad \left.
-14(k_1^4(k_1-k_3)^4+k_1^4k_3^4+(k_1-k_3)^4k_3^4)-28k_1^2(k_1-k_3)^2k_3^2(k_1^2+(k_1-k_3)^2+k_3^2)
\right)
\nonumber \\ && \qquad 
+2(3\alpha+\beta) \left(-(k_1^{10}+(k_1-k_3)^{10}+k_3^{10}) 
\right. \nonumber \\ && \qquad \quad 
+11(k_1^2(k_1-k_3)^8+k_1^2k_3^8+k_1^8(k_1-k_3)^2+k_1^8k_3^2+(k_1-k_3)^2k_3^8+(k_1-k_3)^8k_3^2)
\nonumber \\ && \qquad \quad
-10(k_1^4(k_1-k_3)^6+k_1^4k_3^6+k_1^6(k_1-k_3)^4+k_1^6k_3^4+(k_1-k_3)^4k_3^6+(k_1-k_3)^6k_3^4)
\nonumber \\ && \qquad \quad 
-34k_1^2(k_1-k_3)^2k_3^2(k_1^4+(k_1-k_3)^4+k_3^4)
\nonumber \\ && \qquad \quad \left.
-42k_1^2(k_1-k_3)^2k_3^2(k_1^2(k_1-k_3)+k_1^2k_3^2+(k_1-k_3)^2k_3^2)
\right)
\Biggr]
\nonumber \\ && \qquad \qquad \times
\frac{-1}{2(3\alpha+\beta)(k_1-k_3)^4 + \kappa^{-2}(k_1-k_3)^2}
\left(- k_{2\gamma}k_4^\gamma -2m^2+6\xi \left(k_2-k_4 \right)^2\right)
\nonumber \\&&
%
=\frac{\kappa^2}{48 k_1^2k_3^2} 
\Biggl[
(k_1+k_2)^2(k_1-k_4)^2(k_1-k_3)^2+6k_1^2(k_1+k_2)^2(k_1-k_4)^2-(4k_1^2+m^2)(k_1-k_3)^4
\nonumber \\ &&
+(11k_1^4+4k_1^2m^2-m^4)(k_1-k_3)^2+10k_1^6-12k_1^4m^2-6k_1^2m^4\Biggr]
\nonumber \\ &&
+\frac{\kappa^2}{6}\frac{3k_1^4 (k_1-k_3)^2 -4k_1^2m^2(k_1-k_3)^2-2k_1^4m^2}{(k_1-k_3)^2\left((k_1-k_3)^2-k_1^2\right)}
\nonumber \\ &&
-\kappa^2
\frac{(k_1-k_3)^4 -6 (k_1+k_2)^2(k_1-k_4)^2 
-4(k_1^2-m^2)(k_1-k_3)^2+6(k_1^2-m^2)^2+8k_1^2m^2
}{12 \left(\kappa^2 \beta (k_1-k_3)^2 -1 \right)(k_1-k_3)^2}
\nonumber \\
&&\qquad
+\frac{\xi \kappa^2}{3}
\left[
-\frac{\left( k_1\cdot k_3 \right)^3}{k_1^4} -8\frac{\left( k_1\cdot k_3 \right)^2}{k_1^2} +7 \left( k_1\cdot k_3 \right) - 10 k_1^2
+\frac{12 k_1^4}{\left((k_1-k_3)^2 -k_1^2 \right)}
\right] .
\end{eqnarray}
The sum $\A_s+\A_u+\A_t+\A_d$ is shown in eq.\eq{sum3}.

\section{Derivation of the UV behavior} \label{AppUV}

In the calculation, inner products of $k_\mu$, $l_\mu$, $t_\mu$ and $u_\mu$ appear frequently. 
Before going to the calculation of each scattering amplitude, we summarise the inner products,
\begin{eqnarray}
&&k_1\cdot k_2\sim  -2k^2 - \frac{m_1^2+m^2}{2} + \frac{\left(m_1^2-m^2\right)^2}{8k^2} , \label{kk} \\
&&k_1\cdot k_3\sim -kq\left(1-\cos\theta \right) - \frac{kq}{2} \left( \frac{m_1^2}{k^2}+\frac{m_3^2}{q^2} \right) 
+ \frac{kq}8 \left( \frac{m_1^2}{k^2}-\frac{m_3^2}{q^2} \right)^2 , \\
&&k_1\cdot k_4\sim -kq\left(1+\cos\theta \right) - \frac{kq}{2} \left( \frac{m_1^2}{k^2}+\frac{m^2}{q^2} \right) 
+ \frac{kq}8 \left( \frac{m_1^2}{k^2}-\frac{m^2}{q^2} \right)^2 , \\
&&k_2\cdot k_3\sim -kq\left(1+\cos\theta \right) - \frac{kq}{2} \left( \frac{m^2}{k^2}+\frac{m_3^2}{q^2} \right) 
+ \frac{kq}8 \left( \frac{m^2}{k^2}-\frac{m_3^2}{q^2} \right)^2 , \\
&&k_2\cdot k_4\sim -kq\left(1-\cos\theta \right) - \frac{kq}{2} \left( \frac{m^2}{k^2}+\frac{m^2}{q^2} \right) 
+ \frac{kq}8 \left( \frac{m^2}{k^2}-\frac{m^2}{q^2} \right)^2 , \\
&&k_3\cdot k_4\sim  -2q^2 - \frac{m_3^2+m^2}{2} + \frac{\left(m_3^2-m^2\right)^2}{8q^2} , \\
&&l_1\cdot k_2\sim \frac{1}{m_1} \left[ -2k^2 - \frac{m_1^2+m^2}{2} + \frac{\left(m_1^4+m^4\right)}{8k^2} \right] , \\
&&l_1\cdot k_3\sim \frac{1}{m_1} \left[-kq\left(1-\cos\theta \right) - \frac{kq}{2} \left( -\frac{m_1^2}{k^2} \cos\theta+\frac{m_3^2}{q^2} \right) 
+ \frac{kq}8 \left( -\frac{m_1^4}{k^4}\cos\theta+\frac{m_3^4}{q^4} \right)\right] , \nonumber \\ \\
&&l_1\cdot k_4\sim\frac{1}{m_1} \left[ -kq\left(1+\cos\theta \right) - \frac{kq}{2} \left( \frac{m_1^2}{k^2} \cos\theta+\frac{m^2}{q^2} \right) 
+ \frac{kq}8 \left( \frac{m_1^4}{k^4}\cos\theta +\frac{m^4}{q^4} \right)\right] , \\
&&l_3\cdot k_1\sim \frac{1}{m_3} \left[ -kq\left(1-\cos\theta \right) - \frac{kq}{2} \left( \frac{m_1^2}{k^2}-\frac{m_3^2}{q^2}\cos\theta \right) 
+ \frac{kq}8 \left( \frac{m_1^4}{k^4}-\frac{m_3^4}{q^4} \cos\theta \right)\right] , \nonumber \\ \\
&&l_3\cdot k_2\sim \frac{1}{m_3} \left[ -kq\left(1+\cos\theta \right) - \frac{kq}{2} \left( \frac{m^2}{k^2}+\frac{m_3^2}{q^2}\cos\theta \right) 
+ \frac{kq}8 \left( \frac{m^4}{k^4}+\frac{m_3^4}{q^4}\cos\theta \right)\right] , \nonumber \\ \\
&&l_3\cdot k_4\sim \frac{1}{m_3} \left[  -2q^2 - \frac{m_3^2+m^2}{2} + \frac{\left(m_3^4+m^4\right)}{8q^2} \right] , \\
&&t_1\cdot k_2= t_3\cdot k_4 =0,\\
&&t_1\cdot k_3= q\sin\theta,  \quad 
t_1\cdot k_4= -q\sin\theta  ,\\
&&t_3\cdot k_1= - k\sin\theta,  \quad 
t_3\cdot k_2= k\sin\theta , \\
&&l_1\cdot l_3\sim \frac{1}{m_1m_3} \left[-kq\left(1-\cos\theta \right) + \frac{kq}{2} \left( \frac{m_1^2}{k^2}+\frac{m_3^2}{q^2} \right) \cos\theta 
- \frac{kq}8 \left( \frac{m_1^2}{k^2}-\frac{m_3^2}{q^2} \right)^2 \cos\theta\right] ,\nonumber \\  \\
&&l_1\cdot t_3= - \sqrt{k^2+m_1^2}\sin \theta / m_1 \sim - \left(\frac{ k}{m_1} + \frac{m_1}{2k} - \frac{m_1^3}{8k^3} \right)\sin \theta  , \\
&&l_3\cdot t_1= \sqrt{q^2+m_3^2}\sin \theta / m_3 \sim \left(\frac{ q}{m_3} + \frac{m_3}{2q} - \frac{m_3^3}{8q^3} \right)\sin \theta  , \\
&&t_3\cdot t_1= \cos \theta  , \label{tt}
\end{eqnarray}
where ``$\sim$'' means that we ignore the higher order terms of $k^{-1}$. 
For later convenience, we show also the following relations,
\begin{eqnarray}
&&k_1^2 + 2k_1\cdot k_2 = -4kq -m_1^2-m_3^2-m^2 + \cO (k^{-2}) ,  \\
&&k_1^2 - 2k_1\cdot k_4 = 2kq (1+\cos \theta)+m^2 + \cO (k^{-2}) .
\end{eqnarray}
We are now in a position to obtain the behaviors of the amplitudes in the UV limit, which is the main purpose of this paper.
As we commented in the last paragraph of section \ref{MD}, 
the scattering amplitudes involving both even and odd modes vanish. 
There are thirteen non-zero scattering amplitudes in total. 
We give the analysis of each.

\subsection*{{\bf E.1}\hspace{5mm}$h^{(2,o)}\phi \, \to h^{(2,o)}\phi$}

Since these modes satisfy
\begin{eqnarray}
&&k\cdot \e_1 \cdot k =k\cdot \e_3 \cdot k =0 \qquad  \mbox{ for any $k$}, \nonumber \\
&&k_2\cdot \e_1 \cdot \e_3 \cdot k_4 =0, \quad
k_4\cdot \e_1 \cdot \e_3 \cdot k_2 = \dfrac12 (t_1 \cdot k_4)(t_3 \cdot k_2),  \\
&&\Tr[\e_1 \cdot \e_3] =  (t_1 \cdot t_3)  ,\nonumber
\end{eqnarray}
the scattering amplitude becomes simple, 
\begin{eqnarray}
&&{\A}_s +{\A}_t +{\A}_u +{\A}_d 
\nonumber \\
&&\quad
=
\frac{ \kappa^2 }{8 k_1 \cdot k_3}
\left[-2(k_1\cdot k_2+k_1\cdot k_4) (t_1 \cdot k_4)(t_3 \cdot k_2)
-\left(k_1\cdot k_2+k_1\cdot k_4\right)^2
(t_1 \cdot t_3)
\right]
 \nonumber \\ && \qquad\qquad
-\frac{ \kappa^2 }4(t_1 \cdot k_4)(t_3 \cdot k_2)
+\frac{ \kappa^2 }{8} (k_1 \cdot k_3)(t_1 \cdot t_3)
+\cO\left( k^0 \right)
\nonumber \\
&&\quad
=
- \kappa^2 \frac{(k_1\cdot k_2)}{2 k_1 \cdot k_3}
\left[  (t_1 \cdot k_4)(t_3 \cdot k_2)
+
( k_1\cdot k_4)
(t_1 \cdot t_3)
\right] 
+\cO\left( k^0 \right)
\nonumber \\
&&\quad
=
- \kappa^2 k^2\frac{1+\cos \theta }{1-\cos \theta}
+\cO\left( k^0 \right) .
\end{eqnarray}

\subsection*{{\bf E.2}\hspace{5mm}$h^{(2,o)}\phi \, \to I^{(1,o)}\phi$}

In this case, the modes satisfiy
\begin{eqnarray}
&&k\cdot \e_1 \cdot k =k\cdot \e_3 \cdot k =0 \qquad  \mbox{ for any $k$}, \nonumber \\
&&k_2\cdot \e_1 \cdot \e_3 \cdot k_4 =0, \quad
k_4\cdot \e_1 \cdot \e_3 \cdot k_2 = \dfrac12 (t_1 \cdot k_4)(l_3 \cdot k_2), \\
&&\Tr[\e_1 \cdot \e_3] =  (t_1 \cdot l_3) .\nonumber
\end{eqnarray}
Calculation becomes uncomplicated, 

\begin{eqnarray}
&&{\A}_s +{\A}_t +{\A}_u +{\A}_d 
\nonumber \\
&&=
\frac{ \kappa^2 }{8 k_1 \cdot k_3}
\left[2(k_1\cdot k_2+k_1\cdot k_4)(t_1 \cdot k_4)(l_3 \cdot k_2)
-\left(k_1\cdot k_2+k_1\cdot k_4\right)^2
(t_1 \cdot l_3)
\right]
 \nonumber \\ && \qquad\qquad
-\frac{ \kappa^2 }4(t_1 \cdot k_4)(l_3 \cdot k_2)
+\frac{ \kappa^2 }{8} (k_1 \cdot k_3)(t_1 \cdot l_3)
+\cO\left( k^{-1} \right)
\nonumber \\
&&= \kappa^2 
\frac{(k_1^2+2 k_1\cdot k_2)}{8 k_1 \cdot k_3}
\left[-2 (t_1 \cdot k_4)(l_3 \cdot k_2)
+ 
(k_1^2-2 k_1\cdot k_4)
(t_1 \cdot l_3)
\right] 
+\cO\left( k^{-1} \right)
\nonumber \\
&&= \kappa^2 
\frac{m_3 k  \sin \theta }{2(1-\cos \theta)}
+\cO\left( k^{-1} \right).
\end{eqnarray}

\subsection*{{\bf E.3}\hspace{5mm}$I^{(1,o)}\phi \, \to I^{(1,o)}\phi$}

In this case, $\e_1$ and $\e_3$ satisfy
\begin{eqnarray}
\begin{array}{ll}
k\cdot \e_1 \cdot k =k\cdot \e_3 \cdot k =0 &\qquad  \mbox{ for any $k$}, \\
k_i\cdot \e_1 \cdot \e_3 \cdot k_j = \dfrac12 (l_1 \cdot k_i)(l_3 \cdot k_j) & \qquad \mbox{ for any $k_i$ and $k_j$}, \\
\Tr[\e_1 \cdot \e_3] =  (l_1 \cdot l_3) .&
\end{array}
\end{eqnarray}
Since the initial and the final gravitons are massive gravitons, we have the following relations
\begin{eqnarray}
\begin{array}{ll}
 l_1\cdot l_3 = \dfrac{1}{m_I^2} k_1 \cdot k_3 +  ( 1+ \cos \theta ), \\
l_1 \cdot k_2=l_3 \cdot k_4= \dfrac{1}{m_I} k_1 \cdot k_2+\dfrac14 \dfrac{m_I m_S^2}{k^2} +\cO\left( k^{-4} \right),  \\
l_1 \cdot k_4=l_3 \cdot k_2= \dfrac{1}{m_I} k_1 \cdot k_4+ \dfrac{1-\cos\theta}{2}m_I - \dfrac{1-\cos\theta}{8}\dfrac{m_I^3}{k^2} 
+ \dfrac14 \dfrac{m_I m_S^2}{k^2} +\cO\left( k^{-4} \right) .
\end{array}
\end{eqnarray}
Then, we have
\begin{eqnarray}
&&
 -\frac14 (k_1^2-2k_1\cdot k_4 ) (k_1^2+2k_1\cdot k_2) \mbox{Tr} [e_1\cdot e_3]
+(k_1^2-2k_1\cdot k_4 ) k_2 \cdot e_1 \cdot e_3 \cdot k_4
\nonumber \\ && \qquad \qquad\qquad 
+ (k_1^2+2k_1\cdot k_2) k_4 \cdot e_1 \cdot e_3 \cdot k_2
-2 (k_2 \cdot e_1\cdot k_4 )( k_2 \cdot e_3 \cdot k_4)
\nonumber \\ && \qquad 
-\frac{ k_1^2-2k_1\cdot k_4 }{k_1^2+2k_1\cdot k_2} (k_2 \cdot e_1\cdot k_2 )( k_4 \cdot e_3 \cdot k_4)
-\frac{k_1^2+2k_1\cdot k_2}{ k_1^2-2k_1\cdot k_4 } (k_4 \cdot e_1\cdot k_4 )( k_2 \cdot e_3 \cdot k_2)
\nonumber\\
&&=\frac{1}{m_I^2}\left(k_1  \cdot  k_4 + \frac{m_I^2}{2}\right)\left(k_1  \cdot  k_2 - \frac{m_I^2}{2}\right)\left(k_1  \cdot  k_3 +m_I^2(1+\cos\theta )\right)
\nonumber \\
&&\qquad
-\frac{1}{m_I^2}\left(k_1  \cdot  k_4 + \frac{m_I^2}{2}\right)\left(k_1  \cdot  k_2 + \frac{m_I^2m_S^2}{4k^2}\right)^2
\nonumber \\
&&\qquad
+\frac{1}{m_I^2}\left(k_1  \cdot  k_2 - \frac{m_I^2}{2}\right)\left(k_1 \cdot k_4+ \dfrac{1-\cos\theta}{2}m_I^2 - \dfrac{1-\cos\theta}{8}\dfrac{m_I^2}{k^2} 
+ \dfrac14 \dfrac{m_I^2 m_S^2}{k^2}\right)^2
+\cO\left( k^0 \right)
\nonumber \\
&&=
-\frac{1-\cos \theta}{4} m_I^2 k^2+\cO\left( k^0 \right) .
\end{eqnarray}
Moreover, we have
\begin{eqnarray}
&&\frac18 (k_1^2+k_3^2+4m^2) \mbox{Tr} [e_1\cdot e_3]
-k_2 \cdot e_1 \cdot e_3 \cdot k_4
- k_4 \cdot e_1 \cdot e_3 \cdot k_2
\nonumber \\ &&  \qquad
+\frac{ 2 }{k_1^2+2k_1\cdot k_2} (k_2 \cdot e_1\cdot k_2)( k_4 \cdot e_3 \cdot k_4)
+\frac{2}{ k_1^2-2k_1\cdot k_4 } (k_4 \cdot e_1\cdot k_4 )( k_2 \cdot e_3 \cdot k_2) \nonumber \\
&&= -\frac{1}{2m_I^2}(k_1 \cdot k_2)^2 -\frac{1}{2m_I^2}(k_1 \cdot k_4)^2 + \cO\left(k^2\right) \nonumber \\
&&= -\frac{k^4}{2m_I^2} \left(2^2 +(1+\cos\theta)^2\right) + \cO\left(k^2\right).
\end{eqnarray}
From these equations, the total amplitude is estimated as
\begin{eqnarray}
{\A}_s +{\A}_t +{\A}_u +{\A}_d 
=
-\frac{\kappa^2 m_I^2}{8} \frac{2^2 +(1+\cos\theta)^2 +(1-\cos\theta)^2 }{(1-\cos \theta)^2}+ 
\cO\left( k^{-2} \right).
\end{eqnarray}

\subsection*{{\bf E.4}\hspace{5mm}$h^{(2,e)}\phi \, \to h^{(2,e)}\phi$} 

In this case, we have the following relations,
\begin{eqnarray}
&&k_2\cdot \e_1 \cdot k_2 =k_2\cdot \e_1 \cdot k_4 =k_2\cdot \e_3 \cdot k_4 =k_4\cdot \e_3 \cdot k_4 =0  \nonumber \\
&&k_4\cdot \e_1 \cdot k_4 = \frac1{\sqrt{2}} (t_1 \cdot k_4)^2, \qquad
k_2\cdot \e_3 \cdot k_2 = \frac1{\sqrt{2}} (t_3 \cdot k_2)^2,  \nonumber\\
&&k_2\cdot \e_1 \cdot \e_3 \cdot k_4 = 0, \qquad
k_4\cdot \e_1 \cdot \e_3 \cdot k_2 = \frac12 (t_1 \cdot k_4) (t_3 \cdot k_2) (t_1 \cdot t_3) , \\
&&\Tr[\e_1 \cdot \e_3] = \frac12 \left[ (t_1 \cdot t_3)^2 +1 \right].  \nonumber  
\end{eqnarray}
With these relations, the total amplitude is calculated as 
\begin{eqnarray}
&&{\A}_s +{\A}_t +{\A}_u +{\A}_d 
\nonumber \\
&&=
-\frac{ \kappa^2 }{16 k_1 \cdot k_3}
\Bigl[4 (k_1\cdot k_2+k_1\cdot k_4) (t_1 \cdot k_4) (t_3 \cdot k_2) (t_1 \cdot t_3)
\nonumber \\
&&\qquad
+4(t_1 \cdot k_4)^2(t_3 \cdot k_2)^2
+\left(k_1\cdot k_2+k_1\cdot k_4\right)^2
\left[(t_1 \cdot t_3)^2+1\right]
\Bigr]
 \nonumber \\ && \qquad\qquad
-\frac{ \kappa^2 }4 (t_1 \cdot k_4) (t_3 \cdot k_2) (t_1 \cdot t_3)
+\frac{ \kappa^2 }{16} (k_1 \cdot k_3)\left[(t_1 \cdot t_3)^2+1\right]
\nonumber \\
&& \qquad
+ \kappa^2  \frac {(t_1 \cdot k_4)^2(t_3 \cdot k_2)^2}{2[(k_2-k_3)+m^2]} 
+\cO\left( k^0 \right)
\nonumber \\
&&=
-\frac{ \kappa^2  }{4 k_1 \cdot k_3}
\Bigl[2 ( k_1\cdot k_2) (t_1 \cdot k_4)(t_3 \cdot k_2)(t_1 \cdot t_3) 
\nonumber \\
&&\qquad
+(t_1 \cdot k_4)^2(t_3 \cdot k_2)^2
+( k_1\cdot k_2)(k_1\cdot k_4)
\left[(t_1 \cdot t_3)^2+1\right]
\Bigr] 
\nonumber \\
&& \qquad
- \kappa^2 \frac {(t_1 \cdot k_4)^2(t_3 \cdot k_2)^2}{4k_2\cdot k_3} 
+\cO\left( k^0 \right)
\nonumber \\
&&= \kappa^2 k^2  \frac{1+\cos \theta}{1-\cos \theta}
+
\cO\left( k^0 \right) .
\end{eqnarray}

\subsection*{{\bf E.5}\hspace{5mm}$h^{(2,e)}\phi \, \to I^{(1,e)}\phi$} 

The inner products become
\begin{eqnarray}
&&k_2\cdot \e_1 \cdot k_2 =k_2\cdot \e_1 \cdot k_4 = k_4\cdot \e_3 \cdot k_4 =0 ,\qquad
k_4\cdot \e_1 \cdot k_4 = \frac1{\sqrt{2}} (t_1 \cdot k_4)^2,    \nonumber \\
&&
k_2\cdot \e_3 \cdot k_2 = \sqrt{2} (t_3 \cdot k_2)(l_3 \cdot k_2),\qquad
k_2\cdot \e_3 \cdot k_4 = \frac1{\sqrt{2}} (t_3 \cdot k_2)(l_3 \cdot k_4),  \nonumber\\
&&k_4\cdot \e_1 \cdot \e_3 \cdot k_2 = \frac12 (t_1 \cdot k_4) (l_3 \cdot k_2) (t_1 \cdot t_3)
+\frac12 (t_1 \cdot k_4) (t_3 \cdot k_2) (t_1 \cdot l_3) ,  \\
&&k_2\cdot \e_1 \cdot \e_3 \cdot k_4 = 0,  \qquad \Tr[\e_1 \cdot \e_3] =  (t_1 \cdot t_3)(t_1 \cdot l_3) , \nonumber  
\end{eqnarray}
and then we have the total scattering amplitude, 
\begin{eqnarray}
&&{\A}_s +{\A}_t +{\A}_u +{\A}_d 
\nonumber \\
&&=
-\frac{ \kappa^2 }{8 k_1 \cdot k_3}
\Bigl[2(k_1\cdot k_2+k_1\cdot k_4)\left[ 
(t_1 \cdot k_4) (l_3 \cdot k_2) (t_1 \cdot t_3)
+ (t_1 \cdot k_4) (t_3 \cdot k_2) (t_1 \cdot l_3)\right]
\nonumber \\
&& \qquad\qquad
+4(t_1 \cdot k_4)^2  (t_3 \cdot k_2)(l_3 \cdot k_2)
+\left(k_1\cdot k_2+k_1\cdot k_4\right)^2
(t_1 \cdot t_3)(t_1 \cdot l_3)
\Bigr]
\nonumber \\ && 
-\frac{ \kappa^2 }4\left[
 (t_1 \cdot k_4) (l_3 \cdot k_2) (t_1 \cdot t_3)
+ (t_1 \cdot k_4) (t_3 \cdot k_2) (t_1 \cdot l_3)\right]
\nonumber \\ && 
+\frac {\kappa^2 }8 (k_1 \cdot k_3) (t_1 \cdot t_3)(t_1 \cdot l_3)
+ \kappa^2  \frac {(t_1 \cdot k_4)^2(t_3 \cdot k_2)(l_3 \cdot k_2)}{(k_2-k_3)+m^2}
+\cO\left( k^{-1} \right)
\nonumber \\ && 
=
- \kappa^2 \frac{(k_1^2+ 2k_1\cdot k_2) (t_1\cdot t_3)}{8 k_1 \cdot k_3}\left[
2 (t_1 \cdot k_4)(l_3 \cdot k_2)-(k_1^2 -2 k_1 \cdot k_4)(t_1 \cdot l_3)
\right]
\nonumber \\ && 
- \kappa^2 \frac{(t_1 \cdot k_4)(t_3 \cdot k_2)}{4(k_1 \cdot k_3)(k_2 \cdot k_3)} \left[
(k_1^2+2k_1 \cdot k_2)(t_1 \cdot l_3)(k_2 \cdot k_3)
+2(t_1 \cdot k_4)(l_3 \cdot k_2) \left(k_1 \cdot k_3+k_2 \cdot k_3 \right)
\right]
\nonumber \\ && 
- \kappa^2 \frac{k_3^2(t_1 \cdot k_4)^2 (t_3 \cdot k_2)(l_3 \cdot k_2)}{4(k_2 \cdot k_3)^2}
+\cO\left( k^{-1} \right)
\nonumber \\ && 
= \kappa^2  m_3 k \frac{2-\cos \theta}{2(1-\cos \theta)} \sin \theta
+\cO\left( k^{-1} \right).
\end{eqnarray}

\subsection*{{\bf E.6}\hspace{5mm}$I^{(1,e)}\phi \, \to I^{(1,e)}\phi$} 

The calculation in this case is complicated. Let us see the detail. 
$e_{1\mu\nu}$ and $e_{3\mu\nu}$ satisfy
\begin{eqnarray}
&&k_2\cdot \e_1 \cdot k_2 = k_4\cdot \e_3 \cdot k_4 =0 , \qquad 
k_2\cdot \e_1 \cdot k_4 = \frac1{\sqrt{2}} (t_1 \cdot k_4)(l_1 \cdot k_2) , \nonumber \\
&&
k_4\cdot \e_1 \cdot k_4 = \sqrt{2} (t_1 \cdot k_4)(l_1 \cdot k_4),    \quad
k_2\cdot \e_3 \cdot k_2 = \sqrt{2} (t_3 \cdot k_2)(l_3 \cdot k_2), \nonumber \\
&&
k_2\cdot \e_3 \cdot k_4 = \frac1{\sqrt{2}} (t_3 \cdot k_2)(l_3 \cdot k_4),  \quad
k_2\cdot \e_1 \cdot \e_3 \cdot k_4 = \frac12 (l_1 \cdot k_2) (l_3 \cdot k_4) (t_1 \cdot t_3) , \quad  \\
&&k_4\cdot \e_1 \cdot \e_3 \cdot k_2 = \frac12\Bigl[ (t_1 \cdot k_4) (t_3 \cdot k_2) (l_1 \cdot l_3)
+(l_1 \cdot k_4) (l_3 \cdot k_2) (t_1 \cdot t_3)\nonumber \\
&&\qquad\qquad\qquad\qquad\qquad
+(t_1 \cdot k_4) (l_3 \cdot k_2) (l_1 \cdot t_3)
+(l_1 \cdot k_4) (t_3 \cdot k_2) (t_1 \cdot l_3) \Bigr] , \nonumber \\
&&\Tr[\e_1 \cdot \e_3] =  (t_1 \cdot t_3)(l_1 \cdot l_3)+(t_1 \cdot l_3)(l_1 \cdot t_3) .  \nonumber  
\end{eqnarray}
Since now the masses of the initial and final are the same, the inner products of vectors become
\begin{eqnarray}
&& l_1\cdot l_3 = \frac{1}{m_I^2} \left(- (k^2+m_I^2) (1- \cos \theta)+ m_I^2 \right), \nonumber \\
&&l_1 \cdot k_2=l_3 \cdot k_4= -\frac{k}{m_I} \left( \sqrt{k^2+m_S^2}+\sqrt{k^2+m_I^2} \right), \nonumber \\
&&l_1 \cdot k_4=l_3 \cdot k_2= -\frac{k}{m_I} \left( \sqrt{k^2+m_S^2} +\sqrt{k^2+m_I^2} \cos \theta\right) , \nonumber \\
&&= -\frac{k}{m_I} \left( \sqrt{k^2+m_S^2} +\sqrt{k^2+m_I^2}-\sqrt{k^2+m_I^2} (1- \cos \theta)\right) ,
\nonumber \\
&&t_1 \cdot k_3=t_3 \cdot k_2=-t_1 \cdot k_4=-t_3 \cdot k_1= k \sin \theta ,  \\
&&t_1 \cdot l_3=-t_3 \cdot l_1= \frac{1}{m_I} \sqrt{k^2+m_I^2} \sin \theta ,  \nonumber \\
&&k_1 \cdot k_2 = - \sqrt{k^2+m_I^2} \left(\sqrt{k^2+m_S^2}+\sqrt{k^2+m_I^2} \right) + m_I^2 \nonumber \\
&&k_1 \cdot k_4 = - \sqrt{k^2+m_I^2} \left(\sqrt{k^2+m_S^2}+\sqrt{k^2+m_I^2} \cos \theta \right) + m_I^2 \cos \theta \nonumber  \\
&&= - \sqrt{k^2+m_I^2} \left(\sqrt{k^2+m_S^2}+\sqrt{k^2+m_I^2}\right) +(k^2+m_I^2)(1- \cos \theta)  + m_I^2 \cos \theta.
\nonumber
\end{eqnarray}
Then, we have
\begin{eqnarray}
&&\left(k_2\cdot \e_1 \cdot k_4\right)\left( k_2\cdot \e_3 \cdot k_4\right) 
= -\frac{k^4}{2m_I^2} \sin^2 \theta \left( \sqrt{k^2+m_S^2}+\sqrt{k^2+m_I^2}\right)^2  , \nonumber \\
&&
\left(k_4\cdot \e_1 \cdot k_4 \right)\left(k_2\cdot \e_3 \cdot k_2\right)
= -\frac{2k^4}{m_I^2} \sin^2 \theta \left( \sqrt{k^2+m_S^2}+\sqrt{k^2+m_I^2} \cos \theta\right)^2  , \nonumber \\
&&
k_2\cdot \e_1 \cdot \e_3 \cdot k_4 = \frac{k^2}{2m_I^2} \left( \sqrt{k^2+m_S^2}+\sqrt{k^2+m_I^2}\right)^2 \cos \theta ,  \\
&&k_4\cdot \e_1 \cdot \e_3 \cdot k_2 =  \frac{k^2}{2m_I^2}\Biggl[
\left(\sqrt{k^2+m_S^2}+\sqrt{k^2+m_I^2}\right)^2 \cos \theta 
\nonumber \\
&&\qquad\qquad\qquad\qquad\qquad
- 2\sqrt{k^2+m_I^2} \left(\sqrt{k^2+m_S^2}+\sqrt{k^2+m_I^2}\right) (1+\cos \theta-2\cos^2 \theta ) 
\nonumber \\
&&\qquad\qquad\qquad\qquad\qquad
+(k^2+m_I^2)(1-\cos \theta)(3+\cos \theta-4\cos^2 \theta) -m_I^2 \sin^2\theta \Biggr] ,
\nonumber \\
&&\Tr[\e_1 \cdot \e_3] =  \frac{1}{m_I^2} \left( -(k^2+m_I^2) ( 1+\cos \theta-2\cos^2 \theta) + m_I^2\cos \theta\right) .  \nonumber  
\end{eqnarray}
The inverse of $(k_1^2 -k_1\cdot k_4)$ is estimated as
\begin{eqnarray}
&&(k_1^2 - 2k_1\cdot k_4)^{-1} \nonumber \\
&&\qquad
=\left(2 \sqrt{k^2+m_I^2} \left(\sqrt{k^2+m_S^2}+\sqrt{k^2+m_I^2} \cos \theta \right) - m_I^2 (2\cos \theta+1) \right)^{-1}
\nonumber \\
&&\qquad
=\left(2 \sqrt{k^2+m_I^2} \left(\sqrt{k^2+m_S^2}+\sqrt{k^2+m_I^2} \cos \theta \right) \right)^{-1}
\nonumber \\
&&\qquad\qquad \times
\left(1 - \frac{m_I^2 (2\cos \theta+1) }{2 \sqrt{k^2+m_I^2} \left(\sqrt{k^2+m_S^2}+\sqrt{k^2+m_I^2} \cos \theta \right)}\right)^{-1}
\nonumber \\
&&\qquad
=\left(2 \sqrt{k^2+m_I^2} \left(\sqrt{k^2+m_S^2}+\sqrt{k^2+m_I^2} \cos \theta \right) \right)^{-2}
\nonumber \\
&&\qquad\qquad \times
\Biggl[ 2 \sqrt{k^2+m_I^2} \left(\sqrt{k^2+m_S^2}+\sqrt{k^2+m_I^2} \cos \theta \right) + m_I^2 (2\cos \theta+1)
\nonumber \\
&&\qquad\qquad\qquad
+\frac{\left(m_I^2 (2\cos \theta+1) \right)^2 }{2k^2 (1+\cos \theta)}\Biggr] +\cO\left( k^{-8} \right),
\end{eqnarray}
With these equations, we can calculate the following quantity,
\begin{eqnarray}
&&
 -\frac14 (k_1^2-2k_1\cdot k_4 ) (k_1^2+2k_1\cdot k_2) \mbox{Tr} [e_1\cdot e_3]
+(k_1^2-2k_1\cdot k_4 ) k_2 \cdot e_1 \cdot e_3 \cdot k_4
\nonumber \\ && \qquad \qquad\qquad 
+ (k_1^2+2k_1\cdot k_2) k_4 \cdot e_1 \cdot e_3 \cdot k_2
-2 (k_2 \cdot e_1\cdot k_4 )( k_2 \cdot e_3 \cdot k_4)
\nonumber \\ && \qquad 
-\frac{ k_1^2-2k_1\cdot k_4 }{k_1^2+2k_1\cdot k_2} (k_2 \cdot e_1\cdot k_2 )( k_4 \cdot e_3 \cdot k_4)
-\frac{k_1^2+2k_1\cdot k_2}{ k_1^2-2k_1\cdot k_4 } (k_4 \cdot e_1\cdot k_4 )( k_2 \cdot e_3 \cdot k_2)
\nonumber\\
&&=
-\frac{3}{4} (1-\cos \theta) m_I^2 k^2+\cO\left( k^0 \right) .
\end{eqnarray}
where cancelation occurs in the leading and subleading order. 
We also have
\begin{eqnarray}
&&\frac18 (k_1^2+k_3^2+4m^2) \mbox{Tr} [e_1\cdot e_3]
-k_2 \cdot e_1 \cdot e_3 \cdot k_4
- k_4 \cdot e_1 \cdot e_3 \cdot k_2
\nonumber \\ &&  \qquad
+\frac{ 2 }{k_1^2+2k_1\cdot k_2} (k_2 \cdot e_1\cdot k_2)( k_4 \cdot e_3 \cdot k_4)
+\frac{2}{ k_1^2-2k_1\cdot k_4 } (k_4 \cdot e_1\cdot k_4 )( k_2 \cdot e_3 \cdot k_2) \nonumber \\
&&=-\frac{ k^4}{2m_I^2} \left(2^2 +(1+\cos\theta)^2-2(1-\cos\theta)^2\right) + \cO\left(k^2\right).
\end{eqnarray}
From these equations, the total amplitude is estimated as
\begin{eqnarray}
{\A}_s +{\A}_t +{\A}_u +{\A}_d 
=
-\frac{\kappa^2 m_I^2}{8} \frac{2^2 +(1+\cos\theta)^2 +(1-\cos\theta)^2 }{(1-\cos \theta)^2}+ 
\cO\left( k^{-2} \right).
\end{eqnarray}

\subsection*{{\bf E.7}\hspace{5mm}$h^{(2,e)}\phi \, \to I^{(0)}\phi$} 

In this case, the graviton of the final state is the helicity-0 mode. Its basis is written as
\begin{eqnarray}
\e_{3\mu\nu}^{(0)} &=& \frac2{\sqrt {6}} l_{3\mu}l_{3\nu} - \frac1{\sqrt{6}} \left( t_{3\mu}t_{3\nu}+ u_{\mu}u_{\nu} \right) 
\nonumber \\ 
&=& 
-\frac{2}{\sqrt{6}k_3^2} k_{3\mu}k_{3\nu} + \frac{1}{\sqrt{6}} \left( 2\eta_{\mu\nu} -3 t_{3\mu}t_{3\nu}-3 u_{\mu}u_{\nu} \right) .
\end{eqnarray}
With this expression, we have
\begin{eqnarray}
&&k_2\cdot \e_3 \cdot k_2 =-\frac{2}{\sqrt{6}k_3^2} (k_2\cdot k_3)^2 + 
\frac{1}{\sqrt{6}} \left( 2k_2^2 -3 (k_2 \cdot t_3)^2 \right),   \\
&&k_2\cdot \e_3 \cdot k_4 =-\frac{2}{\sqrt{6}k_3^2} (k_2\cdot k_3)(k_3\cdot k_4) + 
\frac{2}{\sqrt{6}}  (k_2\cdot k_4),  \\
&&
k_4\cdot \e_3 \cdot k_4 =-\frac{2}{\sqrt{6}k_3^2} (k_3\cdot k_4)^2+ 
\frac{2}{\sqrt{6}} k_4^2,   \\
&&\Tr[\e_1 \cdot \e_3] = 
-\frac{2}{\sqrt{6}k_3^2} (k_2\cdot \e_1 \cdot k_2+k_4\cdot \e_1 \cdot k_4-2k_2\cdot \e_1 \cdot k_4) 
\nonumber \\ && \qquad \qquad \qquad \qquad
- \frac{3}{\sqrt{6}} \left( t_3\cdot \e_1 \cdot t_3+u\cdot \e_1 \cdot u \right), \\
&&
k_2\cdot \e_1 \cdot \e_3 \cdot k_4 =
-\frac{2}{\sqrt{6}k_3^2} (k_3\cdot k_4) (k_2\cdot \e_1 \cdot k_2- k_2\cdot \e_1 \cdot k_4)
+ \frac{2}{\sqrt{6}} ( k_2\cdot \e_1 \cdot k_4),   \\
&&
k_4\cdot \e_1 \cdot \e_3 \cdot k_2
=
-\frac{2}{\sqrt{6}k_3^2} (k_2\cdot k_3) (k_2\cdot \e_1 \cdot k_4- k_4\cdot \e_1 \cdot k_4)
\nonumber \\ && \qquad \qquad \qquad \qquad
+ \frac{1}{\sqrt{6}}\left(2 ( k_2\cdot \e_1 \cdot k_4)-3 ( t_3\cdot \e_1 \cdot k_4)( t_3\cdot k_2) \right).  
\end{eqnarray}
Substituting the above equations into the form of the total amplitude \eq{sum1} and using 
\begin{eqnarray}
&&k_2 \cdot k_3 = k_1 \cdot k_4 - \frac12 \left( k_1^2-k_3^2\right), \qquad
k_3 \cdot k_4 = k_1 \cdot k_2 + \frac12 \left( k_1^2-k_3^2\right),  \\
&&k_2 \cdot k_4 = k_1 \cdot k_3 - \frac12 \left( k_1^2+k_3^2\right) - m^2 ,  
\end{eqnarray}
we have
\begin{eqnarray}
&&\A_s+\A_t+\A_u+\A_d \nonumber \\
&&= \frac{ \kappa^2 }{\sqrt{6}( k_1 \cdot k_3)}\left[\frac{k_3^2}{2}( k_2\cdot \e_1 \cdot k_4)
+\frac{ k_3^2(k_1 \cdot k_4) }{4(k_1\cdot k_2)}(k_2 \cdot  \e_1  \cdot k_2)
\right.\nonumber \\ && \qquad \qquad \qquad \left.
+\frac{k_3^2(k_1\cdot k_2)}{4(k_1 \cdot k_4)} (k_4 \cdot \e_1 \cdot k_4) 
-\frac{k_1^2}{2 } ( k_2\cdot \e_1 \cdot k_4) 
\right]
\nonumber \\ &&  
+ \frac{ \kappa^2 }{\sqrt{6}}\frac1{2(k_1\cdot k_3)} \left[
3\frac{(k_1\cdot k_2)}{(k_1\cdot k_4)}(t_3\cdot k_2)^2( k_4 \cdot \e_1 \cdot k_4)
+6(k_1\cdot k_2)(t_3\cdot k_2)( t_3\cdot \e_1 \cdot k_4)
\right.\nonumber \\ && \qquad 
+3(k_1\cdot k_2)(k_1\cdot k_4)( t_3\cdot \e_1 \cdot t_3)
+3(k_1\cdot k_2)(k_1\cdot k_4)( u \cdot \e_1 \cdot u)
\nonumber \\ && \qquad 
+\frac32 \frac{k_1^2 \left((k_1\cdot k_2)+(k_1\cdot k_4)\right)}{(k_1\cdot k_4)^2}(t_3\cdot k_2)^2( k_4 \cdot \e_1 \cdot k_4)
+3k_1^2(t_3\cdot k_2)( t_3\cdot \e_1 \cdot k_4)
\nonumber \\ && \qquad 
-\frac32k_1^2(k_1\cdot k_3)( t_3\cdot \e_1 \cdot t_3) 
-(2k_1^2+2k_3^2+4m^2)( k_2\cdot \e_1 \cdot k_4)
\nonumber \\ && \qquad \left.
+2m^2\frac{k_1\cdot k_4}{k_1\cdot k_2}( k_2\cdot \e_1 \cdot k_2)
+2m^2\frac{k_1\cdot k_2}{k_1\cdot k_4}( k_4\cdot \e_1 \cdot k_4)
 \right]
+ (\mbox{lower order of $k$})
\nonumber \\
&&=  \kappa^2 \frac{3(k_1\cdot k_2)}{2\sqrt{6}( k_1 \cdot k_3)}\Biggl[\frac{(t_3\cdot k_2)^2}{(k_1\cdot k_4)}( k_4 \cdot \e_1 \cdot k_4)
+2(t_3\cdot k_2)( t_3\cdot \e_1 \cdot k_4)
\nonumber \\ && \qquad \qquad \qquad \qquad \qquad 
+(k_1\cdot k_4)( t_3\cdot \e_1 \cdot t_3)
+(k_1\cdot k_4)( u \cdot \e_1 \cdot u)\Biggr]
\nonumber \\
&&
+ \kappa^2 \frac{3k_1^2}{4\sqrt{6}( k_1 \cdot k_3)}
\Biggl[\frac{\left((k_1\cdot k_2)+(k_1\cdot k_4)\right)}{(k_1\cdot k_4)^2}(t_3\cdot k_2)^2( k_4 \cdot \e_1 \cdot k_4)
+2(t_3\cdot k_2)( t_3\cdot \e_1 \cdot k_4)
\nonumber \\ && \qquad \qquad \qquad \qquad \qquad 
-(k_1\cdot k_3)( t_3\cdot \e_1 \cdot t_3) 
-2( k_2\cdot \e_1 \cdot k_4)\Biggr]
\nonumber \\
&&
+ \kappa^2 \frac{3k_3^2}{4\sqrt{6}( k_1 \cdot k_3)}
\Biggl[-2( k_2\cdot \e_1 \cdot k_4)
+\frac{ (k_1 \cdot k_4) }{(k_1\cdot k_2)}(k_2 \cdot  \e_1  \cdot k_2)
+\frac{(k_1\cdot k_2)}{(k_1 \cdot k_4)} (k_4 \cdot \e_1 \cdot k_4)\Biggr]
\nonumber \\
&&
+ \kappa^2 \frac{m^2}{\sqrt{6}( k_1 \cdot k_3)}
\Biggl[-2( k_2\cdot \e_1 \cdot k_4)
+\frac{k_1\cdot k_4}{k_1\cdot k_2}( k_2\cdot \e_1 \cdot k_2)
+\frac{k_1\cdot k_2}{k_1\cdot k_4}( k_4\cdot \e_1 \cdot k_4)
\Biggr]
\nonumber \\
&&
+ (\mbox{lower order of $k$}).
\label{e0}
\end{eqnarray}
Note that we do not fix the basis of the in-state graviton in the above equation. 
Therefore, this equation is applicable to the cases for $h^{(1,e)}\phi \, \to h^{(0)}\phi$ and for $h^{(0)}\phi \, \to h^{(0)}\phi$.

Substituting the basis of the in-state graviton
\begin{eqnarray}
\e_{1\mu\nu}^{(2,e)} =&& \frac1{\sqrt {2}} \left( t_{1\mu}t_{1\nu}- u_{\mu}u_{\nu} \right), 
\end{eqnarray}
into Eq.(\ref{e0}), we have
\begin{eqnarray}
&&\A_s+\A_t+\A_u+\A_d 
\nonumber \\
&&= \kappa^2  \frac{3(k_1\cdot k_2)}{4\sqrt{3}( k_1 \cdot k_3)}\left[\frac{(t_3\cdot k_2)^2}{(k_1\cdot k_4)}( k_4 \cdot t_1)^2
+2(t_3\cdot k_2)( t_3\cdot t_1)(t_1 \cdot k_4)
+(k_1\cdot k_4)( t_3\cdot t_1)^2
-(k_1\cdot k_4)\right]
\nonumber \\
&& \qquad \qquad \qquad \qquad \qquad \qquad
+ \cO(k^0)
\nonumber \\
&&= \kappa^2 \frac{3k^2(k_1\cdot k_2)}{4\sqrt{3}( k_1 \cdot k_3)}\left[\frac{\sin^2 \theta}{-(1+\cos \theta)}\sin^2\theta
+2\sin\theta\,\cos \theta(-\sin \theta)
-(1+\cos \theta)\cos^2 \theta
+(1+\cos \theta)\right]
\nonumber \\
&& \qquad \qquad \qquad \qquad \qquad \qquad
+ \cO(k^0)
\nonumber \\
&&= \cO(k^0).
\end{eqnarray}

\subsection*{{\bf E.8}\hspace{5mm}$I^{(1,e)}\phi \, \to I^{(0)}\phi$} 

Substituting  the basis of the in-state graviton
\begin{eqnarray}
&&\e_{1\mu\nu}^{(1,e)} =\frac1{\sqrt {2}} \left( l_{1\mu}t_{1\nu}+ t_{1\mu}l_{1\nu} \right), 
\end{eqnarray}
into Eq.(\ref{e0}), we have

\begin{eqnarray}
&&\A_s+\A_t+\A_u+\A_d 
\nonumber \\
&&=  \kappa^2 \frac{3(k_1\cdot k_2)}{4\sqrt{3}( k_1 \cdot k_3)}\Biggl[2\frac{(t_3\cdot k_2)^2}{(k_1\cdot k_4)}( k_4 \cdot t_1 )(l_1 \cdot k_4)
+2(t_3\cdot k_2)( t_3\cdot t_1 )(l_1 \cdot k_4)
\nonumber \\ && \qquad \qquad \qquad \qquad \qquad 
+2(t_3\cdot k_2)( t_3\cdot l_1 )(t_1 \cdot k_4)
+2(k_1\cdot k_4)( t_3\cdot t_1 )(l_1 \cdot t_3)
\Biggr]
\nonumber \\
&&
+ \kappa^2 \frac{3k_1^2}{8\sqrt{3}( k_1 \cdot k_3)}
\Biggl[2\frac{\left((k_1\cdot k_2)+(k_1\cdot k_4)\right)}{(k_1\cdot k_4)^2}(t_3\cdot k_2)^2( k_4 \cdot t_1 )(l_1 \cdot k_4)
+2(t_3\cdot k_2)( t_3\cdot t_1 )(l_1 \cdot k_4)
\nonumber \\ && \qquad \qquad \qquad 
+2(t_3\cdot k_2)( t_3\cdot l_1 )(t_1 \cdot k_4)
-2(k_1\cdot k_3)( t_3\cdot t_1 )(l_1 \cdot t_3) 
-2( k_2\cdot l_1 )(t_1 \cdot k_4)\Biggr]
\nonumber \\
&&
+ \kappa^2 \frac{3k_3^2}{8\sqrt{3}( k_1 \cdot k_3)}
\Biggl[-2( k_2\cdot l_1 )(t_1 \cdot k_4)
+2\frac{(k_1\cdot k_2)}{(k_1 \cdot k_4)} (k_4 \cdot  t_1 )(l_1  \cdot k_4)\Biggr]
\nonumber \\
&&
+ \kappa^2 \frac{m^2}{2\sqrt{3}( k_1 \cdot k_3)}
\Biggl[-2( k_2\cdot  l_1 )(t_1  \cdot k_4)
+2\frac{k_1\cdot k_2}{k_1\cdot k_4}( k_4\cdot  t_1 )(l_1  \cdot k_4)
\Biggr]
+ \cO(k^0)
\nonumber \\
&&= 
 \kappa^2 \frac{3(k_1\cdot k_2)}{4\sqrt{3}( k_1 \cdot k_3)}\bigl[2k m_1 \sin \theta \bigr]
+ \kappa^2 \frac{3k_1^2}{8\sqrt{3}( k_1 \cdot k_3)}
\left[-4 \frac{k^3}{m_1} \sin \theta \right]
+ \cO(k^0)
\nonumber \\
&&= 
 \cO(k^0)
\end{eqnarray}

\subsection*{{\bf E.9}\hspace{5mm}$I^{(0)}\phi \, \to I^{(0)}\phi$} 

In this case, we can also use Eq.(\ref{e0}). 
As a preparation, we show the contraction of the basis of the in-state graviton
\begin{eqnarray}
\e_{1\mu\nu}^{(0)} &=& \frac2{\sqrt {6}} l_{1\mu}l_{1\nu} - \frac1{\sqrt{6}} \left( t_{1\mu}t_{1\nu}+ u_{\mu}u_{\nu} \right) 
\nonumber \\
&=&
-\frac{2}{\sqrt{6}k_1^2} k_{1\mu}k_{1\nu} + \frac{1}{\sqrt{6}} \left( 2\eta_{\mu\nu} -3 t_{1\mu}t_{1\nu}-3 u_{\mu}u_{\nu} \right)
\end{eqnarray}
with vectors,
\begin{eqnarray}
&&k_2\cdot \e_1 \cdot k_2 = - \frac2{\sqrt {6}} \frac{(k_1 \cdot k_2)^2}{k_1^2}+ \frac2{\sqrt{6}}k_2^2, \qquad  
k_2\cdot \e_1 \cdot k_4 = -\frac2{\sqrt {6}}\frac{ (k_1 \cdot k_2)(k_1 \cdot k_4)}{k_1^2} + \frac2{\sqrt{6}}k_2 \cdot k_4 ,  \nonumber\\
&&
k_4\cdot \e_1 \cdot k_4 = -\frac2{\sqrt {6}}\frac{(k_1 \cdot k_4)^2}{k_1^2}+ \frac2{\sqrt{6}} k_4^2-\frac3{\sqrt{6}}(k_4 \cdot t_1)^2, 
\nonumber\\
&&
t_3 \cdot \e_1 \cdot t_3  = -\frac2{\sqrt {6}}\frac{(k_1 \cdot t_3)^2}{k_1^2}+ \frac2{\sqrt{6}} -\frac3{\sqrt{6}}(t_1 \cdot t_3)^2,
\\
&&
t_3 \cdot \e_1 \cdot k_4 = -\frac2{\sqrt {6}}\frac{(k_1 \cdot t_3 )(k_1 \cdot k_4)}{k_1^2}-\frac3{\sqrt{6}}(t_1 \cdot t_3)(k_4 \cdot t_1), \qquad 
u \cdot \e_1 \cdot u = -\frac1{\sqrt {6}} \nonumber 
\end{eqnarray}
With these equations, the total amplitude is calculated as
\begin{eqnarray}
&&\A_s+\A_t+\A_u+\A_d \nonumber \\
&&= - \kappa^2 \frac{(k_1\cdot k_2)}{2 k_1^2( k_1 \cdot k_3)}\left[\frac{(t_3\cdot k_2)^2}{(k_1\cdot k_4)}(k_1 \cdot k_4)^2
+2(t_3\cdot k_2)(k_1 \cdot t_3 )(k_1 \cdot k_4)
+(k_1\cdot k_4)(k_1 \cdot t_3)^2\right]
\nonumber \\
&&
+ \kappa^2 \frac{(k_1\cdot k_2)}{4( k_1 \cdot k_3)}\Biggl[\frac{(t_3\cdot k_2)^2}{(k_1\cdot k_4)}( -3)(k_4 \cdot t_1)^2
+2(t_3\cdot k_2)(-3)(t_1 \cdot t_3)(k_4 \cdot t_1)
\nonumber \\ && \qquad \qquad \qquad \qquad \qquad 
+(k_1\cdot k_4)\left(2 -3 (t_1 \cdot t_3)^2\right)
+(k_1\cdot k_4)(-1)\Biggr]
\nonumber \\
&&
-\frac{ \kappa^2 }{4 ( k_1 \cdot k_3)}
\Biggl[\frac{\left((k_1\cdot k_2)+(k_1\cdot k_4)\right)}{(k_1\cdot k_4)^2}(t_3\cdot k_2)^2(k_1 \cdot k_4)^2
+2(t_3\cdot k_2)(k_1 \cdot t_3 )(k_1 \cdot k_4)
\nonumber \\ && \qquad \qquad \qquad \qquad \qquad 
-(k_1\cdot k_3)(k_1 \cdot t_3)^2
-2(k_1 \cdot k_2)(k_1 \cdot k_4)\Biggr]
\nonumber \\
&&
- \kappa^2 \frac{k_3^2}{4( k_1 \cdot k_3)}
\left[-2(k_1 \cdot k_2)(k_1 \cdot k_4)
+\frac{ (k_1 \cdot k_4) }{(k_1\cdot k_2)}(k_1 \cdot k_2)^2
+\frac{(k_1\cdot k_2)}{(k_1 \cdot k_4)} (k_1 \cdot k_4)^2\right]
\nonumber \\
&&
- \kappa^2 \frac{m^2}{3( k_1 \cdot k_3)}
\left[-2(k_1 \cdot k_2)(k_1 \cdot k_4)
+\frac{k_1\cdot k_4}{k_1\cdot k_2}(k_1 \cdot k_2)^2
+\frac{k_1\cdot k_2}{k_1\cdot k_4}(k_1 \cdot k_4)^2
\right]
+ \cO(k^0)
\nonumber \\
&&
=- \kappa^2 
\frac{3(k_1\cdot k_2)}{4( k_1 \cdot k_3)}\left[
\frac{1}{(k_1\cdot k_4)}
\bigl( (t_3\cdot k_2)(k_4 \cdot t_1)+(k_1\cdot k_4)(t_1 \cdot t_3)\bigr)^2
-(k_1\cdot k_4)\right]
\nonumber \\
&& \qquad
- \kappa^2 \frac{(t_3\cdot k_2)^2}{4 ( k_1 \cdot k_3)}
\left[\big((k_1\cdot k_2)+(k_1\cdot k_4)\big)
-(k_1 \cdot k_4)
-(k_1\cdot k_3)
\right]
+ \cO(k^0)
\nonumber \\
&&
= \cO(k^0). 
\end{eqnarray}

\subsection*{{\bf E.10}\hspace{5mm}$I^{(2,e)}\phi \, \to I^{(s)}\phi$} 

Since the basis $e_{1\mu\nu}^{(2,e)}$ for the initial graviton does not involve $l_\mu$, 
it does not affect  the order of $k$. 
Then, we find from eq.~\eq{sum2},  
\begin{eqnarray}
\A_s+\A_t+\A_u+\A_d 
= \cO(k^0). 
\end{eqnarray}

\subsection*{{\bf E.11}\hspace{5mm}$I^{(1,e)}\phi \, \to I^{(s)}\phi$} 

In this case, the basis of the initial graviton satisfies
\begin{eqnarray}
&&k_2 \cdot e_1 \cdot k_2 = 0,  \quad 
k_4 \cdot e_1 \cdot k_4 = \sqrt{2} \left(t_1 \cdot k_4\right)  \left(l_1 \cdot k_4\right),  \nonumber \\
&&k_3 \cdot e_1 \cdot k_3 = - \sqrt{2} \left(t_1 \cdot k_4\right)  \left(l_1 \cdot k_3\right).
\end{eqnarray}
Substitute these equations into eq.~\eq{sum2}, we have
\begin{eqnarray}
\A_s+\A_t+\A_u+\A_d 
&=& \frac{\sqrt{6}\kappa^2}{12} \left( m_S^2+2m^2+12\xi m_S^2 \right)  \left(t_1 \cdot k_4\right)  \left(\frac{ l_1 \cdot k_4} { k_1 \cdot k_4}
-\frac{ l_1 \cdot k_3} { k_1 \cdot k_3}\right) + \cO\left( k^{-1} \right)\nonumber \\
&=&
 \cO\left( k^{-1} \right) .
\end{eqnarray}

\subsection*{{\bf E.12}\hspace{5mm}$I^{(0,e)}\phi \, \to I^{(s)}\phi$}

\begin{eqnarray}
&&k_2 \cdot e_1 \cdot k_2 = -\frac{\sqrt{6}}{3}\frac{(k_1 \cdot k_2)^2}{k_1^2} + \frac{\sqrt{6}}{3}k_2^2  \nonumber \\
&&k_4 \cdot e_1 \cdot k_4 =  -\frac{\sqrt{6}}{3}\frac{(k_1 \cdot k_4)^2}{k_1^2} + \frac{\sqrt{6}}{3}k_4^2 -\frac{\sqrt{6}}{2} (k_4 \cdot t_1)^2,  \nonumber \\
&&k_3 \cdot e_1 \cdot k_3 = -\frac{\sqrt{6}}{3}\frac{(k_1 \cdot k_3)^2}{k_1^2} + \frac{\sqrt{6}}{3}k_3^2 -\frac{\sqrt{6}}{2} (k_4 \cdot t_1)^2.
\end{eqnarray}
Substitute these into eq.~\eq{sum2}, the total amplitude can be estimated as 
\begin{eqnarray}
&&\A_s+\A_t+\A_u+\A_d \nonumber \\
&&\qquad= \frac{\sqrt{2}\kappa^2}{6k_1^2} \left( m_S^2+2m^2+12\xi m_S^2 \right)    
\left(\frac{ (k_1 \cdot k_2)^2} { k_1 \cdot k_2}-\frac{ (k_1 \cdot k_4)^2} { k_1 \cdot k_4}-\frac{ (k_1 \cdot k_3)^2} { k_1 \cdot k_3}\right) + \cO\left( k^{0} \right)\nonumber \\
&&\qquad=
 \cO\left( k^{0} \right) .
\end{eqnarray}

\subsection*{{\bf E.13}\hspace{5mm}$I^{(s)}\phi \, \to I^{(s)}\phi$} 

The UV behavior of this case can be obtained by taking the UV limit of eq.~\eq{sum3} directly,
which gives eq.~\eq{IsIs}.

\end{document}